\documentclass[aps,prd,nofootinbib,showkeys,preprint,floatfix]{revtex4}

\usepackage{amssymb}
\usepackage{amsmath}
\usepackage{amsfonts}
\usepackage{graphicx}
\usepackage{color}
\usepackage{xspace}
\usepackage{ulem}
\usepackage{lscape}
\usepackage{ wasysym }

\usepackage{multirow,rotating}

\def\gsim{\raise0.3ex\hbox{$\;>$\kern-0.75em\raise-1.1ex\hbox{$\sim\;$}}}
\def\lsim{\raise0.3ex\hbox{$\;<$\kern-0.75em\raise-1.1ex\hbox{$\sim\;$}}}

\def\pslash{p \hspace{-0.5em}/\;\:}

\def\znbb{0\nu\beta\beta}
\def\meff{\langle m_{\nu} \rangle}
\def\rpv{R_P \hspace{-1.2em}/\;\:}

\newcommand{\ba}[1]{\begin{eqnarray} \label{(#1)}}
\newcommand{\ea}{\end{eqnarray}}

\newcommand{\AddrAHEP}{
  {\it AHEP Group, Instituto de F\'{\i}sica Corpuscular --
    C.S.I.C./Universitat de Val{\`e}ncia \\
    Edificio de Institutos de Paterna, Apartado 22085,
  E--46071 Val{\`e}ncia, Spain}}

\newcommand{\AddrUFSM}{
Universidad T\'ecnica Federico Santa Mar\'\i a, \\ 
Centro-Cient\'\i fico-Tecnol\'{o}gico de Valpara\'\i so, \\ 
Casilla 110-V, Valpara\'\i so,  Chile}

\newcommand{\AddrSaitama}{
Department of Physics, Saitama University, \\ 
Shimo-Okubo 255, 338-8570 Saitama-Sakura, Japan}

\newcommand{\AddrPuc}{
Departamento de F\'isica, Pontif\'icia Universidade Cat\'olica 
do Rio de Janeiro,\\
Rua Marqu\^es de S\~ao Vicente 225, 22451-900 G\'avea, Rio de Janeiro, Brazil}

\def\gsim{\raise0.3ex\hbox{$\;>$\kern-0.75em\raise-1.1ex\hbox{$\sim\;$}}}
\def\lsim{\raise0.3ex\hbox{$\;<$\kern-0.75em\raise-1.1ex\hbox{$\sim\;$}}}

\begin{document}

\preprint{IFIC/15-02, STUPP-2014-221}

\title{Double beta decay and neutrino mass models}

\author{J.C. Helo} \email{juancarlos.helo@usm.cl}\affiliation{\AddrUFSM}
\author{M. Hirsch} \email{mahirsch@ific.uv.es}\affiliation{\AddrAHEP}
\author{T. Ota}\email{toshi@mail.saitama-u.ac.jp}\affiliation{\AddrSaitama}
\author{F.~A.~Pereira dos Santos}\email{fabio.alex@fis.puc-rio.br}\affiliation{\AddrPuc}



\begin{abstract}
Neutrinoless double beta decay allows to constrain lepton number
violating extensions of the standard model. If neutrinos are Majorana
particles, the mass mechanism will always contribute to the decay
rate, however, it is not a priori guaranteed to be the dominant
contribution in all models. Here, we discuss whether the mass
mechanism dominates or not from the theory point of view. We classify
all possible (scalar-mediated) short-range contributions to the decay
rate according to the loop level, at which the corresponding models
will generate Majorana neutrino masses, and discuss the expected
relative size of the different contributions to the decay rate in each
class.
  Our discussion is general for models based on the SM group 
but does not cover models with an extended gauge. 
We also work out the phenomenology of one concrete 2-loop
model in which both, mass mechanism and short-range diagram, might
lead to competitive contributions, in some detail.
\end{abstract}

\maketitle

\tableofcontents

\section{Introduction}

Experimental limits on half-lives of neutrinoless double beta decay
($0\nu\beta\beta$) give stringent bounds on many Lepton Number 
Violating (LNV) extensions of the Standard Model (SM); for a recent 
review see, for example \cite{Deppisch:2012nb}. Recent experimental 
results give limits for $^{76}$Ge \cite{Agostini:2013mzu} and $^{136}$Xe
\cite{Albert:2014awa,Shimizu:2014xxx,Gando:2012zm} in
excess of $10^{25}$ ys, which place an upper limit on the 
effective Majorana mass $\meff$ \footnote{$\meff$ is defined 
as $\meff = \sum_j U_{ej}^2 m_j$, where the sum runs over all 
light neutrinos. This is equivalent to the $(e,e)$ entry of the 
Majorana neutrino mass matrix in the basis where the charged 
lepton mass matrix is diagonal.} of the order of roughly 
$\meff \lsim (0.2-0.4)$ eV, depending on calculations of 
nuclear matrix element~\cite{Muto:1989cd,Faessler:2012ku,Menendez:2011zza}.

However, from the theoretical point of view it is not a priori clear,
whether the mass mechanism gives indeed the dominant contribution to
the double beta decay rate, and many models have been discussed in the
literature where this might not be the case. The classical example
appears in left-right (LR) symmetric extensions of the SM
\cite{Mohapatra:1980yp,Doi:1985dx} and also 
in R-parity violating ($\rpv$) supersymmetric theories with 
both trilinear $\rpv$~\cite{Mohapatra:1986su,Hirsch:1995zi} 
and bilinear $\rpv$~\cite{Faessler:1997db,Hirsch:1998kc} 
terms. 
Furthermore, leptoquark models~\cite{Hirsch:1996ye} and 
more recently models with colour octet scalars~\cite{Choubey:2012ux} 
or colour sextet diquarks~\cite{Brahmachari:2002xc,Gu:2011ak,Kohda:2012sr} 
have been discussed. 

Given that there is such a large list of possible lepton number
violating models, is it possible to determine which contribution to
the $\znbb$ decay rate is the dominant one? - Perhaps, if double 
beta decay were to be observed in the next round of experiments 
and either KATRIN \cite{Steinbrink:2013ska} or cosmological data 
\cite{Lesgourgues:2006nd,Hannestad:2010kz,Wong:2011ip} also find 
hints for a neutrino mass scale of the order of, say, somewhat larger 
than ${\cal O}(0.1)$ eV, one could claim on the basis of minimality 
that the mass mechanism $\langle m_{\nu} \rangle$ gives (at least) 
the most important contribution to the total decay rate. 
However, once upper limits on the total neutrino mass ($\sum m_{\nu}$) 
placed from cosmology drop below the level of ${\cal O}(0.1)$ eV, 
the question becomes exceedingly difficult to answer.

In that case, from the experimental point of view, there remain only a
few possibilities to make progress. For example, measurements of the
angular correlation between the two electrons from $\znbb$
\cite{Doi:1985dx,Ali:2007ec,Arnold:2010tu} or measuring double beta
plus decays \cite{Hirsch:1994es} offer the possibility to identify the
Lorentz structure (equivalently the chiralities of the emitted
electrons) of the LNV processes.  However, realistically the SuperNEMO
proposal \cite{Arnold:2010tu} could only accumulate the necessary
statistics to identify the Lorentz structure, if the half-live of
$^{82}$Se is below $10^{26}$ ys, while there yet exists no
experimental proposal with a sufficient sensitivity for
$0\nu\beta^+/EC$ decays to make use of the ideas discussed in
\cite{Hirsch:1994es}.

From the theoretical point of view, as discussed in
\cite{Pas:1999fc,Pas:2000vn}, the amplitude of $\znbb$ decay can be
divided into a long-range and a short-range part. Here, short-range
means that all particles appearing in the diagrams for double-beta
decay are heavier than the nuclear Fermi scale, i.e. ${\cal O}(0.1)$
GeV. Current limits from $\znbb$ decay then correspond to lower limits
on the effective scale $\Lambda_{\rm LNV}$ of lepton number violation,
\begin{equation}\label{eq:LNVeff}
\Lambda_{\rm LNV} \equiv 
\Big(\frac{m_S^4m_F}{g_{\rm eff}^4}\Big)^{1/5} \gsim 
(1-3) \hskip1mm {\rm TeV},
\end{equation}
where $g_{\rm eff}$ is some mean of the couplings appearing in the
diagram and $m_{S}$ and $m_{F}$ are the masses of the fields that
mediate the $\znbb$ process, see the next section for details.  This
mass scale is testable, at least in principle, at the LHC, and the
combination of future LHC limits (or a possible discovery, to express
it in a more optimistic way) and double beta decay data might allow to
test many, but maybe not all, of the possible short-range diagrams that
contribute to the decay rate \cite{Helo:2013dla,Helo:2013ika}.

In this paper we take a different approach and study the question,
whether the mass mechanism is dominant or not, from a purely
theoretical point of view.  As described above, current and next
generation $\znbb$ decay experiments will test LNV interactions at the
TeV scale. Such TeV-scale LNV interactions, on the other hand, appear
also in the context of radiative neutrino mass models. In other words,
a new physics (short-range) contribution to $\znbb$ decay will always
also produce a non-zero neutrino mass. In this paper we discuss the
relation between possible models for short-range contribution to
$\znbb$ decay and the neutrino mass-generation mechanism. Our study is
based on the complete list of possible decompositions of the $d=9$
(short-range) double beta decay operator given in
\cite{Bonnet:2012kh,Bonnet:2014kh}. The general decomposition list
given in \cite{Bonnet:2012kh,Bonnet:2014kh} is equivalent, in
principle, to defining all models which can give a contribution to
double beta decay, and the black box theorem, see below, guarantees
that all these models will produce Majorana neutrino masses {\em at or
  below} four-loop order.  Our approach therefore basically consists
in classifying all the possible models contributing to $\znbb$ decay
with respect to the loop level at which they will generate Majorana
neutrino masses. We can then discuss the expected size of the two
contributions to $\znbb$: (1) the $d=9$ short-range contribution
$\mathcal{O}_{d=9}$ and (2) the contribution from the neutrino mass
mechanism $\langle m_{\nu} \rangle$ which is radiatively induced in
the corresponding models and conclude model-by-model, which one of the
two is expected to dominate. Given that the list of
\cite{Bonnet:2012kh} is tree-level complete, our discussion is quite
general and covers actually models of neutrino mass from tree-level
models to 4-loop models.

Before entering into the details of our work, let us briefly comment 
on the well-known relation between short-range $\znbb$ contributions 
and neutrino masses, 
i.e., the black box theorem~\cite{Schechter:1981bd}.\footnote{%
In \cite{Hirsch:2006yk} an extension of the black box
  mechanism with flavour violation has been constructed.}
The theorem guarantees that, once $\znbb$ decay has been observed,
neutrinos are Majorana particles. However, the black box theorem does
not demonstrate that the mass mechanism dominates $\znbb$, since it
only guarantees neutrino masses at the level of four-loop.  Obviously,
a four-loop diagram, additionally suppressed by
$m_u^2m_d^2m_e^2/{\Lambda_{\rm LNV}^5}$, can produce only tiny
neutrino masses, which are many orders of magnitude below of what is
required to explain oscillation data \cite{Duerr:2011zd}.
Nevertheless, the black box theorem, together with the general
decomposition of the $d=9$ double beta decay operator published in
\cite{Bonnet:2012kh}, defines the basic idea of our current paper. Indeed 
we find that all ``models'' listed in \cite{Bonnet:2012kh} produce 
neutrino masses at or below the 4-loop order as demanded by the 
theorem.

\begin{figure}[t]
\hskip10mm\includegraphics[width=0.55\linewidth]{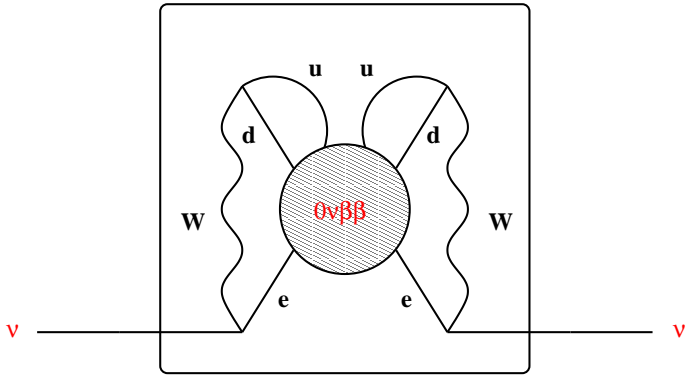}
\caption{Schematic explanation of the black box theorem: 
The theorem guarantees that, once $\znbb$ decay has been observed, 
Majorana neutrino masses will be generated, 
independent of the underlying model, at the latest at the four-loop level. 
By itself, this theorem {\em does not} guarantee that
the mass mechanism is the dominant contribution to $\znbb$ decay.}
\label{fig:blackbox}
\end{figure}

We comment that our work has some overlap with
\cite{Babu:2001ex,deGouvea:2007xp} and \cite{Angel:2012ug}. Babu \&
Leung\cite{Babu:2001ex} have written down all SM invariant $\Delta
L=2$ operators from dimension-5 ($d=5$, the well-known Weinberg
operator \cite{Weinberg:1979sa}) to $d=11$ and showed the relation
between the effective operators and neutrino masses on the basis of
black-box like loop diagrams.  The authors of
\cite{Babu:2001ex,deGouvea:2007xp} discuss then possible ultra-violet
completions for several example operators and give estimates for the
scales $\Lambda_{\rm LNV}$, for which those operators can explain
current neutrino data.  The authors of \cite{Angel:2012ug} provide a
systematic study of these operators, for one-loop and two-loop
neutrino mass models, and discuss also which of these could possibly
be tested at the LHC.  However, our discussion differs from these
papers in that we are mostly interested in double beta decay and its
relation to neutrino mass.

We mention also the work of \cite{delAguila:2012nu}, which pursues the
link between the short-range contribution to $\znbb$ and neutrino
masses, but takes a different approach from ours. The authors focus on
three types of LNV effective interactions which consist only of
leptons and Higgs doublets, and list the models in which those LNV
interactions simultaneously drive both the new physics contribution to
$\znbb$ and neutrino masses. The main difference between
\cite{delAguila:2012nu} 
and our work 
is that 
it is assumed in \cite{delAguila:2012nu} that 
new physics resides in the leptonic sector only. 

In our classification, we also rediscover several models discussed in
the literature previously, like for example leptoquark models
\cite{Hirsch:1996ye}, which can give 1-loop neutrino masses as
discussed in \cite{AristizabalSierra:2007nf} or 2-loop neutrino
masses, as in the model of \cite{Kohda:2012sr} or the one in
\cite{Babu:2011vb}. We do not cover, however, the possible
contributions from light sterile neutrinos. 
There exists a vast amount of
papers on this subject in the literature already
\cite{Blennow:2010th,Mitra:2011qr,LopezPavon:2012zg,Chakrabortty:2012mh,Huang:2013kma,Dev:2013vxa,Pascoli:2013fiz,Merle:2013ibc,Dev:2014xea,Li:2011ss,Girardi:2013zra,Meroni:2014tba,Faessler:2014kka}
and we have nothing new to add on this particular subject.

Since neutrino mass models must not only produce the correct absolute
values of neutrino mass, but also reproduce the observed flavour
structure of the neutrino mass matrix, one also has to pay attention
to constraints from flavour physics observables. In
\cite{Dudley:2008vg} the authors applied the hypothesis of ``minimal
flavour violation'' (MFV) to effective operators that contribute to
$\znbb$ and found that the MFV assumption constrains the effective
operators to be smaller than the detectable level. In this work, we
do not adopt any such theoretical assumption on the flavour structures of
the parameters in the models. Instead we simply consider bounds 
on lepton flavour violating observables as constraints on model 
parameters. We believe this to be the correct approach since any of 
the ``exotic'' contributions to $\znbb$ decay requires the introduction 
of new scalars, not present in the SM, with their own Yukawa interactions 
with SM fermions. Thus the whole concept of MFV is not very well 
funded in any of the models of interest for $\znbb$ decay.

A few disclaimers might also be in order here. Our analysis
concentrates on the true $d=9$ operator, i.e. it covers only the short
range part of the $\znbb$ amplitude. 
Our results thus do not cover, for example, the long-range 
diagrams of R-parity violating SUSY 
\cite{Babu:1995vh, Hirsch:1995cg} or leptoquark models 
\cite{Hirsch:1996ye}.
 Also, we limit ourselves to
scalar exchange, thus models with a coupling
  between new scalars and the SM gauge bosons, such as
\cite{Chen:2006vn,King:2014uha} are not considered. Also, this
restrictions implies that we do not cover models with an extended
gauge group either, especially we do not discuss models with
left-right symmetry.  And, finally, the list of decompositions in
\cite{Bonnet:2012kh} is complete only at tree-level. Thus, we do not
consider cases in which the neutrino mass is generated at some higher
loop level, while the $\znbb$ amplitude appears at one-loop order, as
for example in the recent papers
\cite{Gustafsson:2012vj,Gustafsson:2014vpa}. 

The rest of this paper is organised as follows.  
At the beginning of section \ref{sect:class}, as a preparation,
we will summarise the main results of \cite{Bonnet:2012kh} and discuss
some generalities useful for the latter parts of the paper.  We will
then discuss the classification of the different possible models and
estimate in each case the relative size of the contribution from the
mass mechanism and the short range part of the amplitude.  In section
\ref{sect:exa}, we will discuss one concrete two-loop model of
neutrino mass in more detail.  Section \ref{sect:cncl} summarises and
discusses our main findings. Tables with lists of the different
models, classified as described in section \ref{sect:class} are
deferred to the appendix.

\section{Setup and classification}
\label{sect:class}

In this section, we classify neutrino mass models based on the
decomposition of the $d=9$ $\znbb$ effective operators, according to
the number of loops in the resulting neutrino mass diagrams. In each
class, we will compare the size of the two contributions to the
$\znbb$: (i) $d=9$ operator itself, and (ii) the mass mechanism
$\langle m_{\nu} \rangle$ induced by the $d=9$ operator. 
This
classification therefore allows to identify those models, for 
which one can expect non-standard contributions (beyond the 
ordinary mass mechanism) to be important for $\znbb$ decay. 
Note, that $\langle m_{\nu}\rangle$, is equivalent to the e-e entry 
in the neutrino mass matrix, $(M_{\nu})_{ee}$, in the basis where the 
charged lepton mass matrix is diagonal. Thus, in this section we 
concentrate on the comparison of short-range contributions to the 
size of this entry in $(M_{\nu})$. Of course, a complete fit to all 
neutrino data will need to take into account also all other entries 
in $(M_{\nu})$. A specific example, how this can be done and the 
additional constraints from both, oscillation data and lepton 
flavour violating decays, is discussed in section 3. In the 
discussion in this section, we always keep generation indices 
in the unknown couplings of the different models. Other indices 
could be constrained combining double beta decay with, for 
example, oscillation data. A specific example for this is worked 
out in section 3.

\subsection{Generalities}

\begin{figure}[t]
\hskip-10mm\includegraphics[width=0.5\linewidth]{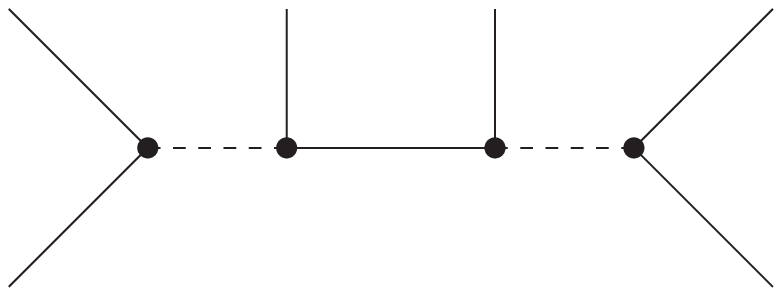}
\hskip10mm\includegraphics[width=0.4\linewidth]{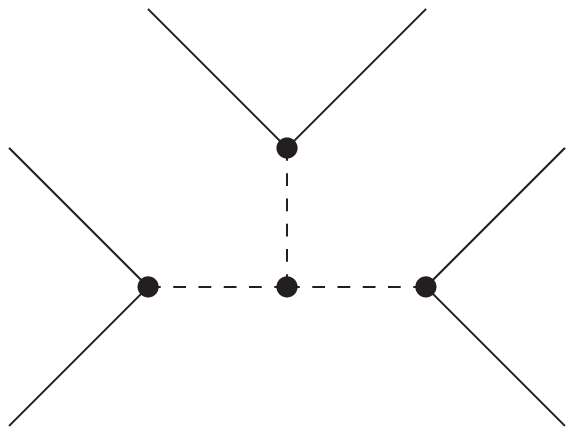}
\caption{The two possible tree-level topologies contributing to 
the $\znbb$ decay rate. The outer lines represent the six SM 
fermions, while for the virtual particles appearing in the inner 
lines we consider only scalar-fermion-scalar (left, topology-I) 
and scalar-scalar-scalar (right, topology-II).}
\label{fig:topos}
\end{figure}

The short-range double beta decay operator
\begin{equation}
\mathcal{O}_{d=9}
\propto 
\bar u \bar u  \, d d \, \bar e \bar e \, 
\label{eq:d9}
\end{equation}
can be generated at tree-level via only two topologies shown in
Fig.~\ref{fig:topos}.  The bosons (depicted with dashed lines in the
diagrams) in these topologies could be either scalars or vectors, but
we will consider only scalar exchange here. Assigning the outer
fermions with either the left (``$L$'') or the right (``$R$'') chirality
in all possible permutations allows to derive the complete list of
``decompositions'' (or proto-models) that can contribute to the
$\znbb$ decay amplitude at tree level~\cite{Bonnet:2012kh}.
The fermion propagator in topology-I contains two terms,
\begin{equation}
\frac{\pslash + m_{\psi}}{p^2 - m_{\psi}^2},
\end{equation}
but in the short-range part of the amplitude the first term is
suppressed relative to the second by a factor of 
$\pslash/m_{\psi} \simeq p_F/m_{\psi}$, where
$p_F$ is the typical Fermi momentum in the
nucleus and the mass $m_{\psi}$ is suppossed to be larger 
than ${\cal O}(100)$ GeV. 
Considering then only decompositions which pick the mass term from
the propagator results in a total of 135 possible decompositions for
Topology-I (T-I in the following), while there are 27 decompositions
in Topology-II (T-II), if we limit ourselves to scalar exchange
\cite{Bonnet:2012kh,Bonnet:2014kh}.
For tables showing the different 
decompositions see the appendix. 

Babu \& Leung \cite{Babu:2001ex} have listed $\Delta L=2$ operators
from $d=5$ to $d=11$. Among the $d=9$ operators in their list, the 
following five are relevant for double beta decay:
\begin{eqnarray}\label{eq:BL}
{\cal O}_{11} = 
 \bar{L}\bar{L}
 \bar{Q}d_R
 \bar{Q}d_R,
& 
{\cal O}_{12} = 
\bar{L}\bar{L} 
\overline{u_R} Q 
\overline{u_R} Q, 
&
{\cal O}_{14} 
= 
\bar{L} \bar{L} 
\overline{u_R} Q 
\bar{Q} d_R,
 \\ \nonumber
{\cal O}_{19} = 
\bar{L} \overline{e_R}
\bar{Q} d_R 
\overline{u_R} d_R,
& 
{\cal O}_{20} 
= \bar{L} \overline{e_R}
\hspace{0.08cm}
\overline{u_R} Q 
\overline{u_R} d_R .
&
\end{eqnarray}
Here, we have suppressed the indices of generation, $SU(3)_{c}$,
$SU(2)_{L}$, and Lorentz spinor, which are contracted appropriately.
In the list of decompositions shown in \cite{Bonnet:2012kh}, there
appears also one operator not given in \cite{Babu:2001ex}, which
is\footnote{As discussed below, this operator induces neutrino mass
  only at the four-loop level.  Probably for that reason it was
  neglected in \cite{Babu:2001ex}.}
\begin{equation}\label{eq:BLX}
{\cal O}_{-} = 
\overline{e_R} 
\hspace{0.08cm}
\overline{e_R}
\hspace{0.08cm}
\overline{u_R} 
d_R
\overline{u_R} 
d_R.
\end{equation}
As the black box theorem demonstrates, one can obtain the Weinberg
operator by connecting the quark legs in these effective operators
with the SM Yukawa interactions. From the effective operator point of
view it seems that neutrino masses are generated at 2-loop level from
${\cal O}_{11}$, ${\cal O}_{12}$ and ${\cal O}_{14}$.  The operators
${\cal O}_{19}$ and ${\cal O}_{20}$ need an additional SM Yukawa
interaction with a charged lepton to generate a neutrino mass term,
thus they end up with 3-loop diagrams.  The operator ${\cal O}_{-}$
leads to neutrino masses only at the 4-loop level, which is equivalent
to the original black box diagram.  However, the classification of the
neutrino mass models with respect to the number of loops, which we
discussed in the introduction, is modified from this naive
expectation, once the decomposition of the operators are specified. In
fact, as shown below, many decompositions of the operators in
Eq. (\ref{eq:BL}) contain automatically the particle content (and
interactions) such that neutrino masses are generated at lower order,
i.e., both tree-level and 1-loop neutrino mass models are found.  And,
surprisingly, also the opposite case exists: If we restrict ourselves
to decomposing the operators only with scalar and fermion mediators,
none of the decompositions of the operator $\mathcal{O}_{14}$
generates a {\it genuine} neutrino mass diagram at the 2-loop
level. We will come back to this important point later in more detail.
{\it Genuineness} is one of the key concepts in our classification
method.  The term, {\it genuine $n$-loop neutrino mass model}, is
defined as the model in which the neutrino mass term is generated at
the $n$-loop level and for which, simultaneously, diagrams with loop
level lower than $n$ are guaranteed to be absent, see
\cite{Sierra:2014rxa} for more details.

\subsection{Tree-level neutrino mass models}

\begin{figure}[t]
\hskip-10mm\includegraphics[width=0.5\linewidth]{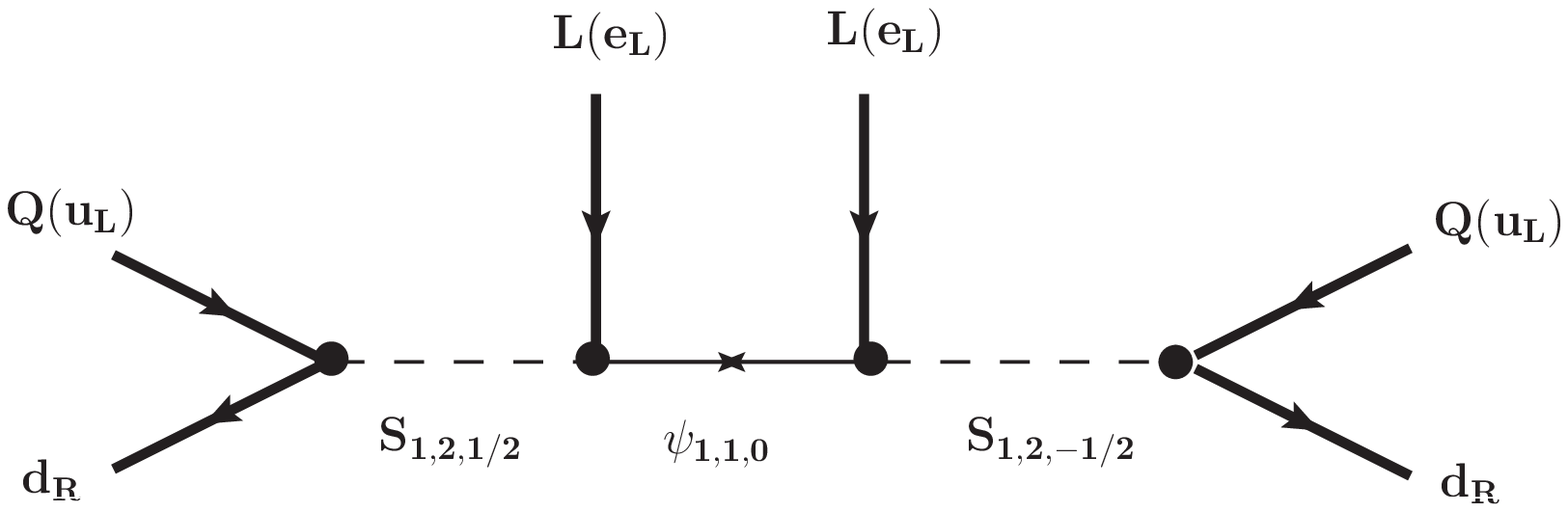}
\hskip10mm\includegraphics[width=0.3\linewidth]{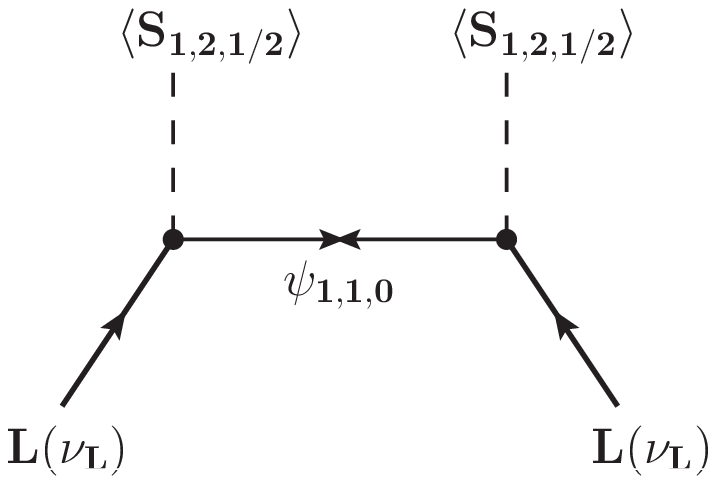}
\caption{To the left: Diagram for $\znbb$ decay via charged scalar
  exchange for Babu-Leung operator ${\cal O}_{11}$ (BL\#11). To the
  right: Tree-level neutrino mass generated via seesaw type-I, using
  the same vertices as in the diagram on the left. Here and in all
  Feynman diagrams below, arrows on fermion lines indicate the flow of
  particle number, not the chirality of the fermion.}
\label{fig:SSI}
\end{figure}

Let us start with a rather trivial example, illustrated in
Fig.~\ref{fig:SSI}, in which the $d=9$ contribution to $\znbb$ is
related to the Majorana neutrino mass at the tree level. Here, as
everywhere else in this paper, subscripts on fields denote their
transformation properties (or charge in case of $U(1)_Y$) under the SM
group, $SU(3)_c\times SU(2)_L\times U(1)_Y$.  A new scalar field is
denoted by the symbol $S$, and a fermion field, which is understood as
either a vector-like fermion or Majorana fermion, is denoted as
$\psi$. Thus, $\psi_{1,1,0}$ has the same quantum numbers as a
right-handed neutrino, while $S_{1,2,1/2}$ is equivalent to a (copy
of) the SM Higgs doublet.

The Lagrangian producing the left diagram of Fig.~\ref{fig:SSI}
necessarily contains the following terms:
\begin{equation}\label{eq:LagSSI}
{\cal L} = (Y_{SQd})_{ij} \ \overline{Q}_i \cdot S_{1,2,1/2} d_{R,j}  
         +  (Y_{S\nu\psi})_i \ \overline{L_i}\cdot S_{1,2,1/2}^{\dagger}\psi_{1,1,0}  
         +  m_{\psi} \overline{(\psi_{1,1,0})^{c}} \psi_{1,1,0}  + {\rm H.c.}.
\end{equation} 
Here, the singlet fermion field $\psi_{1,1,0}$ is allowed 
to have a Majorana mass $m_{\psi}$. 
The dot ($\cdot$) denotes the anti-symmetric tensor (${\rm i} \tau^{2}$) 
for $SU(2)_{L}$.
This Lagrangian generates an effective operator for 
a short-range contribution to $\znbb$,
\begin{align}
 \mathcal{L}_{d=9}
 =
 \frac{((Y_{SQd})_{11} (Y_{S\nu\psi})_{e})^2}{m_{S_{1,2,1/2}}^4 m_{\psi}}
 \left(
 \overline{Q} d_{R}
 \right)
 \cdot
 \overline{L}  
 L^{c} 
 \cdot
 \left(
 \overline{Q} d_{R}
 \right)
+ {\rm H.c.},
\end{align}
which corresponds to ${\cal O}_1^{SR}$ in the notation of
\cite{Pas:2000vn}. (Here we use the notation ${\cal O}_i^{SR}$ for the
five relevant short-range operators, defined in \cite{Pas:2000vn}, to
distinguish them from the lepton number violating operators 
${\cal O}_j$, $j=11,12,14,19,20$ and ``$-$''.)  
$m_{S_{1,2,1/2}}$ is the mass of $S_{1,2,1/2}$.  
Following the method adopted in \cite{Deppisch:2012nb,Bonnet:2012kh} 
and using the experimental 
bound~\cite{Albert:2014awa,Shimizu:2014xxx,Gando:2012zm}
\begin{align}
T_{1/2}^{\znbb} ({}^{136}{\rm Xe}) > 1.6 \cdot 10^{25} \text{ [ys]} 
\label{eq:Thalf-bound-Xe}
\end{align}
one finds the bound on the coefficient of the $d=9$ operator as
\begin{equation}
\frac{1}{8} 
\frac{((Y_{SQd})_{11} (Y_{S\nu\psi})_{e})^2}{m_{S_{1,2,1/2}}^4 m_{\psi}}
=\frac{G_{F}^{2}}{2m_{P}} \epsilon_{1}^{\{RR\}R}
\lesssim
\frac{G_{F}^{2}}{2m_{P}} 2.6 \cdot 10^{-7},
\label{eq:limeps1}
\end{equation}
which can be interpreted as a constraint on the Yukawa coupling:
\begin{equation}
(Y^2_{S\nu\psi})_{e} \lsim 
1.5 \cdot 10^{-6}
\left(\frac{1.0}{Y_{SQd}^2}\right)
\left(\frac{m_{S_{1,2,1/2}}}{\rm 100 \hskip1mm [GeV]}\right)^4
\left(\frac{m_{\psi}}{\rm 100 \hskip1mm [GeV]}\right). 
\label{eq:limYnubb}
\end{equation}
If the scalar mediator $S_{1,2,1/2}$ acquires a vacuum expectation
value (vev), the right diagram in Fig.~\ref{fig:SSI}, which has the
same topology as the type-I seesaw mechanism, will contribute to
neutrino mass(es):
\begin{equation}\label{eq:SSI}
\meff = \frac{(Y^2_{S\nu\psi})_{e} \langle S\rangle^2}{m_{\psi}}
\end{equation}
The experimental bound on the effective neutrino mass $\meff \lesssim
0.3$ eV, which is found from Eq.~\eqref{eq:Thalf-bound-Xe}
under the assumption of the mass mechanism being dominant,  
gives
\begin{equation}
(Y^2_{S\nu\psi})_{e} \lsim 
 1.0 \cdot 10^{-12}
 \left(\frac{m_{\psi}}{\rm 100 \hskip1mm [GeV]}\right) 
 \left(\frac{v_{\rm SM}}{\langle S\rangle}\right)^2
  \left(\frac{\meff}{\rm 0.3 \hskip1mm [eV]}\right), 
\label{eq:limSSI}
\end{equation}
where we have used $v_{\rm SM} \simeq 174$ GeV.
If $S_{1,2,1/2}$ is identified as 
the SM Higgs doublet $H$, as in the ordinary type-I seesaw,  
the constraint on $Y_{S\nu\psi}$ shown in Eq.~\eqref{eq:limSSI} 
is obviously much stronger than Eq.~\eqref{eq:limYnubb}.\footnote{%
If $S_{1,2,1/2}$ is the SM Higgs doublet,  
the coupling $Y_{SQd}$ is identified as the down quark Yukawa 
coupling, $Y_d \sim 3 \times 10^{-5}$, 
and the constraint shown in Eq.~\eqref{eq:limYnubb} actually becomes 
less stringent than even the ordinary perturbativity bound.}
In other words: The mass mechanism dominates the contribution 
to $\znbb$ decay by far, if we consider SM Higgs exchange. 

However, $S_{1,2,1/2}$ is not necessarily the SM Higgs, it could be an
additional new state, such as appear, for example, in multi-Higgs
doublet models. In this case, neutrino masses would still be generated
through the type-I seesaw mechanism, with the vev $\langle S \rangle$
of the scalar $S_{1,2,1/2}$ independent of the SM vev $v_{\rm SM}$, 
such as occurs, for example, in the neutrinophilic neutrino 
mass model of \cite{Ma:2000cc,Ma:2001mr,Haba:2010zi,Haba:2011fn}.
For this case one finds that if the relaxed constraint
\begin{equation}
\langle S\rangle \lsim 0.14 
 \hskip1mm {\rm [GeV]} \hskip1mm  
 \left(
  \frac{(Y_{SQd})_{11}}{1.0}
  \right)^{2}
 \left(
  \frac{\rm 100 \hskip1mm [GeV]}{m_{S}}
 \right)^2 
\label{eq:limSSIvev}
\end{equation}
holds, Eq.~\eqref{eq:limYnubb} becomes more stringent than
Eq.~\eqref{eq:limSSI}, i.e., the short-range diagram will be the
dominant contribution to the $\znbb$ decay amplitude in this 
case.

If $S_{1,2,1/2}$ has exactly zero vev, in the literature often called
the ``inert doublet'', we can no longer directly relate the the relative
size of the $d=9$ operator with the mass mechanism. The only conclusion
one can derive in this particular case is the trivial constraint that
the standard model Higgs coupling with $L$ and $\psi_{1,1,0}$ 
must obey Eq.~\eqref{eq:limSSI}. 

As is well-known, there are only three types of tree-level mass
generation mechanisms (seesaw mechanisms) called
type-I~\cite{Minkowski:1977sc,Yanagida:1979as,GellMann:1980vs,Mohapatra:1979ia},
type-II~\cite{Magg:1980ut,Schechter:1980gr,Wetterich:1981bx,Lazarides:1980nt,Mohapatra:1980yp,Cheng:1980qt}
and type-III~\cite{Foot:1988aq}.  These are mediated by the singlet
Majorana fermion $\psi_{1,1,0}$ (type-I), the triplet scalar
$S_{1,3,1}$ (type-II), and the triplet Majorana fermion
$\psi_{1,3,0}$ (type-III).  From the complete list of decompositions
given in \cite{Bonnet:2012kh}, one can find that T-I-1-i, 2-i-b,
2-ii-b, 2-iii-a, 4-i, and 5-i contain the relevant fermion mediators,
and T-I-1-ii-a, 1-ii-b, 3-ii, 3-iii, T-II-1, and T-II-3 do contain the
scalar mediator. From the discussion above, we can conclude that for
all of these the short-range contribution will be much less important
than the neutrino mass mechanism, unless the vev of the new scalars
$S_{1,2,1/2}$ or $S_{1,3,1}$ are heavily suppressed compared to
$v_{\rm SM}$.

\subsection{1-loop models}

\begin{figure}[t]
\includegraphics[width=0.3\linewidth]{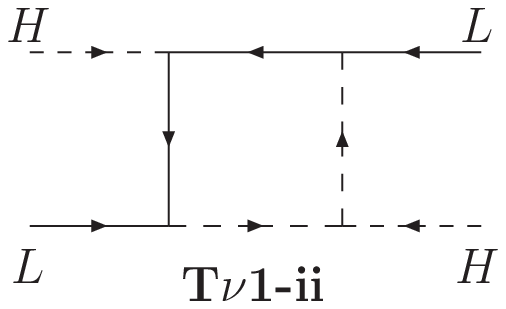}
\hskip5mm
\includegraphics[width=0.3\linewidth]{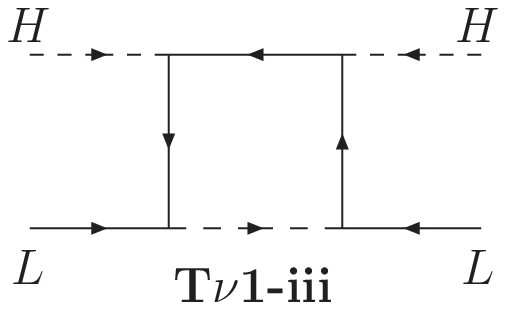}
\hskip5mm
\includegraphics[width=0.3\linewidth]{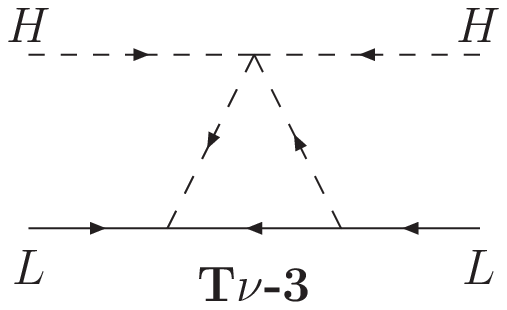}
\caption{Three different 1-loop diagrams for neutrino
  mass~\cite{Bonnet:2012kz}, which can appear in one of the $\znbb$
  decay decompositions. For a discussion see text.}
\label{fig:1lp}
\end{figure}
As shown in \cite{Bonnet:2012kz} (see also \cite{Ma:1998dn}), there
exist a total of four genuine 1-loop diagrams, which can contribute to
the $d=5$ Weinberg operator at the renormalizable
level.\footnote{There are also three more {\em non-genuine} diagrams,
  discussed in \cite{Bonnet:2012kz}, which can be understood as
  one-loop generated vertices for one of the three tree-level
  seesaws.} Three diagrams, shown in Fig.~\ref{fig:1lp}, can be related
to $\znbb$ decay decompositions.\footnote{The remaining diagram
  T${\nu}$-1-i can be understood as opening-up of the quartic scalar
  vertex in T${\nu}$-3, by inserting an additional scalar.}  Here, we
have added a ``$\nu$'' to the naming conventions of
\cite{Bonnet:2012kz}, in order not to confuse the 1-loop neutrino mass
diagrams with the double beta decay topologies.

Let us discuss the relation between the neutrino mass diagram
T${\nu}$-1-ii, which is shown as the left-most diagram in
Fig.~\ref{fig:1lp}, and the decomposition of the $d=9$ $\znbb$ diagram
$ (\overline{u_{L}}d_{R}) (\overline{e_{L}} d_{R}) 
(\overline{u_{L}}\hspace{0.08cm}\overline{e_{L}})$, which is
classified with the ID number T-II-2 ${\cal O}_{11}$ in
~\cite{Bonnet:2012kh}, as an example of this class of models.  The
decomposition leads to a Lagrangian, which contains the following
terms:
\begin{eqnarray}\label{eq:LagLQ}
 {\cal L} 
 &=& 
 (Y_{LdS})_{ij} 
 \overline{L}_i \cdot (S_{3,2,1/6})^{\dagger} d_{R,j}
 + 
 (Y_{LQ S})_{ij} 
 \overline{Q}_j \cdot L^c _iS_{3,1,-1/3}
 \\ 
 &+& 
  (Y_{QdS})_{ij} 
  \overline{Q}_i S_{1,2,1/2}
  d_{R,j}
  + 
 \mu_{SSS} 
 (S_{3,1,-1/3})^{\dagger} 
 S_{3,2,1/6} 
 S_{1,2,1/2}^{\dagger}
 +
 {\rm H.c.}
  \nonumber
  \\  \nonumber
 &+&
  m_{S_{3,2,1/6}}^{2} 
  (S_{3,2,1/6})^{\dagger} 
  S_{3,2,1/6}
  +
  m_{S_{3,1,-1/3}}^{2} 
  (S_{3,1,-1/3})^{\dagger} 
  S_{3,1,-1/3}.
  +
  m_{S_{1,2,1/2}}^{2} 
  (S_{1,2,1/2})^{\dagger} 
  S_{1,2,1/2}.
\end{eqnarray}
The lepton number violation can then be assigned to be due to the
presence of the coupling $\mu_{SSS}$. With this Lagrangian,
Eq.~\eqref{eq:LagLQ}, the effective $d=9$ Lagrangian that contributes
to $\znbb$ process as short-range effects is given as:
\begin{align}
 \mathcal{L}_{d=9}
 =&
 \frac{(Y_{LdS})_{e 1} (Y_{Q d S})_{11} (Y_{LQS})_{1e}\mu_{SSS} }
 {m_{S_{3,2,1/6}}^{2} m_{S_{3,1,-1/3}}^{2} m_{S_{1,2,1/2}}^2}
 (\overline{L} d_{R})
 \cdot
 (\overline{Q} d_{R}) 
 (\overline{Q} \cdot L^{c})
 +
 {\rm H.c.}
 \nonumber
 \\
 \supset&
 -
 \frac{(Y_{LdS})_{e 1} (Y_{Q d S})_{11} (Y_{LQS})_{1e}\mu_{SSS} }
 {m_{S_{3,2,1/6}}^{2} m_{S_{3,1,-1/3}}^{2} m_{S_{1,2,1/2}}^2}
 \frac{1}{16}
 (\mathcal{O}_{1}^{SR})_{\{RR\}R}
\end{align}
The experimental bound, see Eq.~\eqref{eq:Thalf-bound-Xe}, on $\znbb$
decay can then be interpreted again as an upper limit on the new
leptonic Yukawa interactions as:
\begin{align}
 (Y_{LdS})_{e 1} (Y_{LQS})_{1e}
 <
 0.15
 \left(
  \frac
 {m_{S_{3,2,1/6}}^{2} m_{S_{3,1,-1/3}}^{2} m_{S_{1,2,1/2}}^2}{1.0 [\text{TeV}^{6}]}
 \right)
 \left(
 \frac{1.0 [\text{TeV}]}{\mu_{SSS}}
 \right)
 \left(
 \frac{1.0}{(Y_{Qd S})_{11}}
 \right).
\label{eq:0n2b-bound-d9-1loop}
\end{align}

With the interactions shown in Eq.~\eqref{eq:LagLQ}, neutrinos acquire
Majorana masses through the diagram {\bf T$\nu$-1-ii}.  This neutrino
mass generation mechanism through the leptoquark-Higgs coupling was
first proposed in \cite{Hirsch:1996qy} and discussed in detail in
\cite{Mahanta:1999xd,AristizabalSierra:2007nf}.

Assuming the coupling $\mu_{SSS}$ is smaller than the average of the
leptoquark quark masses $m_{\rm LQ}=(m_{S_{3,2,1/6}}+m_{S_{3,1,-1/3}})/2$, we
can roughly estimate the neutrino mass as
\begin{equation}
(m_{\nu})_{\alpha\beta} 
 \simeq
 \frac{1}{16 \pi^2} 
 \frac{m_{d_k} \mu_{SSS}}{m_{\rm LQ}^2} \langle S_{1/2} \rangle 
 \left[
  (Y_{LQS}^{\dagger})_{\beta k} 
  (Y_{LdS}^{\dagger})_{k \alpha}
  + (\beta \leftrightarrow \alpha) 
 \right],
\label{eq:mnuLQ}
\end{equation}
where $m_{d_k}$ is the mass of the down-type quark (of generation $k$),
which enters this estimation as the vertex at the left-upper corner of
{\bf T$\nu$-1-ii} of Fig.~\ref{fig:1lp}. Applying the bound on the
effective Majorana mass $\langle m_{\nu} \rangle < 0.3$ eV (which is
obtained from Eq.~\eqref{eq:Thalf-bound-Xe} with the assumption of the
mass mechanism dominance) to Eq.~\eqref{eq:mnuLQ}, we have
\begin{align}
(Y_{LdS})_{ek} (Y_{LQS})_{ek}
 <
 0.034 
 \left(
 \frac{4.0\text{[GeV]}}{m_{d_k}}
 \right)
 \left(
 \frac{m_{\rm LQ}}{1.0\text{[TeV]}}
 \right)^{2}
 \left(
 \frac{1.0[\text{MeV}]}{\mu_{SSS}}
 \right)
\left(
 \frac{174 [\text{GeV}]}{\langle  S_{1/2} \rangle }
 \right)
 \left(
 \frac{\langle m_{\nu} \rangle}{0.3 \text{[eV]}}
 \right).
\label{eq:0n2b-bound-massmech-1loop}
\end{align}
Assuming that the flavour structure of the new Yukawa interactions in
Eq.~\eqref{eq:LagLQ} is not strongly hierarchical, one concludes that
third generation quarks give 
the largest contribution to the neutrino
mass. If the vev $\langle S_{1/2} \rangle$ is as large as the SM Higgs
vev, this constraint is more than six (three) orders of magnitude more
stringent than Eq. (\ref{eq:0n2b-bound-d9-1loop}) for $k=3$ ($k=1$).
Note again that if the vev $\langle S_{1/2} \rangle$ vanishes
then, as in the tree-level case, 
the $d=9$ contribution to $\znbb$ and the mass mechanism 
in this class of models are independent of each other.

If the leptoquark mass is set to $\mathcal{O}(1)$ TeV and Yukawa
couplings are taken to be ${\cal O}(0.1)$, 
the trilinear coupling $\mu_{SSS}$ must
be ${\cal O}(100)$ keV to reproduce ${\cal O}(0.05)$ eV of neutrino
masses, which is the minimum value necessary to reproduce data on
atmospheric neutrino oscillations.  Such a small value of the coupling
$\mu_{SSS}$ can (obviously) be probed only in a $\znbb$ process
dominated by the mass mechanism, since there is no other LNV process, 
for which experiments have even remotely comparable sensitivity.

To finish this discussion of the one-loop neutrino mass case, recall
the possible two-loop contributions to neutrino masses in this model.
As shown in the next subsection (and the list in the appendix), the
model described with Eq.~\eqref{eq:LagLQ} generates neutrino masses at
the two-loop level, even if the value of $\langle S_{1/2} \rangle$ is
identical with zero.  However, the one-loop contribution discussed in
this subsection can easily (even with a value of $ \langle S_{1/2}
\rangle$ smaller than MeV) dominate over the two-loop diagrams, as can
be seen from Eqs. \eqref{eq:mnuLQ} -
\eqref{eq:0n2b-bound-massmech-1loop}.  Therefore, we classify this
type of the models separately from the {\it genuine two-loop models},
which will be defined and explained in detail in the next subsection.

Very similar arguments can be applied to the other two topologies,
{\bf T$\nu$-1-iii} and {\bf T$\nu$-3}, shown in Fig.~\ref{fig:1lp}.
Such one-loop neutrino mass models appear in quite a large number of
decompositions.  We give the complete list of this class of models in
Table~\ref{Tab:1lp} in the appendix, together with the additional
interaction that is required (and is allowed by the SM gauge symmetries)
to generate the corresponding one-loop diagram.

\subsection{2-loop models}
\label{Sec:Class-2loop}

The effective operators $\mathcal{O}_{11}$, $\mathcal{O}_{12}$, and
$\mathcal{O}_{14}$ contain two lepton doublets, and thus naively one
expects them to generate neutrino masses at 2-loop level,\footnote{%
When the two lepton doublets are anti-symmetric in
  $SU(2)_{L}$, an additional loop is necessary to obtain a neutrino
  mass term, i.e., the resulting neutrino mass diagram contains
  three loops, which correspond to $\mathcal{O}_{11a}$, $\mathcal{O}_{12b}$
  and $\mathcal{O}_{14a}$ in the list of \cite{deGouvea:2007xp}. In 
  $\znbb$ decay only the operators $\mathcal{O}_{11b}$, $\mathcal{O}_{12a}$
  and $\mathcal{O}_{14b}$, which contain symmetric pieces in $SU(2)_{L}$, 
can appear. These are called $\mathcal{O}_{11}$, $\mathcal{O}_{12}$
  and $\mathcal{O}_{14}$ for brevity here and in \cite{Bonnet:2012kh}.}
when their quark legs are connected with the SM Yukawa interactions.
However, this naive picture has to be modified, once the possible
decompositions of the effective operators are taken into account.

In this subsection, we will first demonstrate why some of the
decompositions of the operators $\mathcal{O}_{11}$,
$\mathcal{O}_{12}$, and $\mathcal{O}_{14}$ do not generate neutrino
masses {\em genuinely} at the 2-loop level. 
Here, we use the terminology {\it genuine 2-loop diagrams}, 
following  \cite{Sierra:2014rxa}, to imply that the 
corresponding model (or decomposition in our case) generates 
neutrino masses at 2-loop order and that no lower order diagram 
exists.

\begin{figure}[t]
\unitlength=1cm
\begin{picture}(12.5,4.5)
 \thicklines
 \put(0,0){\includegraphics[width=6cm]{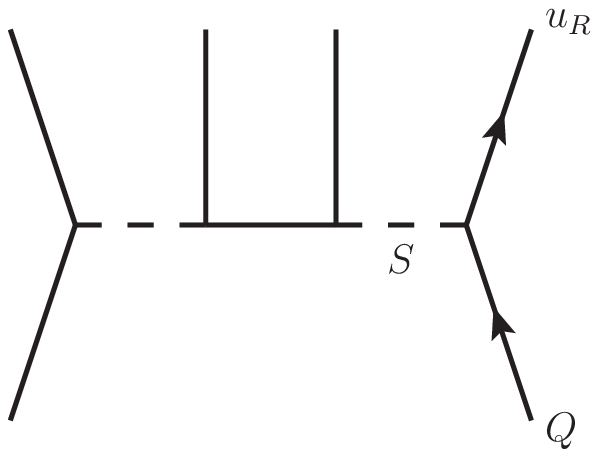}}
 \put(6.2,2){\vector(1,0){2}}
 \put(8.5,1){\includegraphics[width=4.5cm]{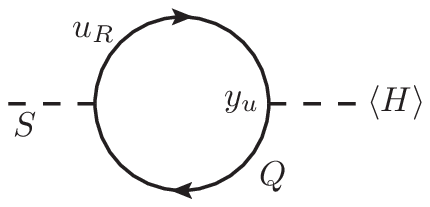}}
\end{picture}
\\
\begin{picture}(12.5,4.5)
 \thicklines
 \put(0,0){\includegraphics[width=6cm]{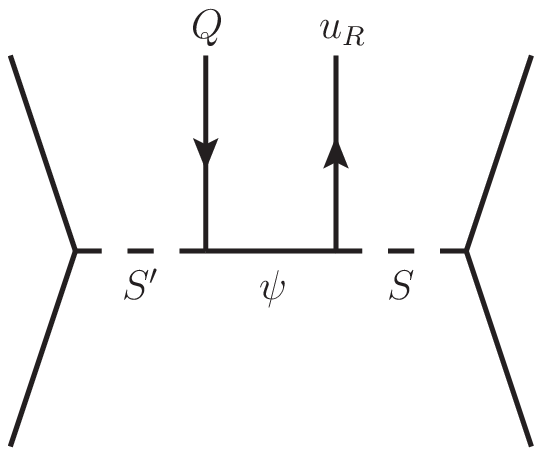}}
 \put(6.2,2){\vector(1,0){2}}
 \put(8.5,1){\includegraphics[width=4.5cm]{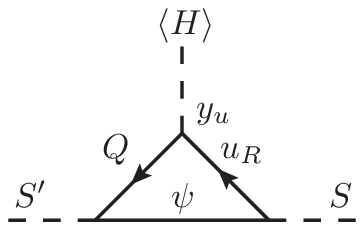}}
\end{picture}
\caption{Decompositions of the Babu-Leung operator \#14,
  $\mathcal{O}_{14} \propto \overline{L} \hspace{0.08cm}
  \overline{L} \hspace{0.08cm} \overline{Q} d_{R} \overline{u_{R}} Q$,
  with two scalar ($S$ and $S'$) and one fermion ($\psi$) mediators.
  For a neutrino mass diagram the quark legs $\overline{u_{R}}$ and
  $Q$ must be connected via the SM Yukawa interaction $y_{u}$.  The
  resulting loops (shown on the right) are infinite one-loop
  corrections to the scalar mass terms, that therefore must be
  contained in the tree-level Lagrangians of the corresponding
  models.}
\label{Fig:BL14}
\end{figure}

Let us first discuss the decomposition of $\mathcal{O}_{14}$. 
This operator contains:
\begin{align}
 \overline{L}_{\dot{a}}, \quad
 \overline{L}_{\dot{a}}, \quad
 \overline{u_{R}}^{a}, \quad
 Q_{a}, \quad
 \overline{Q}_{\dot{a}}, \quad
 {d_{R}}^{\dot{a}}.
\end{align}
Here we explicitly wrote the 2-component spinor indices for the
fermions: dotted for a right-handed field (complex conjugate of
left-handed field), undotted for a left-handed field.  As everywhere
else in this paper, we restrict the discussion to fermions and scalars
as mediators. For $\mathcal{O}_{14}$ this implies that the spinor
indices on $Q$ and $\overline{u_{R}}$ must be contracted for the
effective operator being a Lorentz scalar.  There are only two choices
to assign these two quarks to the outer legs of a $d=9$ tree diagram:
(i) They form a Yukawa interaction with a scalar mediator, i.e.,
$(\overline{u_{R}} Q \cdot S)$.  This is shown as the upper diagram of
Fig.~\ref{Fig:BL14}.  Or: (ii) Each of these quarks forms a Yukawa
interaction with a fermion mediator $\psi$ and one of the scalar
mediators, i.e., $(\overline{u_{R}} \psi S) (\overline{\psi} Q S')$.
This is shown as the lower diagram of Fig.~\ref{Fig:BL14}.  When the
loops are closed, as shown on the right of Fig.~\ref{Fig:BL14}, via
the SM Yukawa interaction $y_{u} (\overline{Q} \cdot H^{*} u_{R})$,
the resulting quark loop is divergent. In other words, these loops are
infinite corrections for (i) a mass term mixing the scalar mediator
and the SM Higgs doublet $m_{SH}^{2} S H^{*}$, or (ii) a corresponding
term mixing two scalar mediators $m_{SS'}^{2} SS'$. Therefore, the
original tree-level Lagrangian generating these operators must
contain these scalar mass terms as counter terms for the infinities,
and the quark loop appearing in neutrino mass diagrams must be
substituted with those scalar mass terms.

This simple argument actually holds for any decomposition of
$\mathcal{O}_{14}$.  In short, one cannot construct a {\it genuine}
(irreducible) 2-loop diagram for neutrino masses from the
decomposition of $\mathcal{O}_{14}$, if the mediators are restricted
to scalars and fermions. For this reason, in the appendix the
decompositions of $\mathcal{O}_{14}$ are instead either listed under
``tree-level'' or ``1-loop'', depending on the additional interaction
necessary.

\begin{figure}[t]
 \unitlength=1cm
 \begin{picture}(13,4)
  \put(0,0){\includegraphics[width=6cm]{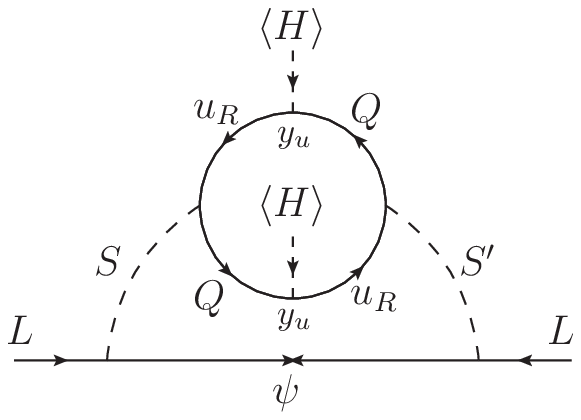}}
  \put(7,0.5){\includegraphics[width=6cm]{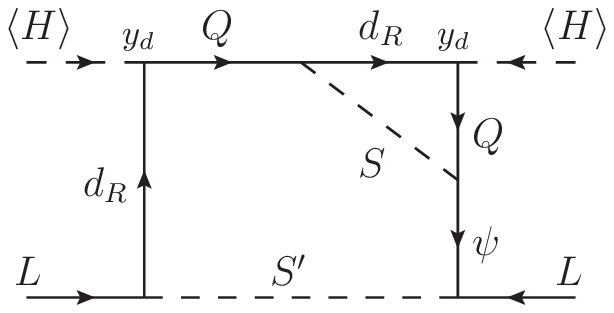}}  
 \end{picture}
\caption{Examples of non-genuine 2-loop neutrino mass diagrams, based
  on the decompositions of T-I-1-i $\mathcal{O}_{12}$ (left) and
  T-I-2-ii-a $\mathcal{O}_{11}$ (right). For a discussion see text.}
\label{Fig:non-genuine-2loop}
\end{figure}
The next question we must address is: Can all the decompositions of
$\mathcal{O}_{11(b)}$ and $\mathcal{O}_{12(a)}$ genuinely generate
neutrino masses at 2-loop level?  The answer is no, and the argument
for those cases is very similar to the one presented above for
$\mathcal{O}_{14}$. Two concrete examples are shown in
Fig.~\ref{Fig:non-genuine-2loop}. Consider first the diagram on the
left, based on decomposition T-I-1-i $\mathcal{O}_{12}$. The inner
loop in these classes of neutrino mass diagrams actually corresponds
to an infinite (1-loop) correction to the scalar quartic interaction
$\lambda_{SS'HH} SS' HH$. In the diagram on the right, the ``inner''
loop involving $S$ generates a Yukawa interaction $y_{Q \psi H}
\bar{\psi} Q \cdot H$. Again, this correction is infinite, thus
requiring a tree-level counter term which must be contained in the
original Lagrangian. Given these additional (but required)
interactions, the models contain neutrino mass diagrams at 1-loop
order. Note that, the left diagram in Fig.~\ref{Fig:non-genuine-2loop}
corresponds to Diagram (A) of Fig.~14 in \cite{Angel:2012ug} (with
appropriate Higgs insertions) and also Diagram (c) of Fig.~5 in
\cite{Farzan:2012ev}.  The right diagram is Diagram (B) in
\cite{Angel:2012ug} and (d) in \cite{Farzan:2012ev}. Quite a 
number of possible decompositions of $\mathcal{O}_{11}$ and 
$\mathcal{O}_{12}$ follow this pattern and are thus listed in 
the tables in the appendix as 1-loop models, together with the 
additional-but-necessary interactions.

After filtering out all decompositions that result in non-genuine
2-loop neutrino mass diagrams, we have found that for all remaining
decompositions there are only three types of genuine 2-loop diagrams,
all of them based on ${\cal O}_{11}$. These are shown in
Fig.~\ref{Fig:genuine-2loop}. The naming scheme in this figure follows
\cite{Sierra:2014rxa}.  In the appendix, we present the complete list
of the genuine 2-loop neutrino mass models and specify the class of
neutrino mass diagrams, into which each model falls.
In Table \ref{Tab:2lp}, two of the
  decompositions based on the BL operator ${\cal O}_{19}$ are also
  listed.  These appear in the table for two-loop models due to the
  fact that the intermediate fermion is of Majorana type, i.e., for
  these decompositions, the ``asymmetric'' operator ${\cal O}_{19}
  \propto \bar{L} \overline{e_R}$ is always accompanied by the
  ``symmetric'' operator ${\cal O}_{11} \propto \bar{L}\bar{L}$, and
  the associated operator generates neutrino masses at the two-loop
  level.  The catalogue of the effective operators appearing with
  their associated operators is given in the tables of
  \cite{Bonnet:2012kh}.

\begin{figure}[t]
 \unitlength=1cm
 \hspace*{-1cm}
 \begin{picture}(16,3)
  \put(0,0){\includegraphics[width=5.5cm]{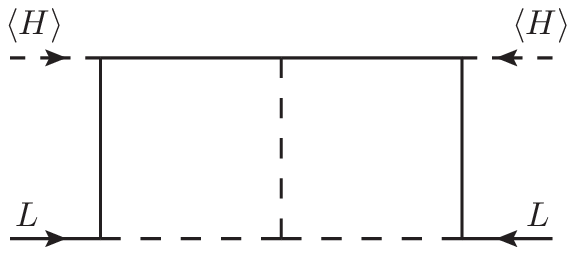}}
  \put(5.5,0){\includegraphics[width=5.5cm]{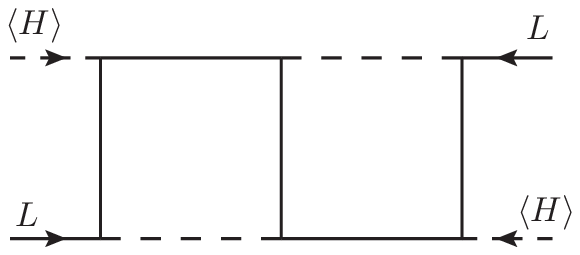}}
  \put(11,0){\includegraphics[width=5.5cm]{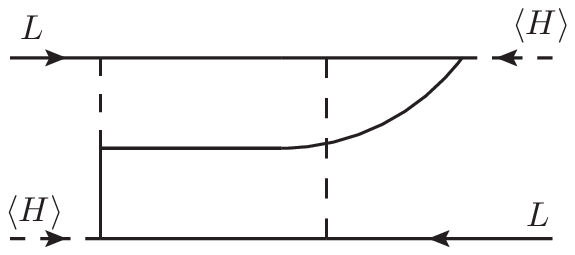}}
 \end{picture}
 \caption{%
 Genuine 2-loop neutrino mass diagrams based on
 decomposition of the Babu-Leung operator ${\cal O}_{11}$.  From
 left to right the diagrams are identified as CLBZ-1, PTBM-1 and
 PTBM-4, following the classification of \cite{Sierra:2014rxa}.  The
 left diagram (CLBZ-1) was discussed in the context of a neutrino
 mass model in \cite{Kohda:2012sr}, and corresponds to the diagram
 in Fig.~10 in \cite{Angel:2012ug} and (e) in \cite{Farzan:2012ev}.
 The diagram (PTBM-1) in the middle was discussed in
 \cite{Angel:2013hla}, and corresponds to Diagram (D2) in Fig.~14 in
 \cite{Angel:2012ug} and (f) in Fig.~5 in \cite{Farzan:2012ev}.  The
 diagram on the right corresponds to Diagram (C) in
 \cite{Angel:2012ug} and (f) in Fig.~5 \cite{Farzan:2012ev}. A model
 based on this diagram will be discussed in Sec.~\ref{sect:exa}.}
 \label{Fig:genuine-2loop}
\end{figure}
Among these three genuine diagrams, neutrino mass models based on the
CLBZ-1 and the PTBM-1 diagrams have already been studied in the
context of the decomposition of the $d=9$
operators~\cite{Kohda:2012sr,Angel:2013hla}.  Therefore, we will
discuss a 2-loop neutrino mass model that is associated with the
remaining possibility, i.e., the PTBM-4 diagram.  
Here, as in the previous subsections, we compare the $d=9$
contribution to $\znbb$ with the mass mechanism contribution and
postpone the detailed discussion on phenomenology of this model till
Sec.~\ref{sect:exa}.

The example we choose is based on T-I-4-ii-b, ${\cal O}_{11}$. 
The Lagrangian for this decomposition contains the terms
\begin{eqnarray}\label{eq:lag4iib}
 \mathcal{L} &=&
  (Y_{QQS})_{ij} 
  (\overline{Q}_{i} \vec{\tau} \cdot Q^{c}_{j}) 
  \hat{\vec{S}}_{6,3,1/3}
  + 
  (Y_{L\psi S})_{\alpha}
  (\overline{L}_{\alpha}
  \vec{\tau}
  \psi_{6,2,1/6})
  \vec{S}_{6,3,1/3}^{\dagger}
  \\
 & + &
 (Y_{\psi dS})_{i} 
 (\overline{{\hat \psi_{6,2,1/6}}} d_{R, i})S_{3,2,1/6}
 + 
 (Y_{LdS})_{\alpha i} 
 (\overline{L}_{\alpha} d_{R, i})
 \cdot S_{3,2,1/6}^{\dagger}.
\nonumber 
\end{eqnarray}
We use the notation $\hat{\vec{S}}_{6,3,1/3} = (\vec{S}_{6,3,1/3})_{X}
(T_{\bar{\bf 6}})^{X}_{IJ}$ and $\hat{\psi}_{6,2,1/6} =
(\psi_{6,2,1/6})_{X} (T_{\bar{\bf 6}})^{X}_{IJ}$.  The tensors
$T_{{\bf 6}}$ and $T_{\bar{\bf 6}}$ in the $SU(3)_{c}$ are given in
\cite{Bonnet:2012kh}.  Here, $\vec{\tau}$ is the Pauli matrix vector
for a triplet of $SU(2)_{L}$.  The effective $d=9$ operator resulting
from this Lagrangian can be written with the following linear
combination of the basis operators
$\mathcal{O}_{i\in\{1\text{-}5\}}^{SR}$ of the short-range
contributions to $\znbb$ decay as:
\begin{align}
 \mathcal{L}_{\text{eff}}
 =&
 -
 \frac{(Y_{QQS})_{11} (Y_{L\psi S})_{e} (Y_{\psi d S})_{1} {(Y_{LdS})_{e1}}}
 {m_{S_{6,3,1/3}}^{2} m_{S_{3,2,1/6}}^{2} m_{\psi}}
 \left[
 (\overline{Q}_{1} T_{\bf \bar{6}}
 \tau^{a}
 \cdot Q^{c}_{1})
 (\overline{L}_{e}
 \tau^{a})
 (d_{R} T_{\bf 6})
 \cdot ({L_{e}}^{c} d_{R})
 \right]
 +{\rm H.c.}
 \nonumber 
 \\
 \supset &
 \frac{(Y_{QQS})_{11} (Y_{L\psi S})_{e} (Y_{\psi d S})_{1} {(Y_{LdS})_{e1}}}
 {m_{S_{6,3,1/3}}^{2} m_{S_{3,2,1/6}}^{2} m_{\psi}}
 \left[
 \frac{1}{16}
 (\mathcal{O}_{1}^{SR})_{\{RR\}R}
 -
 \frac{1}{64}
 (\mathcal{O}_{2}^{SR})_{\{RR\}R}
 \right]
 \label{eq:Leff-2loop-example}
\end{align}
and the experimental bound Eq.~\eqref{eq:Thalf-bound-Xe}
constrains a combination of the coefficients
\begin{align}\label{eq:Lim-2loop-example}
 (Y_{QQS})_{11} (Y_{L\psi S})_{e} (Y_{\psi d S})_{1} {(Y_{LdS})_{e1}}
 <
 6.3\cdot 10^{-3}
 \left(
 \frac
 {m_{S_{3,2,1/6}}^{2} m_{S_{6,3,1/3}}^{2} m_{\psi}}{1.0 [\text{TeV}^{5}]}
 \right).
\end{align}
On the other hand, the neutrino mass generated from the 2-loop diagram
based on the effective operator Eq.~\eqref{eq:Leff-2loop-example} also
contributes to $\znbb$ through the mass mechanism.  The size of 2-loop
neutrino mass diagram can be roughly estimated
as~\cite{Babu:2001ex,deGouvea:2007xp,Angel:2012ug,Sierra:2014rxa}
\begin{align}
 (m_{\nu})_{\alpha \beta}
 \simeq
 \frac{N_c}{(16\pi^{2})^{2}}
 \frac{m_{b}^{2}}{\Lambda_{\rm LNV}}
 \left[
 (Y_{QQS})_{33} (Y_{L\psi S})_{\alpha} 
 (Y_{\psi d S})_{3} {(Y_{LdS})_{\beta 3}}
 +
 (\beta \leftrightarrow \alpha)
 \right].
 \label{eq:mNu-2loop-estimate}
\end{align}
Applying the experimental bound from $\znbb$ to
Eq.~\eqref{eq:mNu-2loop-estimate} and 
substituting $N_c=6$ because 
of the colour sextet combination in the loop, 
we can place the
bound on the couplings of the third generation quarks:
\begin{align}\label{eq:lim2lpmnu}
 (Y_{QQS})_{33} (Y_{L\psi S})_{e} (Y_{\psi d S})_{3} {(Y_{LdS})_{e3}}
 < 
 7.2 \cdot 10^{-5}
 \left(
 \frac
 {4.18 [\text{GeV}]}{m_b}
 \right)^2
 \left(
 \frac
 {\Lambda_{\rm LNV}}{1.0 [\text{TeV}]}
 \right)
  \left(
 \frac
 {\langle m_{\nu} \rangle}{0.3 [\text{eV}]}
 \right).
\end{align}
As this rough estimation shows, the mass mechanism and the short-range
part of the amplitude give similar contributions to $\znbb$ decay.
Note that assuming flavour democratic Yukawa couplings, the
constraints Eq.~\eqref{eq:Lim-2loop-example} and
Eq.~\eqref{eq:lim2lpmnu} become equally strong if the mass scale of
the new particles is taken to be roughly $\sim 300$ GeV.  For larger
mass values, the short range contribution can dominate only if Yukawas
with index ``3'' are smaller than those with index ``1'', otherwise
the mass mechanism dominates.  A more detailed discussion using the
full expression for the two-loop neutrino mass integral will be
presented in Sec.~\ref{sect:exa}.

\subsection{3-loop models}

From the point-of-view of effective operators,
the Babu-Leung operators $\mathcal{O}_{19}$ and $\mathcal{O}_{20}$
require three SM Yukawa interactions to generate neutrino masses:
two quark Yukawa interactions and one charged-lepton Yukawa
interaction, to convert $e_{R}$ in the effective operators to $L$
for a neutrino mass.  This fact leads us to three-loop neutrino mass
models.  However, some of the possible decompositions of
$\mathcal{O}_{19}$ and $\mathcal{O}_{20}$ contain the ingredients to
generate neutrino masses at a level lower than three-loop.  In such a
case, the lower loop contributions can easily dominate neutrino masses 
and make the contribution from a three-loop diagram sub-dominant. This
can happen for two reasons. First, there are decompositions based on
${\cal O}_{19}$ or ${\cal O}_{20}$, in which ${\cal O}_{11}$
necessarily also appears. We call this ``associated'' operators and
classify those decompositions in the class corresponding to those
lower loop levels, which usually will dominate over the 3-loop
contribution.  
Decompositions T-I-2-iii-a and
T-I-5-i of $\mathcal{O}_{19}$ with the Majorana fermion
$\psi_{8,1,0}$, which are listed in the table of two-loop models
(Tab.~\ref{Tab:2lp}), are categorised in this class.  
And, second,
there are decompositions for ${\cal O}_{19}$ or ${\cal O}_{20}$, in
which the 3-loop diagrams are not genuine in the sense that one of the
sub-diagrams corresponds to the 1-loop generation of a certain
vertex. We will discuss this case in a bit more detail.

\begin{figure}[t]
\unitlength=1cm
\begin{picture}(18,3) 
\put(0,0){\includegraphics[width=0.5\linewidth]{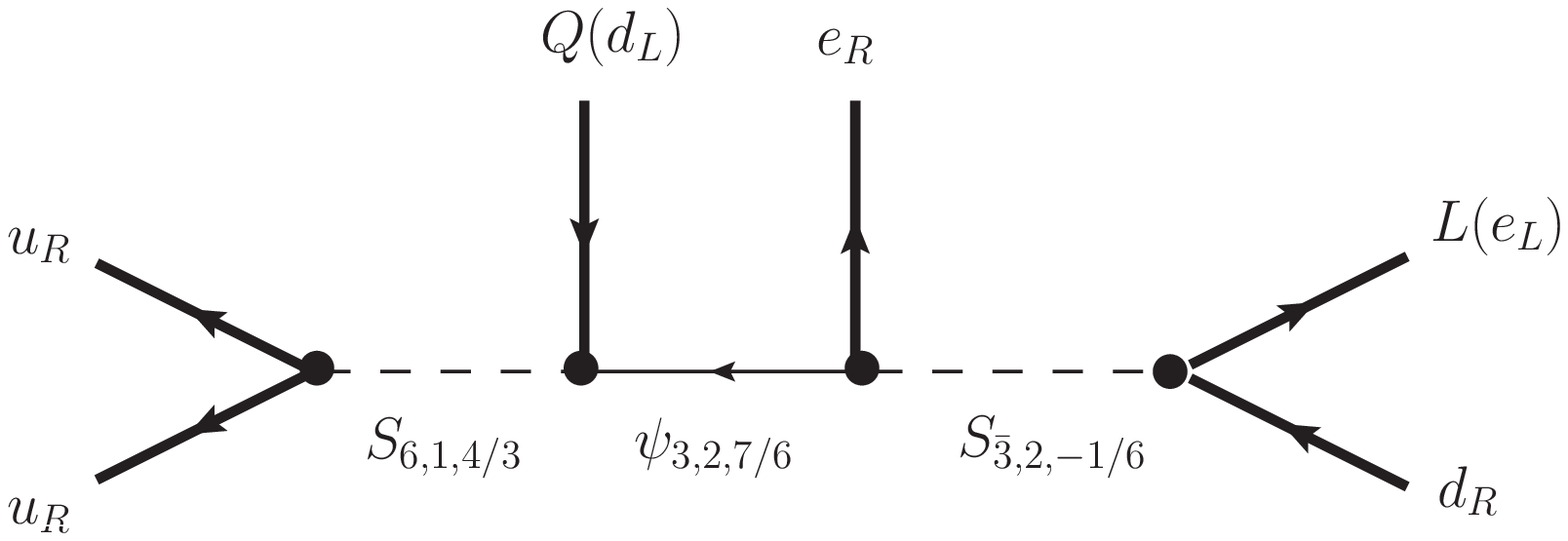}}
\put(9,0){\includegraphics[width=0.5\linewidth]{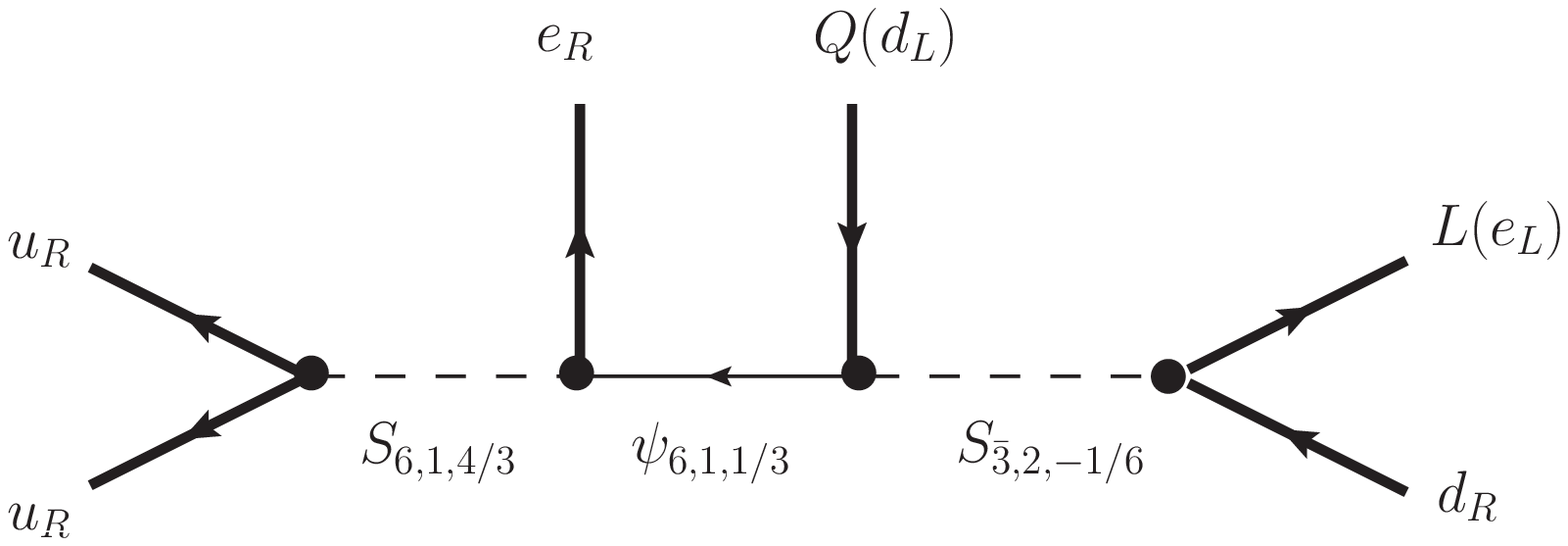}}
\end{picture}
\caption{The Feynman diagrams for the decompositions T-I-4-ii-a (left)
  and T-I-4-ii-b (right) for BL operator ${\cal O}_{20}$. The former
  leads to 2-loop $d=7$ neutrino masses, while the latter is an
  example of a 3-loop neutrino mass model, see text.}
\label{fig:3lpbb}
\end{figure}
\begin{figure}[t]
\unitlength=1cm
\begin{picture}(18,8) 
\put(-0.1,0){\includegraphics[width=0.5\linewidth]{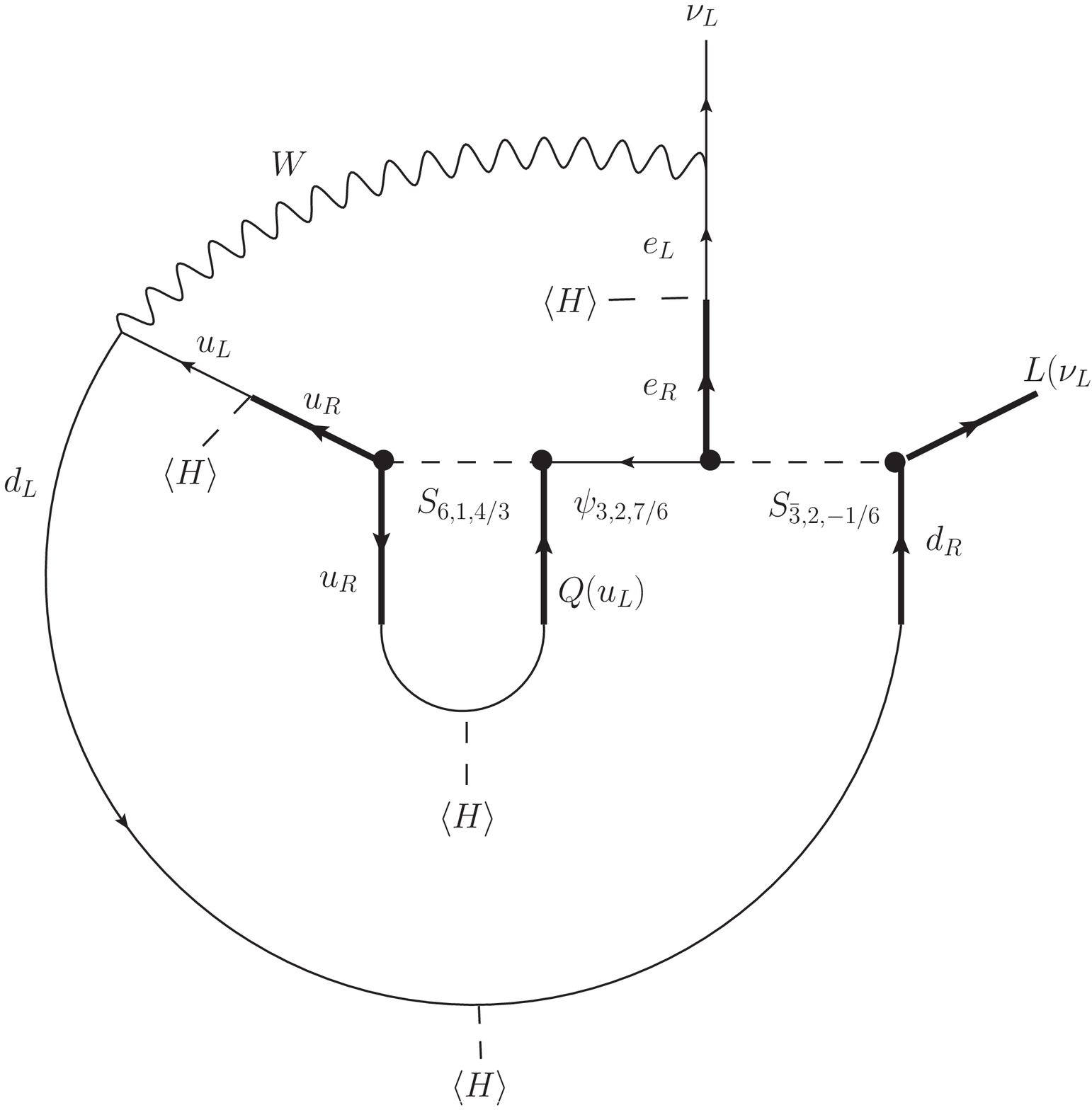}}
\put(8.2,0){\includegraphics[width=0.5\linewidth]{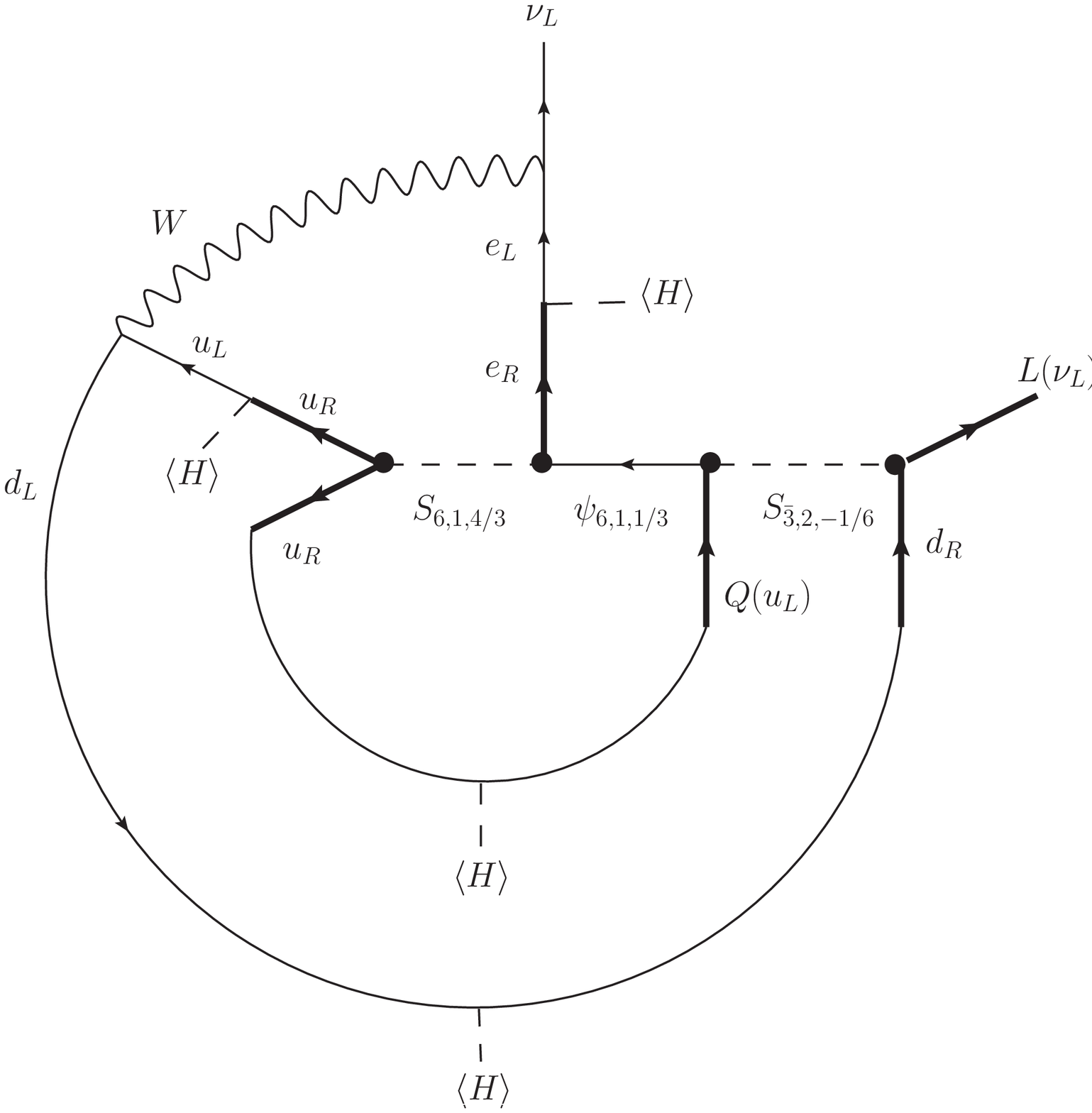}}
\end{picture}
\caption{Examples of 3-loop diagrams for the decompositions  T-I-4-ii-a (left)
  and T-I-4-ii-b (right) for BL operator ${\cal O}_{20}$, 
see Fig.~\ref{fig:3lpbb}. The diagram to the right corresponds to 
a genuine 3-loop model, while the one to the left is not. See text.}
\label{fig:3lp}
\end{figure}
\begin{figure}[t]
\includegraphics[width=0.5\linewidth]{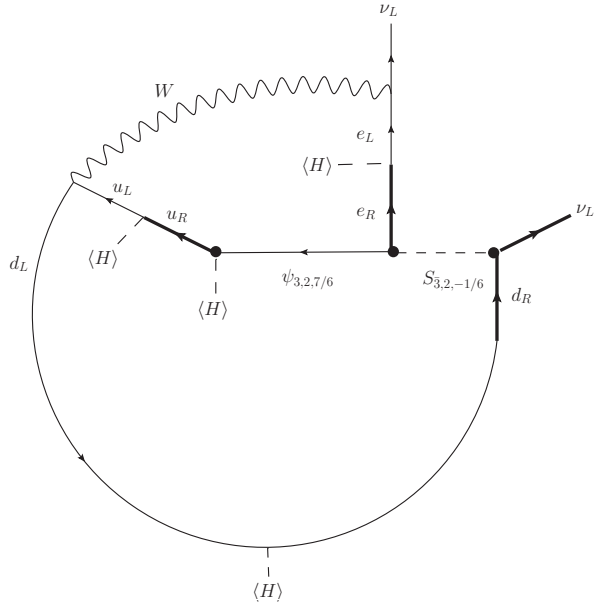}
\caption{Example of a 2-loop $d=7$ diagram for the decomposition  
  T-I-4-ii-a.
 This diagram corresponds to 
 Diagram (a) in Fig.~\ref{fig:2loop-dim7}.} 
\label{fig:2lpd7}
\end{figure}

Take the examples of the decompositions T-I-4-ii-a and T-I-4-ii-b, 
both ${\cal O}_{20}$. The Feynman diagrams are given in Fig.~\ref{fig:3lpbb},
and show 
T-I-4-ii-a $\to $ $(\overline{u_R}\hspace{0.08cm}\overline{u_R})
(d_L)(\overline{e_R})(\overline{e_L} d_R)$ and 
T-I-4-ii-b $\to $ $(\overline{u_R}\hspace{0.08cm}\overline{u_R})
(\overline{e_R})(d_L)(\overline{e_L} d_R)$ graphically. 
As we will see, 
despite the similarity between these two cases, 
T-I-4-ii-b will lead to a genuine 3-loop model, 
while T-I-4-ii-a will not.
Consider the examples of 3-loop diagrams for these two decompositions
shown in Fig.~\ref{fig:3lp}.  
First of all,
note that the loop diagrams shown in Fig.~\ref{fig:3lp} should be
understood as examples only, because there might be more than one
diagram contributing to the full neutrino mass matrix for each
decomposition.  
In the diagram for T-I-4-ii-a (left), one sees that
the innermost loop effectively generates the vertex 
$\overline{u_R} \psi_{3,2,7/6}H^\dag$ at 1-loop order.  
This one-loop sub-diagram is
infinite and, therefore, a tree-level counter term is necessarily to
be included in the Lagrangian to absorb the infinity.
In fact, the quantum numbers of the particles
involved in the loop are such that the necessary vertex actually
cannot be forbidden at tree-level by the SM gauge symmetry.  
This tree-level coupling has a value that is not fixed 
by the $\znbb$ decay amplitude, but the 2-loop ($d=7$) diagram 
that results from this coupling, see Fig.~\ref{fig:2lpd7}, 
can easily dominate over the 3-loop diagram.\footnote{%
Naively, the 2-loop $d=7$ contribution becomes more 
important than the 3-loop contribution, 
when the tree-level coupling $Y_{u \psi_H}$ 
of the necessary interaction 
$\overline{u_R} \psi_{3,2,7/6}H^\dag$ is 
larger than $Y_{uuS} Y_{Q \psi S}/(16 \pi^2) \times$(some
logarithmic factor).}  We have classified therefore all these cases
of ${\cal O}_{19}$ and ${\cal O}_{20}$, where such a tree-level vertex
is allowed, as ``2-loop $d=7$'' (Tab.~\ref{Tab:2lp-7}) in Appendix.
%
The loop diagram based on the decomposition
T-I-4-ii-b, which is shown as the right diagram of
Figure~\ref{fig:3lp}, does not contain such an inner loop and, thus,
such a construction is not possible for this decomposition.
Decompositions of this type are therefore classified as
genuine 3-loop models in the appendix.
 
We give here a rough estimate of the size of the neutrino mass
generated by the 3-loop diagram based on T-I-4-ii-b in order to
present some general arguments on the relative size of the $d=9$
contributions and the mass mechanism contributions to $\znbb$ in this
class of models.  This example, in which the $d=9$ is mediated by a
di-quark, $S_{6,1,4/3}$, a leptoquark, $S_{3,2,1/6}$, and an exotic
colour-sextet fermion, $\psi_{6,1,1/3}$, leads to a Lagrangian that
contains the terms:
\begin{eqnarray}\label{eq:lag3lp}
 {\cal L} 
 &=& 
 (Y_{uuS})_{ij} \
 \overline{u_{R,i}} \
 {\hat S}_{6,1,4/3} \ 
 {u_{R,j}}^{c} 
 +
 (Y_{e \psi S})_i \
 \overline{\psi_{6,1,1/3}}\
 e_{R,i}  \
 S_{6,1,4/3}
 \nonumber 
 \\ 
 &+& 
 (Y_{Q \psi S})_{i} \
 \overline{Q_i} \  \hat{\psi}_{6,1,1/3} 
 \cdot (S_{3,2,1/6})^{\dagger}
 + 
 (Y_{L d S})_{ij} \
 \overline{d_{R,j}} \
 L_i \cdot  
 S_{3,2,1/6}
 +
 {\rm H.c.}.
\end{eqnarray}
As above, we use the notation $\hat{S}_{6,1,4/3} =
(S_{6,1,4/3})_{X} (T_{\bar{\bf 6}})^{X}_{IJ}$ and
$\hat{\psi}_{6,1,1/3} = (\psi_{6,1,1/3})_{X} (T_{\bar{\bf
    6}})^{X}_{IJ}$.  Together, the $Y_{e\psi S}$, $Y_{Q\psi S}$, and
$Y_{LdS}$ terms necessarily violate lepton number by two units.  All
generation indices in the couplings in Eq.~\eqref{eq:lag3lp} have been
suppressed for simplicity.  The contribution to the neutrino mass
matrix can be roughly estimated as
\begin{eqnarray}
(m_{\nu})_{\alpha\beta} 
\simeq
\frac{N_c}{(16 \pi^2)^3}
\left[
\frac{m_{t}^{2}  m_{b} m_{e_\alpha} }
{\Lambda_{\rm LNV}^{3}} 
(Y_{uu S})_{33} (Y_{e \psi S})_{\alpha}
(Y_{Q \psi S})_{3} (Y_{L d S})_{3 \beta} 
+ (\alpha \leftrightarrow \beta)
\right],
\label{eq:mnu3lp}
\end{eqnarray}
where $\Lambda_{\rm LNV} \simeq m_{S_{6,1,4/3}} \simeq m_{S_{3,2,1/6}}
\simeq m_{\psi}$ is the mass scale of the heavy states, which is
typically taken to be TeV. $N_c$ is a colour factor. Here, we assumed
that all the SM fermion masses are much smaller than  
$\Lambda_{\rm LNV}$.  Putting all the Yukawa couplings in
Eq.~\eqref{eq:mnu3lp} equal to unity and $\Lambda_{\rm LNV}=1$ TeV and
$N_c=6$ (for a colour sextet combination), one finds\footnote{Using
  $\meff\le 0.3$ eV, we can formally write the constraint on the
  Yukawa couplings in the form of:
\[
 (Y_{uuS})_{33} (Y_{e\psi S})_{e} (Y_{Q\psi S})_{3} {(Y_{LdS})_{e3}}
 <
 3 \cdot 10^{3}
 \left(
 \frac
 {\Lambda_{\rm LNV}^{3}}{1.0 [\text{TeV}^{3}]}
 \right),
\]
 which is much worse than even the trivial constraint 
 derived from perturbativity.}
\begin{eqnarray}\label{eq:3lpNum}
(m_{\nu})_{ee} \sim  1 \times 10^{-5}\hskip1mm {\rm eV},
&
(m_{\nu})_{\mu\mu} \sim 2 \times 10^{-2}\hskip1mm {\rm eV},
&
(m_{\nu})_{\tau\tau} \sim 0.3 \hskip1mm {\rm eV}.
\end{eqnarray}
This implies that the mass mechanism contribution to $\znbb$ is
guaranteed to be sub-dominant in this class of models.  Also,
Eq.~(\ref{eq:3lpNum}) shows that 3-loop models can potentially explain
neutrino oscillation data only if all of the involved Yukawa couplings 
are set to be ${\cal O}(1)$.  
Thus, we expect such models to be quite constrained from upper limits on
flavour violating decays of charged leptons.  We will not discuss this
class of models in more detail here, since their detailed
phenomenology is outside the scope of this paper.

The effective $d=9$ operator resulting from the Lagrangian
Eq.~(\ref{eq:lag3lp}) can be written with the following linear
combination of the basis operators $\mathcal{O}_{i\in\{1\text{-}5\}}^{SR}$
of the short-range contributions to $\znbb$ decay as:
\begin{align}
 \mathcal{L}_{\text{eff}}
 =&
 \frac{(Y_{uuS})_{11} (Y_{e\psi S})_{e} (Y_{Q \psi S})_{1} {(Y_{LdS})_{e1}}}
 {m_{S_{1/3}}^{2} m_{S_{1/6}}^{2} m_{\psi}}
 \left[
 \frac{1}{16 i}
 (\mathcal{O}_{4}^{SR})_{\{RR\}R}
 -
 \frac{1}{16}
 (\mathcal{O}_{5})^{SR}_{\{RR\}R}
 \right]
 \label{eq:Leff-3loop-example}
\end{align}
and the experimental bound Eq.~\eqref{eq:Thalf-bound-Xe}
constrains a combination of the coefficients to be:
\begin{align}\label{eq:Lim-3loop-example}
 (Y_{uuS})_{11} (Y_{e\psi S})_{e} (Y_{Q\psi S})_{1} {(Y_{LdS})_{e1}}
 <
 1.5\cdot 10^{-2}
 \left(
 \frac
 {m_{S_{1/6}}^{2} m_{S_{1/3}}^{2} m_{\psi}}{1.0 [\text{TeV}^{5}]}
 \right).
\end{align}
The difference in the short-range bounds, 
Eq.~(\ref{eq:Lim-2loop-example}) and Eq.~(\ref{eq:Lim-3loop-example}), 
is due to the different values of nuclear matrix elements entering the
transition operator.  All other three-loop models will have
constraints similar to the ones discussed here.  They are listed in
Table~\ref{Tab:3lp} in the appendix.

\subsection{4-loop models}

\begin{figure}[t]
\includegraphics[width=0.3\linewidth]{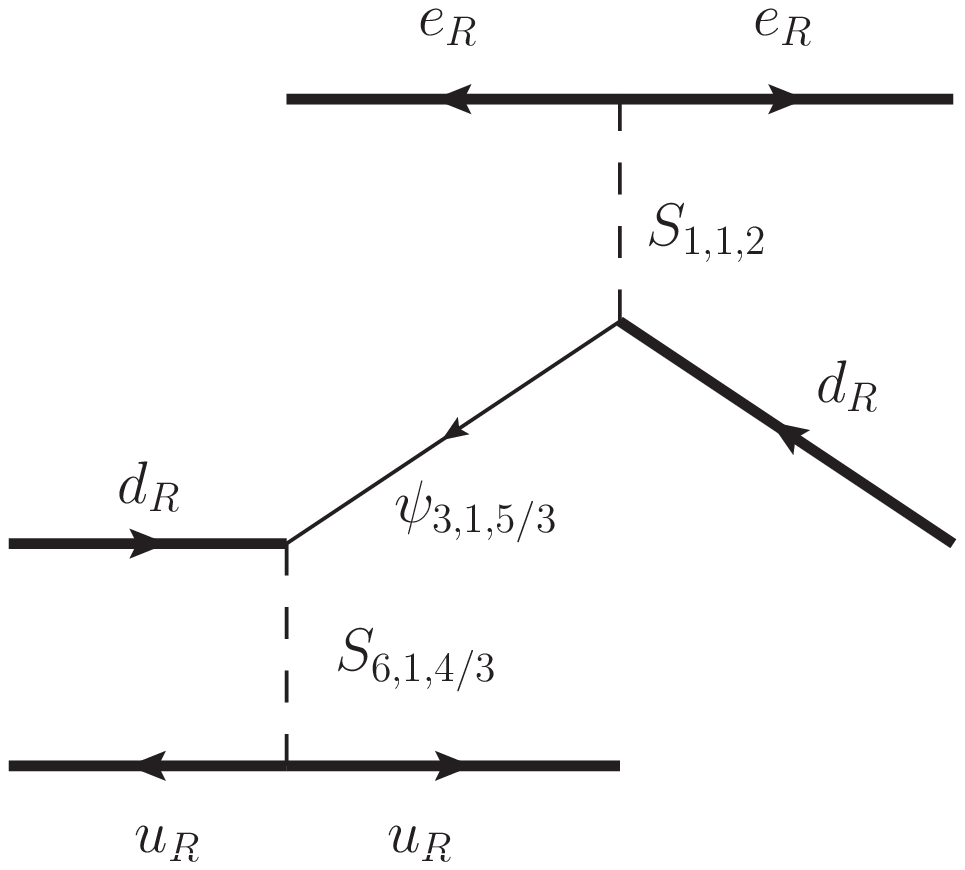}
\hskip5mm
\includegraphics[width=0.6\linewidth]{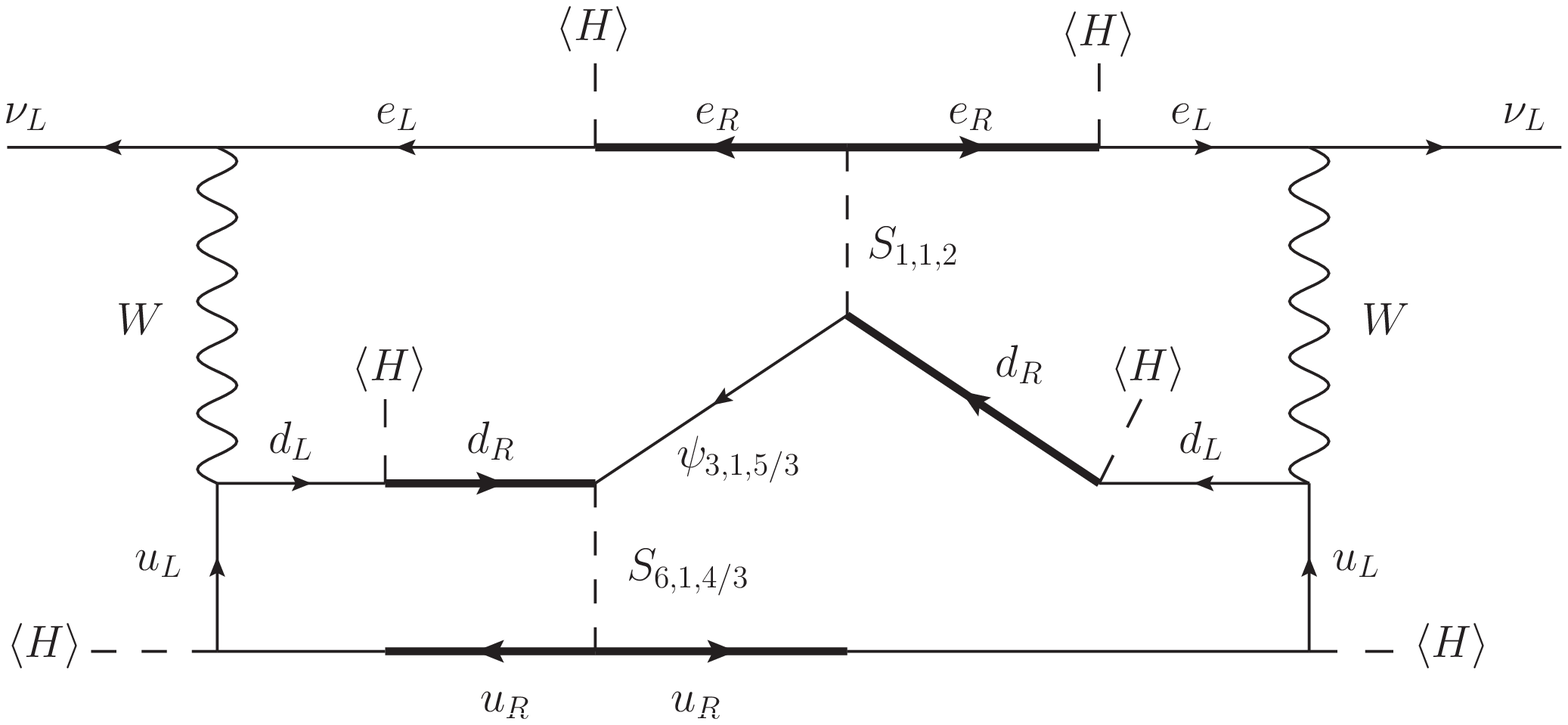}
\caption{An example of a four-loop neutrino mass model. To the left: 
$\znbb$ decay via 
the $d=9$ operator 
$(\overline{u_R}\hspace{0.08cm}\overline{u_R})
(d_R)(d_R)(\overline{e_R} \hspace{0.08cm} \overline{e_R})$. 
To the right: four-loop $d=9$ neutrino mass, see text.}
\label{fig:4lp}
\end{figure}

Finally, 
all operators ${\cal O}_{-} = 
\overline{e_R} \hspace{0.08cm} \overline{e_R}
\hspace{0.08cm}\overline{u_{R}} d_R \overline{u_{R}} d_R$, with
exception of decomposition T-I-5-i (see Table~\ref{Tab:0lp} in the
appendix), will lead to four-loop neutrino mass models.  The simplest
possibility to construct a four-loop diagram for these operators is to
use a SM charged current interaction.  We estimate that this
gives the dominant contribution to the neutrino mass.  Here we show
an example of the decompositions of the $\znbb$ decay operator
$\mathcal{O}_{-}$ in Fig.~\ref{fig:4lp}, which is based on decomposition
T-I-3-ii $(\overline{u_R}\hspace{0.08cm}\overline{u_R})(d_R)(d_R)
(\overline{e_R}\hspace{0.08cm}\overline{e_R})$.  The four-loop
neutrino mass diagram based on this decomposition is also shown on the
right.
%
%
Taking the limit $m_{\psi_{3,1,5/3}} \sim m_{S_{6,1,4/3}} \sim
m_{S_{1,1,2}} \gg m_W, m_t$ one can estimate the order of magnitude of
this four-loop diagram, which is,
\begin{eqnarray}
(m_{\nu})_{\alpha\beta} 
&\sim &
\frac{g^{4}}{(16 \pi^2)^4}
\frac{ m_{e_{\alpha}}m_{e_{\beta}} 
m_{u_i} m_{u_j} m_{d_i} m_{d_j} }
{m_{\psi_{3,1,5/3}}m_{S_{6,1,4/3}}^2 m_{S_{1,1,2}}^2}
(Y_{\psi d S})_{i} 
(Y_{e e S})_{\alpha \beta}
(Y_{uu S})_{ij}
(Y_{\psi d S})_{j}.
\label{eq:4lp}
\end{eqnarray}
The expression Eq.~\eqref{eq:4lp} shows that this four-loop
contribution would yield only $(m_{\nu})_{\tau\tau} \sim {\cal
  O}(10^{-10})$ eV for $m_{\psi_{3,1,5/3}} \sim m_{S_{6,1,4/3}} \sim
m_{S_{1,1,2}} \sim 1 $ TeV, even when choosing all SM fermion masses
to be third generation.  Since this is obviously many orders of
magnitude below the values of neutrino masses required from
oscillation experiments, models of this category by themselves cannot
be considered realistic.  Of course, neutrinos could be quasi-Dirac
particles, explaining oscillation data by Dirac mass terms (using
additionally introduced right-handed neutrinos), while $\znbb$ decay
is dominated by the short-range diagrams such as the one shown in
Fig.~\ref{fig:4lp}.  However, constraints on Yukawa couplings will be
similar to those derived in the previous subsections in
Eq.~\eqref{eq:Leff-2loop-example} and
Eq.~\eqref{eq:Leff-3loop-example}, with the exact value depending on
the decomposition under consideration.  All four-loop cases are listed
in Table~\ref{Tab:4lp} in the appendix.


\section{A concrete 2-loop example}
\label{sect:exa}

In this section we will discuss one concrete genuine 2-loop neutrino
mass model in some more detail. The example we choose is based on
the decomposition T-I-4-ii-b of the Babu-Leung operator ${\cal O}_{11}$, 
which has not been discussed in the literature before. 
However, all $\znbb$ decompositions that generate 2-loop
neutrino masses behave quite similarly, in what concerns fits for
neutrino oscillation data and constraints from lepton flavour
violation searches. 
Thus, most of the discussion presented below can be applied 
qualitatively also to all other 2-loop decompositions.

Any model of neutrino mass must not only generate the correct neutrino
mass scale, but also be able to explain the observed neutrino mixing 
angles. For a recent update of all oscillation data, see, for example,
\cite{Forero:2014bxa}. In addition, since the neutrino mass matrix has
a non-trivial flavour pattern, one also expects that low-energy 
models\footnote{By ``low-energy'' we mean TeV-scale, as in contrast to
  ``high-scale'' seesaw models.} of neutrino mass are constrained by
charged lepton flavour violation (LFV) searches. Here we will discuss only
$\mu\to e \gamma$, since the experimental upper limit on this process
provides usually the most stringent constraints in many models.  We
note that the authors of \cite{Angel:2013hla} present a 2-loop model,
which corresponds to the decomposition T-I-5-i and discuss also the 
constraints from other LFV searches, which we expect are very similar 
in our example.

Below we will discuss two variations of the model based on T-I-4-ii-b.
First (in Sec.~\ref{Sec:one-psi}), we introduce only one copy of the
exotic fermion $\psi_{6,2,-1/6}$ for simplicity.  Next (in
Sec.~\ref{Sec:three-psi}), we will allow to have three copies of these
fermions, which allows to fit also quasi-degenerate neutrinos.

\subsection{General formulas for neutrino masses and $\mu\to e \gamma$}

The Yukawa part of the Lagrangian describing the interactions between 
the exotic diquark, $S_{6,3,1/3}$, the leptoquark, $S_{3,2,1/6}$, and 
the coloured vector-like fermion, 
$\psi_{6,2,-1/6}$ can be written as:
\begin{eqnarray}
 \mathcal{L} =&
 (Y_{QQS})_{ij} 
 (\overline{Q}_{i} \vec{\tau} \cdot Q^{c}_{j}) 
 \hat{\vec{S}}_{6,3,1/3}
 +
 (Y_{L\psi S})_{\alpha k} 
 (\overline{L}_{\alpha} \vec{\tau} \psi_{k})
 \vec{S}_{6,3,1/3}^{\dagger}
 \\
 &
 + (Y_{\psi dS})_{ki} 
 (\overline{\hat{\psi}_{k}}
 d_{R,i}) 
 S_{3,2,1/6}
 + 
 (Y_{LdS})_{\alpha i}
 (\overline{L}_{\alpha}{d_{R,i}})
 \cdot 
 S_{3,2,1/6}^{\dagger}
\label{eq:lag}
\nonumber 
\end{eqnarray}
Here, $i,j$ are generation indices for quarks, we use Greek indices 
for lepton generations and $k$ runs over the number of copies of 
$\psi_{6,2,-1/6}$. 
\begin{figure}[t]
\includegraphics[width=10cm]{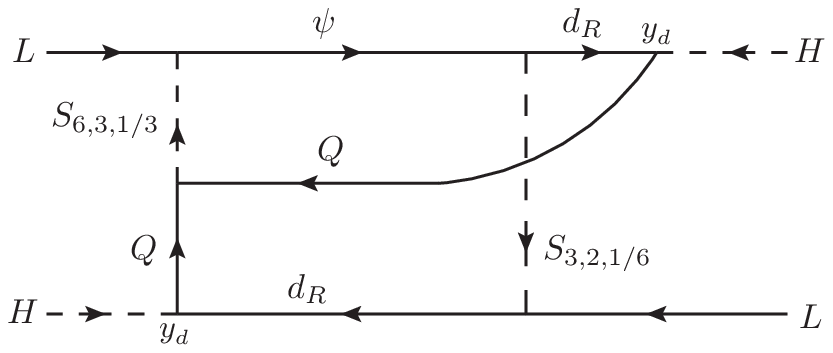}
\caption{Two loop diagram for neutrino masses generated 
by the Lagrangian in Eq.~\eqref{eq:lag}.}
\label{fig:2loop-example}
\end{figure}
This Lagrangian generates a 2-loop diagram which corresponds to PTBM-4
according to classification by \cite{Sierra:2014rxa}. Following 
the general formulas from \cite{Sierra:2014rxa},
the neutrino mass matrix can be expressed as:
\begin{eqnarray}
(m_{\nu})_{\alpha\beta}&=&\frac{N_c  m_{\psi_{k}}}{(16\pi^2)^{2}}
\left[
(Y_{QQS})_{ij}
(Y_{L\psi S})_{\alpha k}
(Y_{\psi dS})_{ki}
(Y_{LdS})_{\beta j}
+
(Y_{QQS})_{ij}
(Y_{L\psi S})_{\beta k}
(Y_{\psi dS})_{ki}
(Y_{LdS})_{\alpha j}
\right]
\nonumber
\\ 
 &  & \times
F(m_{\psi_{k}},m_{S_{3,2,1/6}},m_{d_i},m_{S_{6,3,1/3}},m_{d_j})
\label{eq:mnu-2loop-example}
\end{eqnarray}
where $N_c$ is a colour factor, with $N_c=6$ for this model. Summation
over all flavour indices $i,j,k$ is implied.
$F(m_{\psi_{k}},m_{S_{3,2,1/6}},m_{d_i},m_{S_{6,3,1/3}},m_{d_j})$ is a
loop integral defined as:
\begin{eqnarray}
\label{eq:int1}
F(m_{\psi_{k}},m_{S_{3,2,1/6}},m_{d_i},m_{S_{6,3,1/3}},m_{d_j}) 
 = \frac{m_{d_i}m_{d_j}}{\pi^4}\\ \nonumber
\times \int d^4q \int d^4k 
 \frac{1}{(q^2 - m^2_{\psi_{k}})(q^2 - m^2_{S_{3,2,1/6}})(k^2 - m^2_{d_i}) 
(k^2 - m^2_{S_{6,3,1/3}})((q+k)^2 - m^2_{d_j})}.
\end{eqnarray}
Due to the strong hierarchy in down-type quark masses, the integral 
in Eq.~(\ref{eq:int1}) is completely dominated by the contributions 
from bottom quarks, unless the couplings $Y_{QQS}^{ij}$, $Y_{\psi dS}^{ki}$ 
and $Y_{LdS}^{\beta j}$ follow an equally strong inverse hierarchy. 
We have thus taken into account only the contributions from 
bottom quark exchange 
in our numerical evaluation. 
Since it is convenient to rewrite Eq.~(\ref{eq:int1}) 
in terms of dimensionless 
parameters, we define $z,\,r$ and $t_k$ as 
\begin{eqnarray}
z\equiv\frac{m^2_{S_{3,2,1/6}}}{m^2_{b}},
\quad\quad 
r\equiv\frac{m^2_{S_{6,3,1/3}}}{m^2_{b}},
\quad\quad \mathrm{and}\quad\quad 
t_{k}\equiv\frac{m^2_{\psi_{k}}}{m^2_{b}}.
\label{eq:def}
\end{eqnarray}
Rescaling the loop momenta, the integral can then be written as:
\begin{eqnarray}
{\hat I}(t_k,z,1,r)=\frac{1}{\pi^4}\int d^4q \int d^4k 
\frac{1}{(q^2 - t_k)(q^2 - z)(k^2 - 1)(k^2 - r)((q+k)^2 - 1)}.
\label{eq:int2}
\end{eqnarray}
This integral has been analytically calculated several times in
literature. We follow the procedure outlined in \cite{Sierra:2014rxa},
based on the calculations of \cite{Angel:2013hla}.
We will fit the neutrino mass calculated with
  Eq.~\eqref{eq:mnu-2loop-example} to neutrino oscillation data.  The
  discussion depends on the number of copies of the fermion mediator
  $\psi_{6,2,-1/6}$; as mentioned above we will discuss two different
  scenarios in the following subsections.

The rate of the LFV process $\mu\rightarrow e \gamma$ has also
been calculated several times in literature. 
We adapt the general formulas shown in \cite{Lavoura:2003xp} 
for our particular case.  
The amplitude for $\mu\rightarrow e \gamma$ decay is given by
\begin{eqnarray}
{\cal{M}}(\mu \rightarrow e \gamma)=
e \sigma_R 
\epsilon^*_{\alpha}
q_{\beta} 
\bar{u}(p_e)
i \sigma^{\alpha\beta}
u(p_{\mu}),
\end{eqnarray}
where $e$ is the electric charge, 
$\epsilon_{\alpha}$ is the photon
polarization vector, 
$q_{\beta}$ is the momentum of photon,
and 
$\sigma^{\alpha \beta} \equiv (i/2) [\gamma^{\alpha}, \gamma^{\beta}]$.
There are two contributions to the coefficient $\sigma_R$
in the model we are discussing; 
one is the one-loop diagram with the diquark and the exotic fermion, 
the other is that with a bottom quark and the leptoquark. 
The total $\sigma_R$ is given by
\begin{eqnarray}
\sigma_R=
i 
\frac{m_{\mu}}{16\pi^{2}}
\left[
18
\sum_{k}
(Y_{L\psi S})_{\mu k} (Y_{L\psi S}^{\dagger})_{k e}
\frac{F_{2} (x_{k})}{m_{S_{6,3,1/3}}^{2}}
+
(Y_{LdS})_{\mu 3} (Y_{LdS}^{\dagger})_{3 e}
\frac{2F_{2} (x_{S}) - F_{1}(x_{S})}{m_{S_{3,2,1/6}}^{2}}
\right],
\label{eq:sigmaigmaR}
\end{eqnarray}
where 
$x_{k} \equiv \frac{m^2_{\psi_k}}{m^2_{S_{6,3,1/3}}}$ and
$x_{S}\equiv\frac{m^2_{b}}{m^2_{S_{3,2,1/6}}}$. 
The functions $F_{1}(x)$ and $F_{2}(x)$ are defined as
\begin{align}
F_{1}(x)=&
 \frac{x^{2} - 5 x -2}{12(x-1)^{3}} + \frac{ x \ln x}{2 (x-1)^{4}},
 \\
F_{2}(x)=&
 \frac{2 x^{2} + 5 x -1}{12 (x-1)^{3}}
 -
 \frac{x^{2} \ln x}{2 (x-1)^{4}},
\end{align}
which are presented in Eqs.~(40) and (41) in \cite{Lavoura:2003xp}.
The branching ratio for the $\mu\rightarrow e\gamma$ process, 
neglecting the electron mass, can then be expressed with 
the coefficient $\sigma_{R}$ as
\begin{eqnarray}\label{eq:brmueg}
\mathrm{Br}(\mu\rightarrow e \gamma)
\simeq
\frac{48 \pi^{3} \alpha \left| \sigma_{R} \right|^{2}}{G_{F}^{2} m_{\mu}^{2}},
\end{eqnarray}
where $\alpha$ is the fine-structure constant.

\subsection{One generation of $\psi_{6,2,1/6}$}
\label{Sec:one-psi}
\begin{center}
\begin{figure}[t]
\includegraphics[width=0.5\linewidth]{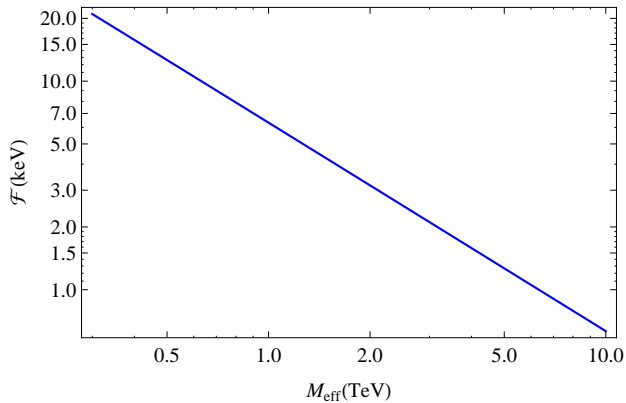}
\caption{The prefactor ${\cal F}$, defined in Eq.~(\ref{eq:prefac}),
  for $(Y_{QQS})_{33}=(Y_{\psi dS})_{3}=1$ in units of keV as a
  function of $M_{\text{eff}}$ in TeV. Here, $M_{\text{eff}} =
  m_{\psi} \simeq m_{S_{3,2,1/6}}\simeq m_{S_{6,3,1/3}}$.}
\label{fig:prefac}
\end{figure}
\end{center}

The analysis presented in this section uses very similar methods to
the one in ref. \cite{Choubey:2012ux}, where double beta decay and LFV is discussed 
in a 1-loop neutrino mass model containing colour octets.
We will first consider a variant of the model, in which there is only
one copy of the fermion mediator $\psi_{6,2,1/6}$.  The expression for
the neutrino mass matrix in this case is given by suppressing the
index for $\psi$ in Eq.~\eqref{eq:mnu-2loop-example}, which gives
\begin{eqnarray}
(m_{\nu})_{\alpha\beta}&=& 
\left[
(Y_{L\psi S})_{\alpha}
(Y_{LdS})_{\beta 3}
+
(Y_{L\psi S})_{\beta}
(Y_{LdS})_{\alpha 3}
\right] {\cal F},
\label{eq:mnu2}
\end{eqnarray}
where 
\begin{equation}
{\cal F}= 
\frac{N_c  m_{\psi}}{(16\pi^2)^{2}}
(Y_{QQS})_{33}
(Y_{\psi dS})_{3}
{\hat I}(t,z,1,r).
\label{eq:prefac}
\end{equation}
Since $\det(m_{\nu})=0$ in this case, this version of the model can
fit only to the hierarchical neutrino mass spectra (both of the normal
and the inverse type), but not to the degenerate spectrum.\footnote{%
  However, we remind that this is true, only when contributions to
  neutrino masses from the first and the second generation quarks are
  negligible.}  The eigenvalues of Eq.~\eqref{eq:mnu2} can be easily
found to be:
\begin{equation}
m_{\nu_{1(3)}} = 0,
\quad 
m_{\nu_{2,3(1,2)}} = 
\left[
\sum_{\alpha} (Y_{L\psi S})_{\alpha} (Y_{LdS})_{\alpha 3} 
\mp 
\sqrt{
\sum_{\alpha}\left|(Y_{L\psi S})_{\alpha}\right|^{2} 
\sum_{\alpha}\left|(Y_{LdS})_{\alpha 3} \right|^{2}
}
\right]
{\cal F}
\label{eq:mnu-eigenvalues-one-psi}
\end{equation}
for normal hierarchy (inverted hierarchy).  In Fig.~\ref{fig:prefac},
we give typical values for the common factor ${\cal F}$, which are
calculated with the assumption of a nearly degenerate spectrum of
heavy particles with the mass scale $M_{\text{eff}} \equiv m_{\psi}
\simeq m_{S_{3,2,1/6}} \simeq m_{S_{6,3,1/3}}$ and
$(Y_{QQS})_{33}=(Y_{\psi dS})_{3}=1$.  From
Eq.~\eqref{eq:mnu-eigenvalues-one-psi} and Fig.~\ref{fig:prefac}, one
can estimate the constraints from neutrino masses on the size of the
Yukawa couplings.  In order to reproduce the neutrino mass suggested
by atmospheric neutrino oscillation ($m_{\nu_{3}} \sim 0.05$ eV),
keeping the common mass scale $M_{\text{eff}}$ at 1 TeV, the Yukawa
couplings $Y_{L\psi S}^{\alpha}$ and $Y_{LdS}^{\beta}$ must be set
typically to $\mathcal{O}(10^{-2})$.
Although the eigenvectors of Eq.~\eqref{eq:mnu2} can be calculated
analytically, numerical exercises might be more helpful to grasp
phenomenological aspects of the model.  In the following, we will
generate random sets of Yukawa couplings $(Y_{L\psi S})_{\alpha}$ and
$(Y_{LdS})_{\beta 3}$ under the condition that they reproduce the
latest neutrino oscillation data \cite{Forero:2014bxa} within 3
$\sigma$ C.L.  We will only show plots with the Yukawa couplings that
fit the normal hierarchical neutrino spectrum, because plots for the
inverse hierarchical case look qualitatively similar.

Let us start the discussion with double beta decay.  The half-life of
$\znbb$ induced by the Majorana mass of neutrino is proportional to
the inverse-square of the effective neutrino mass:
\begin{equation}
T_{1/2}^{\znbb} 
\propto 
\Big[ (m_{\nu})_{ee} \Big]^{-2}
\end{equation}
For the normal hierarchy case, the effective mass is roughly given as
$(m_{\nu})_{ee} \sim s_{12}^{2} \sqrt{\Delta m_{21}^{2}} \sim 3 \times
10^{-3}$ eV, which results in half-lives of the order of
$T_{1/2}^{\znbb} \sim 10^{29}$ ys.  For the inverse hierarchy case,
one finds $(m_{\nu})_{ee} \sim \sqrt{\Delta m_{31}^{2}} \sim 5 \times
10^{-2}$ eV, which leads to $T_{1/2}^{\znbb} \sim 10^{27}$ ys.  The
current experimental limits to the half-life of $^{136}$Xe and
$^{76}$Ge are of the order of $T_{1/2}^{\znbb} \sim (1-2)\times
10^{25}$ ys~\cite{KamLANDZen:2012aa,Albert:2014awa,Agostini:2013mzu},
while the next round of experiments could reach eventually
$T_{1/2}^{\znbb} \sim 10^{27}$ ys.  Therefore, only the inverse
hierarchical case can result in measurable half-lifes.

\begin{center}
\begin{figure}[t]
\includegraphics[width=0.5\linewidth]{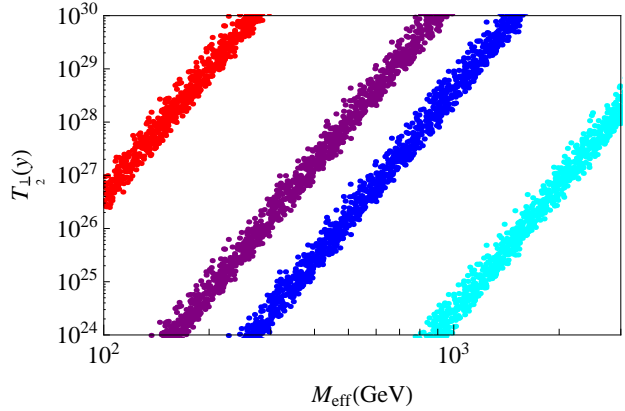}
\caption{Calculated half-lives for $\znbb$ decay of $^{136}$Xe,
considering only the short range contribution to the decay rate.
The different colours correspond to (from left to right) 
$\eta_{31}=1$, $5$, $10$ and $50$. If the third and first 
generation couplings are of the same order, $\znbb$ decay 
will have an immeasurably large half-life in the variant of the 
model with only one copy of $\psi_{6,2,1/6}$.}
\label{fig:tsrrnd}
\end{figure}
\end{center}
The short range-contribution due to the $d=9$ operator
(cf. Eq.~\eqref{eq:Leff-2loop-example}) is proportional to the
following combinations of the parameters:
\begin{equation}
T_{1/2}^{\znbb} \propto \left[ 
\frac{(Y_{QQS})_{11} (Y_{L\psi S})_{e} (Y_{\psi d S})_{1}
{(Y_{LdS})_{e1}}}
{M_{\text{eff}}^{5}}
       \right]^{-2},
\end{equation}
i.e., while the neutrino mass matrix is dominated by Yukawa couplings
of the third quark generation, double beta decay is sensitive only
to the Yukawa couplings that couple to the first generation quarks.
To discuss the relation between these two contributions to $\znbb$, we
introduce a scaling factor
\begin{equation}
\eta_{31} \equiv
\left[
\frac{(Y_{QQS})_{11} (Y_{\psi d S})_{1} {(Y_{LdS})_{e1}}}
{(Y_{QQS})_{33} (Y_{\psi d S})_{3} {(Y_{LdS})_{e3}}}
\right]^{1/3},
\label{eq:eta31}
\end{equation}
i.e., $\eta_{31}=1$ corresponds to quark flavour universality in
the Yukawa couplings.  In Fig.~\ref{fig:tsrrnd}, we calculate
half-lives induced from the short-range contribution with randomly
generated Yukawa couplings, assuming different values of $\eta_{31}
\in\{1,5,10,50\}$.
Taking $\eta_{31}=1$, we find quite long half-lives, too
  large to be measured in realistic experiments.  On the other hand,
with $\eta_{31}=10$, we find a lower limit on $M_{\text{eff}}$, which
is approximately $M_{\text{eff}} \gtrsim 400$ GeV.
This is still not competitive with leptoquark searches 
at the LHC, which places constraints on the masses 
of leptoquarks at $m_{S_{3,2,1/6}} \sim (600-1000)$ GeV 
(depending on generation) already in the first 
run~\cite{CMS:2014qpa,CMS:2012zva,Khachatryan:2014ura}
Thus, it is reasonable to conclude that, as for the mass mechanism
contribution, the short-range contribution to the half-life is also
expected to be too long to be measured in the near future in this
variant of the model, unless $\eta_{31}$ is very large
(i.e., for highly inverse hierarchical Yukawa
  couplings in terms of the quark generations).

\begin{center}
\begin{figure}[t]
\includegraphics[width=0.5\linewidth]{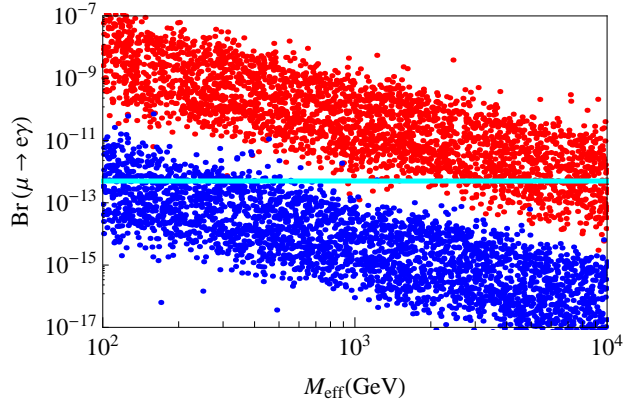}
\caption{Branching ratio Br($\mu\to e\gamma$) as a function of 
$M_{\text{eff}}$ in GeV. Red (blue) points have been calculated with 
$(Y_{QQS})_{33}=(Y_{\psi dS})_{3}=10^{-2}$ ($10^{-1}$). 
The horizontal line is the experimental upper limit from 
the MEG experiment~\cite{Adam:2013mnn}.}
\label{fig:muegamrnd}
\end{figure}
\end{center}

Finally, we discuss briefly the LFV process $\mu\to e\gamma$.  While
the neutrino mass matrix is proportional to the combination of the
Yukawa couplings $(Y_{L\psi S})_{\alpha} (Y_{LdS})_{\beta 3}
(Y_{QQS})_{33} (Y_{\psi dS})_{3}$, the branching ratio of Br($\mu\to
e\gamma$) depends only on $|(Y_{L\psi S})_{2} (Y_{L\psi
  S}^{\dagger})_{1}|^2$ and $|(Y_{LdS})_{2 3} (Y_{LdS}^{\dagger})_{1
  3}|^2$.  In Fig.~\ref{fig:muegamrnd}, we show Br($\mu\to e\gamma$)
for two different choices of the set of $(Y_{QQS})_{33}$ and $(Y_{\psi
  dS})_{3}$, as a function of $M_{\text{eff}}$, assuming again that
the mass spectra of heavy particles are nearly degenerate for
simplicity.
With the choice $(Y_{QQS})_{33}=(Y_{\psi dS})_{3}=10^{-2}$ the LFV
process can place a bound on $M_{\text{eff}}$ of roughly $M_{\text{eff}} \gsim$
TeV. However, the bound depends strongly on the exact choice of the
remaining Yukawa couplings $Y_{L \psi S}$ and $Y_{L d S}$.
On the other hand, the LFV process can exclude only few
  parameter points in the case of $(Y_{QQS})_{33}=(Y_{\psi
    dS})_{3}=10^{-1}$, and no useful limit on $M_{\text{eff}}$ can be
  derived.

As we have seen in this subsection, this variant of the model can
reproduce oscillation data without running into conflict with LFV
searches.  However, it is interesting to note that 
the $(Y_{QQS})_{33}$ interaction with the size required for
reproducing neutrino masses will result in sizeable decay rates of
the diquark into third generation quarks (both tops and bottoms),
which should be testable at the LHC.

\subsection{Three generations of $\psi_{6,2,1/6}$}
\label{Sec:three-psi}

Next, we examine the model with more than one copy of
the fermion mediator $\psi_{6,2,-1/6}$, 
which can fit not only hierarchical neutrino mass spectra,
but also a quasi-degenerate spectrum.
Here, we introduce three copies of $\psi_{6,2,-1/6}$, 
motivated by the observed generations of SM fermions. 

\begin{figure}[t]
\centering
\includegraphics[width=0.45\linewidth]{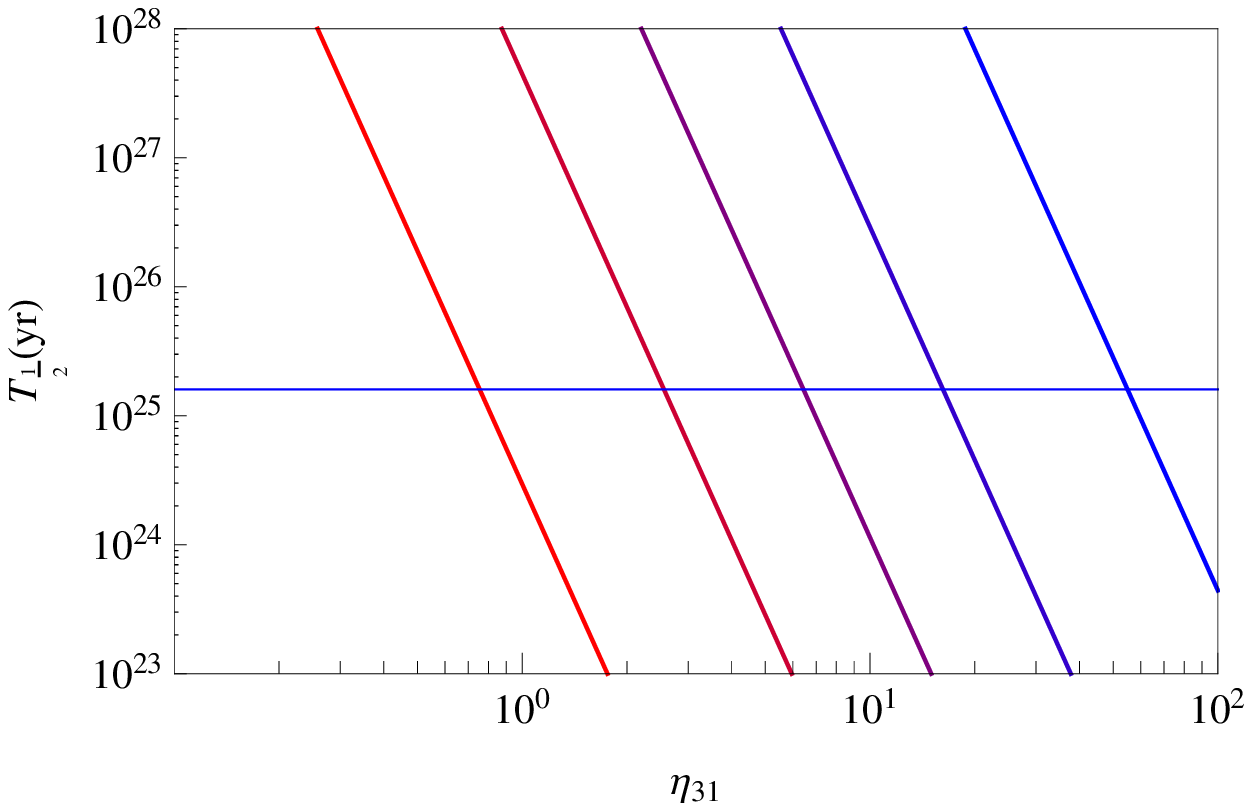}
\includegraphics[width=0.45\linewidth]{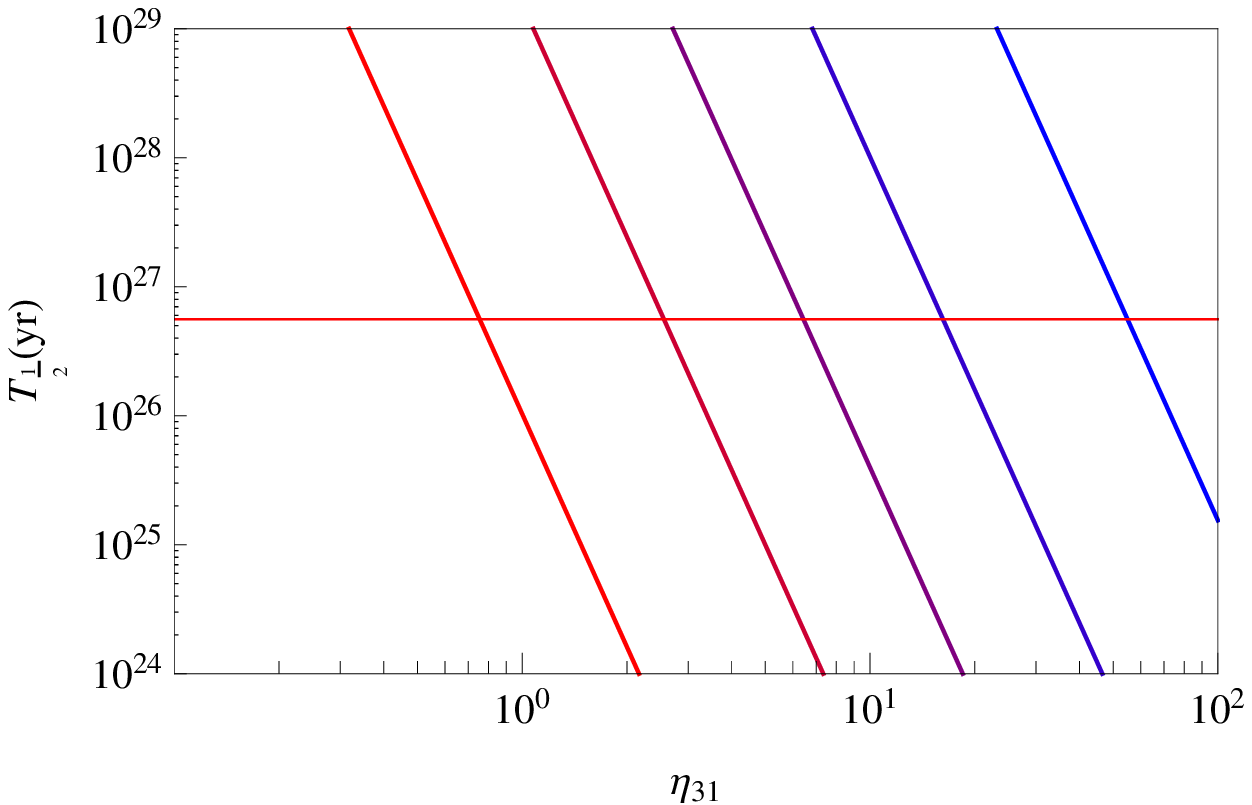}
\includegraphics[width=0.45\linewidth]{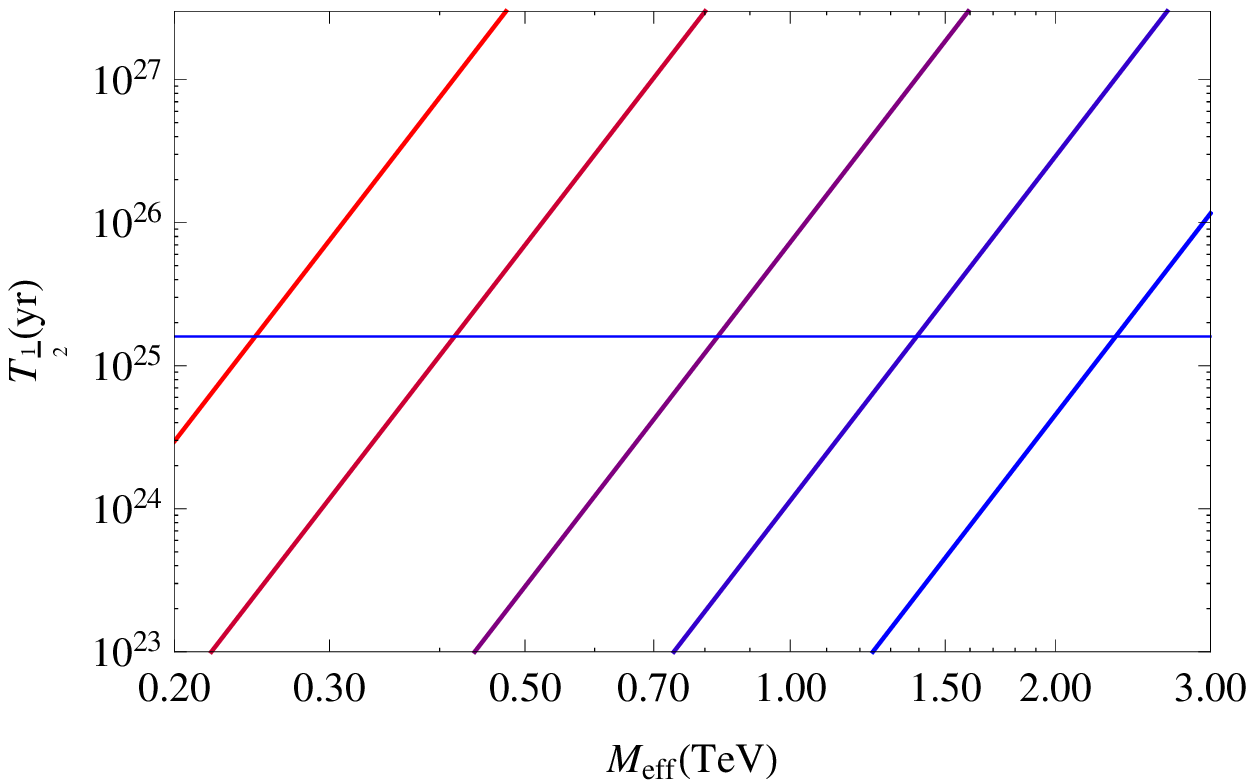}
\includegraphics[width=0.45\linewidth]{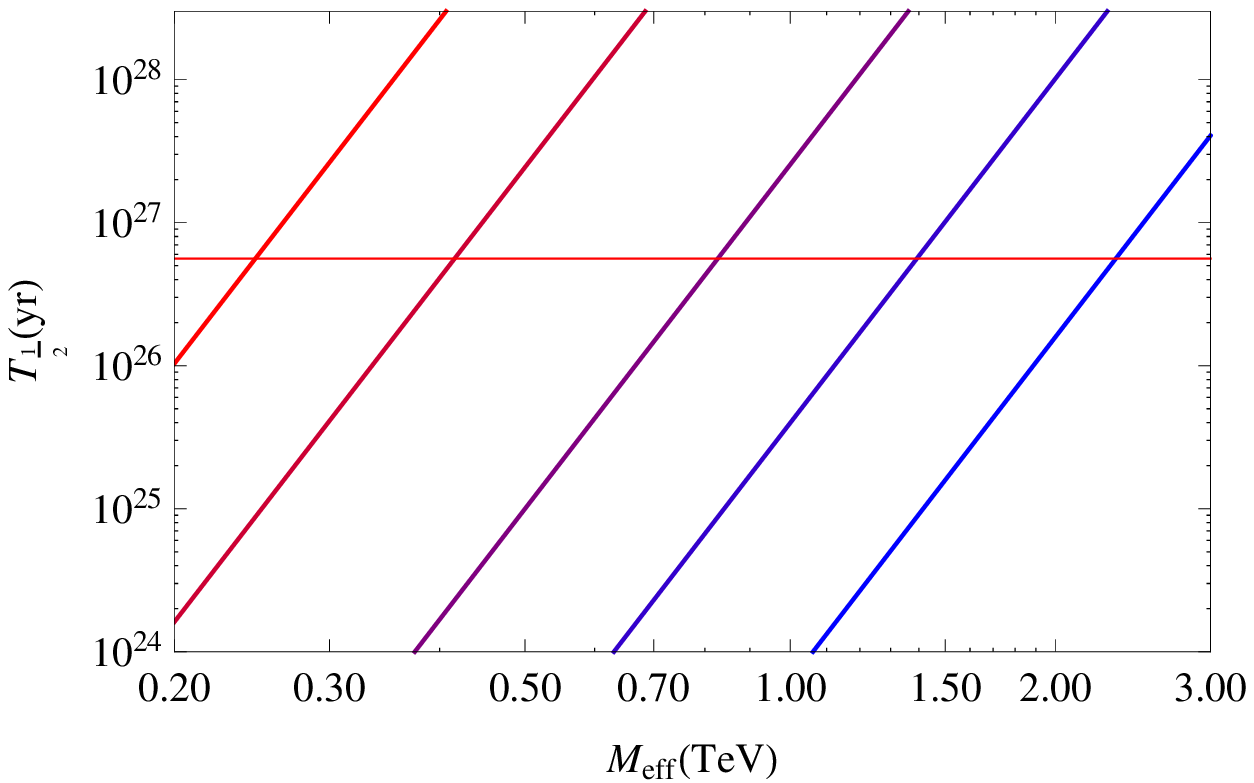}
\includegraphics[width=0.45\linewidth]{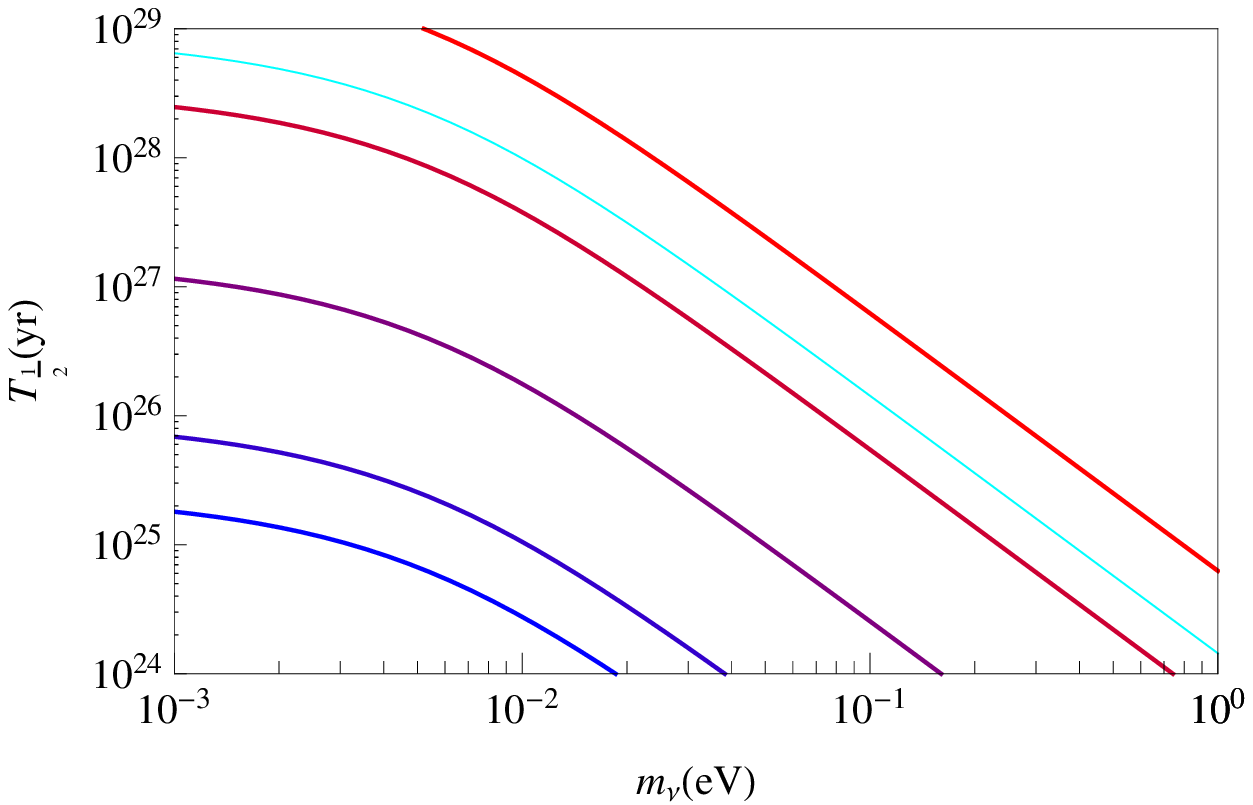}
\includegraphics[width=0.45\linewidth]{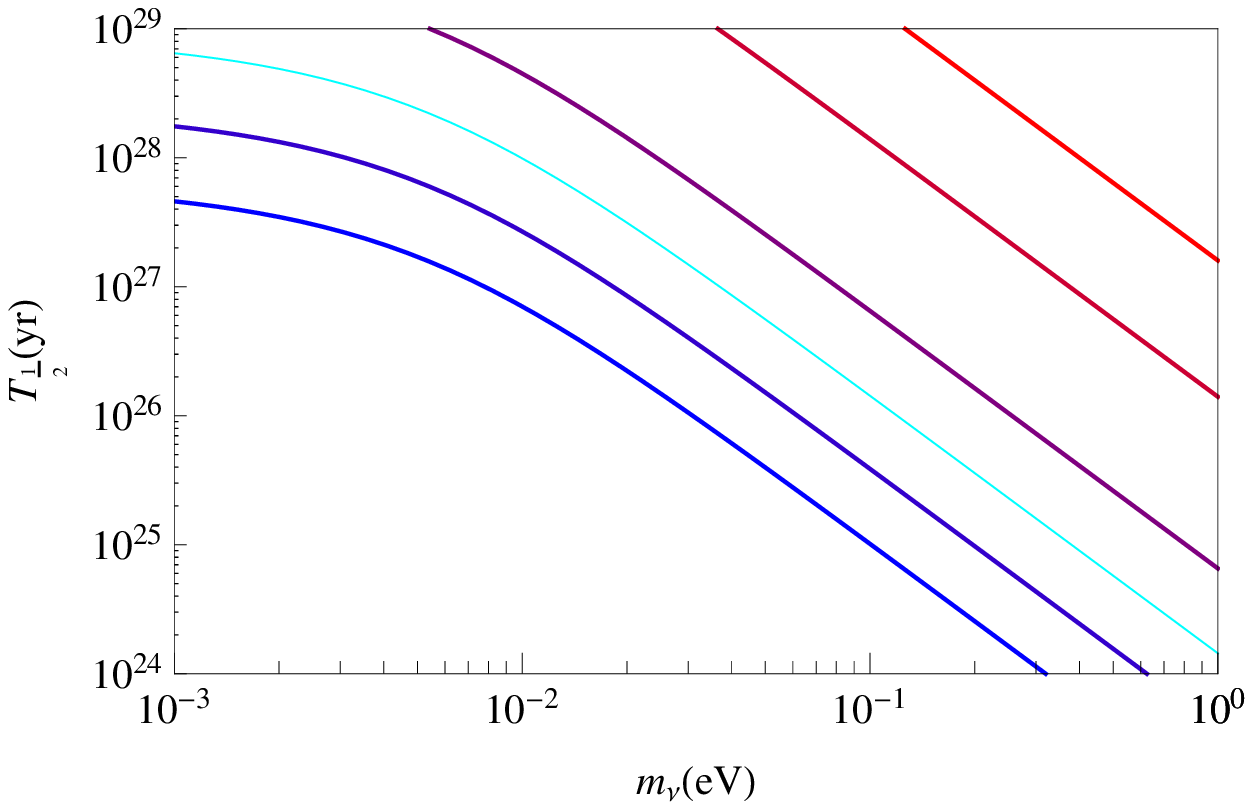}
\caption{Calculated half-lives for $\znbb$ decay of $^{136}$Xe,
  considering only the short-range contribution to the decay rate.
  The various plots show from top to bottom: $T_{1/2}$ versus
  $\eta_{31}$, $M_{\text{eff}}$ and $m_{\nu_1}$, 
 for a fixed set of neutrino
 oscillation parameters and different choices of $\eta_{31}$,
 $M_{\text{eff}}$ and $m_{\nu_1}$ as follows: 
 In the top plots to the left (right) $m_{\nu_1}=0.3$ eV (0.05 eV), 
 different lines show different
 choices of $M_{\text{eff}}$; from left to right: 
 $M_{\text{eff}} = 0.2$, $0.5$,
 $1$, $2$ and $5$ TeV. In the middle panel, to the left (right)
 $m_{\nu_1}=0.3$ eV (0.05 eV), different lines show different choices
 of $\eta_{31}$; from left to right: $\eta_{31}=1$, $2$, $5$, $10$
 and $20$.  In the lower panel, to the left (right): $M_{\text{eff}}=0.5$
 TeV (1 TeV), different lines are for different choices for
 $\eta_{31}$; from top to bottom: $\eta_{31}=2$, $3$, $5$, $8$ and
 $10$. For comparison we also show the half-lives for the neutrino
 mass mechanism as horizontal lines in the top and middle panel and
 as cyan lines in the lower panel. Oscillation parameters are
 $s_{13}^2$, $\Delta m_{31}^2$ and $\Delta m_{21}^2$ at
 their best-fit values, while $s_{23}^2=1/2$ and
 $s_{12}^2=1/3$ for the case of normal hierarchy.}
\label{fig:tsr}
\end{figure}

To simplify the following discussion,
we adopt the following ansatz in the flavour structure of the 
Yukawa couplings:\footnote{%
This ansatz can be justified 
by introducing a flavour symmetry
with flavour-charged scalar (flavon) fields.
}
\begin{equation}
(Y_{LdS})_{\alpha 3}
(Y_{\psi dS})_{k 3} 
= y (Y_{L\psi S})_{\alpha k}.
\end{equation}
With this ansatz, all the flavour structure relevant to phenomenology
can be represented with only one vector (apart from a possible
normalization factor $y$).  The neutrino mass matrix can then be cast
into the form:
\begin{eqnarray}
(m_{\nu})_{\alpha\beta} 
=  
(\Lambda)_{\alpha k}
\hat{I}_{k}
(\Lambda^{T})_{k \beta},
\label{eq:mnuCI}
\end{eqnarray}
where the $\Lambda$ is defined as
\begin{eqnarray}
\label{lambda}
\Lambda_{\alpha k} \equiv (Y_{L\psi S})_{\alpha k}
=
\frac{1}{y}
(Y_{LdS})_{\alpha 3}
(Y_{\psi dS})_{k 3},
\end{eqnarray}
and ${\hat I}$ is given as
\begin{eqnarray}
{\hat I}_{k}
=
\frac{2N_c m_{\psi_k}}{(16\pi^2)^{2}} y (Y_{QQS})_{33} I(z,1,t_k,r)
\end{eqnarray}
Comparing Eq.~(\ref{eq:mnuCI}) with the neutrino mass and mixing
matrix, we can find the direct relation between ${\Lambda}$ and the
measured neutrino data.  Following the procedure originally developed
by Casas and Ibarra for seesaw type-I~\cite{Casas:2001sr}, we
parametrize $\Lambda$ as
\begin{eqnarray}
\label{eq:CI}
\left( \Lambda^{T} \right)_{k \alpha}
=
\left( \sqrt{{\hat I}^{-1}} \right)_{k} 
R_{ki} 
\left( \sqrt{\hat{m}_{\nu}} \right)_{i}
\left( U_{\nu}^{\dagger} \right)_{i \alpha}.
\end{eqnarray}
Here, $\hat{m}_{\nu}$ is the matrix of eigenvalues of $m_{\nu}$, 
which is diagonalized with the neutrino mixing matrix $U_{\nu}$ via
\begin{equation}
(U_{\nu}^{T})_{i \alpha} \, 
(m_{\nu})_{\alpha \beta} \, 
(U_{\nu})_{\beta j} 
\equiv \hat{m}_{\nu}=
\text{diag}
\begin{pmatrix}
 m_{\nu_{1}}
 &
 m_{\nu_{2}}
 &
 m_{\nu_{3}}
\end{pmatrix},
\end{equation}
for which we use the following standard parametrization
\begin{equation}
\label{eq:mixing}
U_{\nu}=
\left(
\begin{array}{ccc}
 c_{12}c_{13} & s_{12}c_{13}  & s_{13}e^{i\delta}  \\
-s_{12}c_{23}-c_{12}s_{23}s_{13}e^{-i\delta}  & 
c_{12}c_{23}-s_{12}s_{23}s_{13}e^{-i\delta}  & s_{23}c_{13}  \\
s_{12}s_{23}-c_{12}c_{23}s_{13}e^{-i\delta}  & 
-c_{12}s_{23}-s_{12}c_{23}s_{13}e^{-i\delta}  & c_{23}c_{13}  
\end{array}
\right) 
\left(
\begin{array}{ccc}
e^{i\alpha_1} & 0 & 0 \\
0 & e^{i\alpha_2}  & 0 \\
0 & 0 & 1
\end{array}
\right)
\end{equation}
$c_{ij} =\cos \theta_{ij}$, $s_{ij} = \sin \theta_{ij}$
with the mixing angles $\theta_{ij}$,
$\delta$ is the Dirac phase and $\alpha_1$, $\alpha_2$ are Majorana
phases. 
Finally, $R$ is a complex orthogonal matrix which satisfies
the condition $R^{T} R=1$. 
We use the following parametrization for the $R$ matrix in
terms of three complex angles $\theta_1, \theta_2,$ and $\theta_3$ 
as
\begin{equation}
R\ =\ \left( \begin{array}{ccc} 
c_{2} c_{3} & -c_{1} s_{3}-s_1 s_2 c_3& s_{1} s_3- c_1 s_2 c_3\\ 
c_{2} s_{3} & c_{1} c_{3}-s_{1}s_{2}s_{3} & -s_{1}c_{3}-c_1 s_2 s_3 \\ 
s_{2}  & s_{1} c_{2} & c_{1}c_{2}\end{array} \right).
\end{equation}
After fitting the neutrino oscillation data with the parametrization
shown above, there remain $y$, $(Y_{QQS})_{33}$ and the masses
$m_{S_{6,3,1/3}}$, $m_{S_{3,2,1/6}}$, $m_{\psi_k}$ as free parameters.
For the calculation of the short-range contribution to the $\znbb$
decay, we also have the parameter $\eta_{31}$.  For simplicity, we set
$y=1$ and assume again a nearly degenerate spectrum for heavy
particles, which is parameterized with $M_{\text{eff}}$.  We can then
calculate half-lives $T_{1/2}^{\znbb}$ for both neutrino mass
mechanism and the short-range contribution, as a function of
$m_{\nu_1}$, $M_{\text{eff}}$ and $\eta_{31}$.
In Fig.~\ref{fig:tsr}, we fixed the oscillation parameters  
$s_{13}^2$, $\Delta m_{31}^2$ and $\Delta m_{21}^2$ 
at their best-fit values, while $s_{23}^2=1/2$ and
$s_{12}^2=1/3$ and set $\delta$ as well as the Majorana 
phases to zero, just to sketch out some phenomenological 
aspects of this example.
Each panel shows 
the half-life of the short-range contribution
for $\znbb$ decay of $^{136}$Xe as a function of 
$\eta_{31}$ (top panel), $M_{\text{eff}}$ (middle panel) 
and $m_{\nu_1}$ (bottom panel). 
In each panel,
we examine several choices for the remaining parameters,
which are explained in the figure caption. 
The corresponding half-lives induced from the mass mechanism 
are also indicated. 
As shown in Fig.~\ref{fig:tsr}, half-lives can vary over many orders
of magnitude with the choice of parameters.  The amplitudes induced
from the mass mechanism becomes the same order as that from the
short-range, when $\eta_{31} \sim 2.7$ ($6.5$) for
$M_{\text{eff}}=0.5$ TeV ($1$ TeV).  As in the case with only one
generation of $\psi_{6,2,1/6}$, the mass mechanism dominates the
$\znbb$, if the ratio $\eta_{31}$ is taken to be unity and the heavy
mass scale $M_{\text{eff}}$ is given at the typical LHC search
sensitivities.  However, since the three-generation case can fit the
quasi-degenerate neutrino spectrum, $\znbb$ decay half-lives can be
much shorter than in the one generation case and can saturate the
experimental bound.

\begin{figure}[t]
\centering
\includegraphics[width=0.45\linewidth]{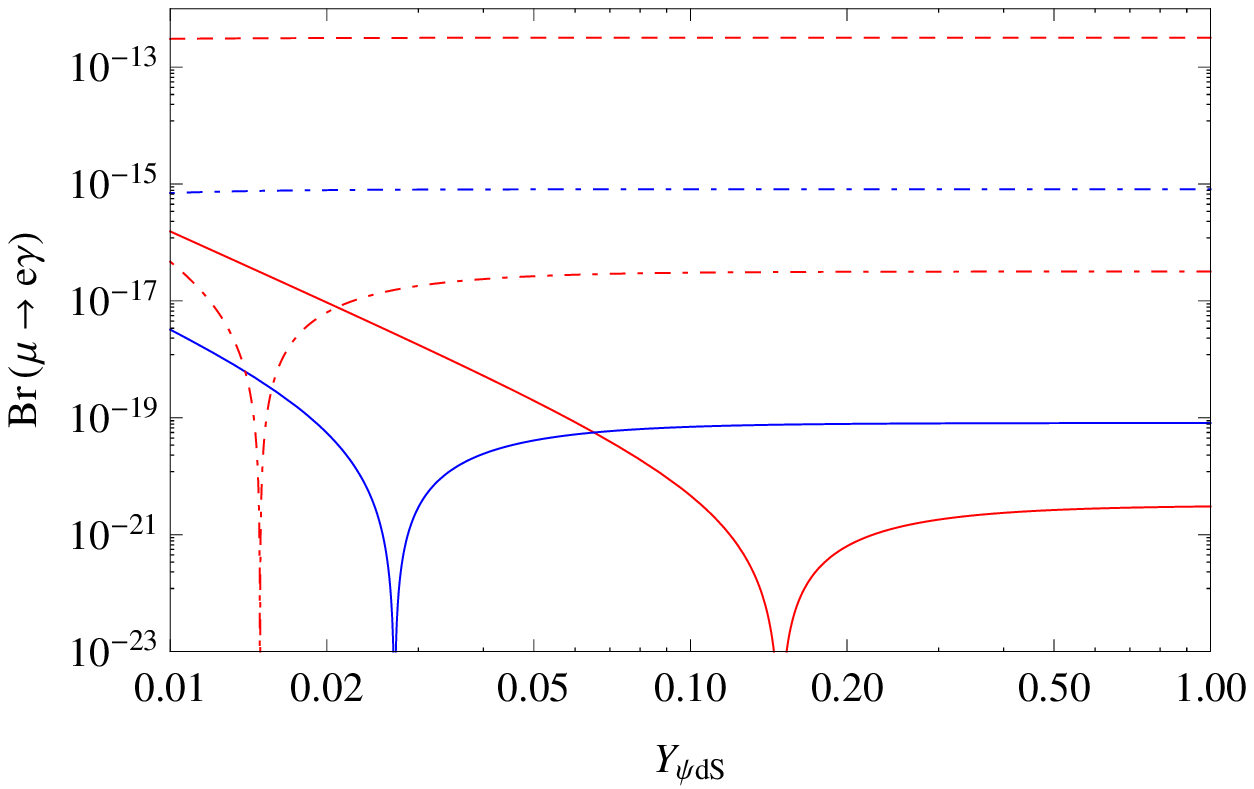}
\includegraphics[width=0.45\linewidth]{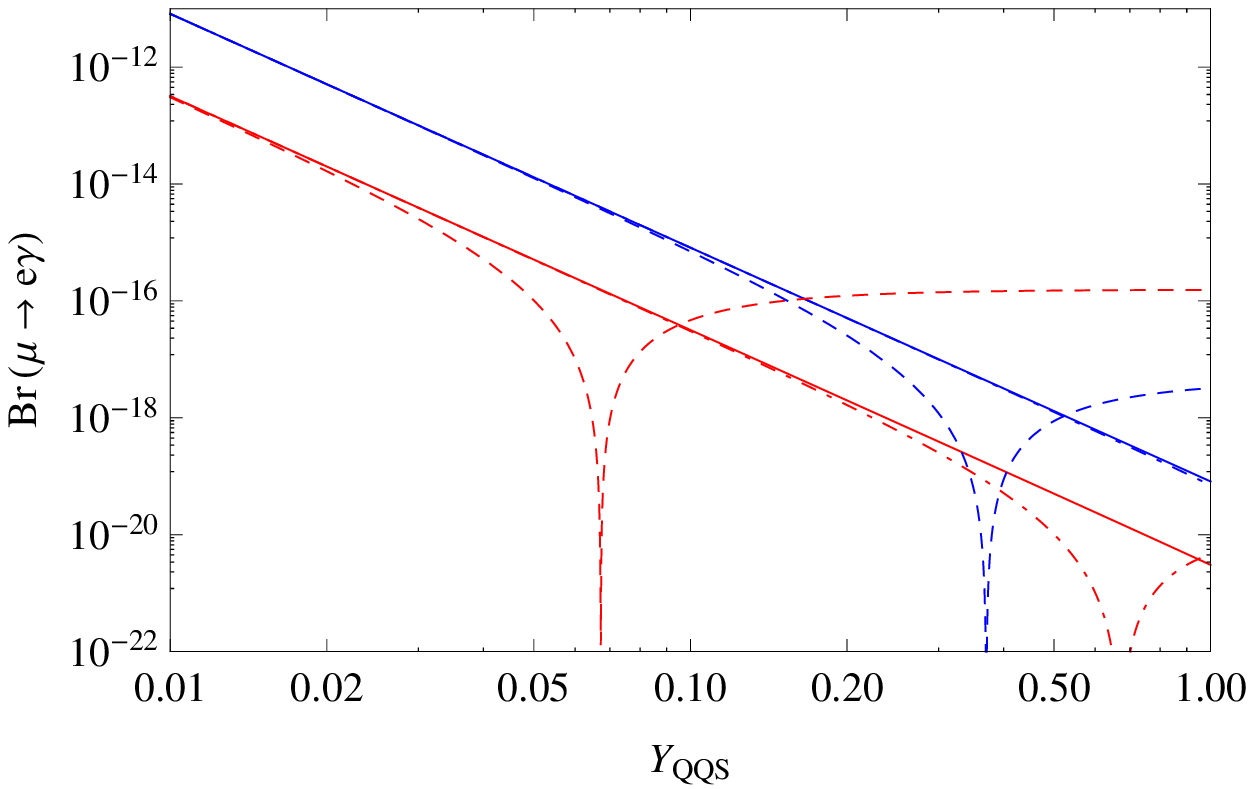}
\caption{Br($\mu\to e\gamma$) versus $Y_{\psi dS} \equiv 
 (Y_{\psi dS})_{13}$ (left) and $Y_{QQS}\equiv(Y_{QQS})_{33}$ (right), 
 for a fixed choice of $M_{\text{eff}}=1$ TeV, 
 neutrino oscillation parameters as in
 Fig.~\ref{fig:tsr} and $m_{\nu_1}=0.05$ eV (red lines) and
 $m_{\nu_1}=0.3$ eV (blue lines). Full, dot-dashed and dashed lines
 are for $Y_{\psi dS}$ (right) and $Y_{QQS}$ (left) equal to $1$,
 $10^{-1}$ and $10^{-2}$ respectively. Br($\mu\to e\gamma$) can 
saturate the experimental bound only for small values of these 
couplings, since smaller values of  $Y_{\psi dS}$ and $Y_{QQS}$ 
require larger values for $Y_{L\psi S}$ and $Y_{LdS}$, in order 
to fit neutrino data.}
\label{fig:mueg}
\end{figure}
We now turn to Br($\mu\to e\gamma$).  Again, as in the one generation
case, the neutrino mass matrix depends on 
Yukawa couplings, but is not directly related to Br($\mu\to e\gamma$).  
Therefore, we have always the freedom to adjust $(Y_{\psi dS})_{k3}$ and 
$(Y_{QQS})_{33}$ so as to fit the neutrino masses.  The other Yukawa 
couplings are then fixed by the neutrino data (and the choice of 
$M_{\text{eff}}$), and we can use them to calculate Br($\mu\to e\gamma$).  
Fig.~\ref{fig:mueg} shows some examples with a value of $M_{\text{eff}}=1$ 
TeV. The plots show that constraints from
Br($\mu\to e\gamma$) can be easily fulfilled. For this choice of
$M_{\text{eff}}$, only if both $(Y_{QQS})_{33}$ and $(Y_{\psi dS})_{13}$ 
are set to order $\mathcal{O}(10^{-2})$ or lower, the
predicted Br($\mu\to e\gamma$) can saturate the experimental
bound.

\section{Conclusions and discussion}
\label{sect:cncl}

We have discussed the relation between the $d=9$ short-range
contributions to the $\znbb$ decay amplitude with neutrino mass
models.  All contributions to $\znbb$ decay violate lepton number and,
therefore, generate also Majorana neutrino masses. We have classified
all possible (scalar-mediated) short-range contributions to the decay
rate according to the loop level, at which the corresponding models
will generate Majorana neutrino masses.  Possibilities range from
tree-level to 4-loop neutrino masses.  For each case we have discussed
one example briefly and given estimates of the typical constraints
imposed by both the short-range contribution and the mass mechanism.
Generally, one expects that for models with tree- or 1-loop neutrino
masses, the short-range $\znbb$ decay amplitude will be sub-dominant
to the mass mechanism. For 2-loop models short-range $\znbb$ decay
amplitude and mass mechanism can be comparable, while for 3-loop and
4-loop models the short-range part of the amplitude will dominate.

We have also discussed one particular example of a 2-loop model in
more detail. Here, we have shown different parts of parameter space
where mass mechanism or short-range amplitude dominant can each be
dominant. In the study, we have taken recent neutrino oscillation data
and constraints from LFV experiments into consideration.

In the appendix we give the full list of decompositions, classified
according to our scheme, in tabular form.

\medskip
\centerline{\bf Acknowledgements}

\medskip
T.O. is grateful to Prof. Junji Hisano for insightful 
comments on flavour structure of the effective interactions. 
J.C.H. thanks the IFIC for hospitality during his stay.  Work
supported by the Spanish grants FPA2014-58183-P and Multidark
CSD2009-00064 (MINECO), and PROMETEOII/2014/084 (Generalitat
Valenciana), by Fondecyt (Chile) under grants 11121557 and by CONICYT
(Chile) project 791100017.
The research of T.O. is supported by JSPS Grants-in-Aid for Scientific
Research on Innovative Areas {\it Unification and Development of the
Neutrino Science Frontier} {\sf Number 2610 5503}.
The research of F.A.P.S.  is supported by the Brazilian Research
Council CNPq.

\section{Appendix}

Here, we give tables in which all possible scalar short-range
decompositions are classified according to the loop level at which
they will generate neutrino masses.  The decompositions which generate
neutrino masses at tree, 1-loop, 2-loop, 3-loop and 4-loop level are
listed in Tables \ref{Tab:0lp} to \ref{Tab:4lp}.  The identification
number given to each decomposition is defined in
\cite{Bonnet:2012kh,Bonnet:2014kh}.  The notation T-I and T-II refers
to the two possible topologies of the decompositions of $d=9$ $\znbb$
effective operators, and the BL number is the Babu-Leung
classification of the effective neutrino mass operator, given in
\cite{Babu:2001ex}.  The columns ``Add. Int.''  specify additional
interactions, with respect to those appearing in the
decomposition. While they do not appear directly in the $\znbb$
diagram, these additional interaction can not be forbidden by any
symmetry, without forbidding the corresponding $\znbb$ decay
decomposition at the same time. Once present, they generate a neutrino
mass diagrams at the quoted loop level. The columns ``Diagram''
specify the topology of the neutrino mass diagram, for example
``type-I'' for seesaw type-I and so forth. The identification numbers
for the 1-loop and 2-loop neutrino mass diagrams are taken from the
general topology classification given in \cite{Bonnet:2012kz} and
\cite{Sierra:2014rxa}, respectively.

Here, we briefly comment on ``associated operators''.  As
discussed in Ref.~\cite{Bonnet:2012kz}, some of the decompositions
generate not only the original effective operator but also necessarily
generate other operators, when all possible contractions are carried
out.  We call this associated operators.  For example, the
decomposition T-I-2-iii-a of the BL~\#19 operator consists of the
following fundamental interactions,
\begin{align}
 \mathcal{L}_{\text{T-I-2-iii-a}}
 =&
 Y_{LdS}
 (\overline{L} d_{R})
 \cdot
 S_{\bar{3},2,-1/6}
 +
 Y_{Q\psi S}
 \left(\overline{Q}
 \vec{\lambda} 
 \vec{\psi}\right)
 \cdot 
 S_{\bar{3},2,-1/6}^{\dagger}
 \nonumber 
 \\
 &+
 Y_{\psi dS}
 S_{\bar{3},1,1/3}
 \left(\vec{\overline{\psi}} \vec{\lambda} d_{R}\right)
 +
 Y_{ueS}
 \left(
 \overline{u_{R}}
 {e_{R}}^{c}
 \right)
 S_{\bar{3},1,1/3}^{\dagger}
+
 {\rm H.c.},
\label{eq:example-associated-op}
\end{align}
where $\vec{\lambda}$ is the Gell-Mann matrices.
The first two interactions, together with the Majorana mass of the
fermion $\psi_{8,1,0}$ result in the BL~\#11 operator $(\overline{L}
d_{R}) (\overline{Q}) (\overline{Q}) (\overline{L} d_{R})$.  In the
same way, the last two interactions lead to the $d=9$ lepton number
violating effective operator $(\overline{u_{R}} \overline{e_{R}})
(d_{R}) (d_{R}) (\overline{u_{R}} \overline{e_{R}})$, which is
$\mathcal{O}_{-}$ in Eq.~\eqref{eq:BLX}.  All the decompositions
accompanied by associated operators were listed in tables of
Ref.~\cite{Bonnet:2012kz}.  We take into account the associated
effective operators in our classification scheme.  In short, if the
associated operator generates neutrino masses at a lower loop level
than the original one, we classify the decomposition with the loop
level of the associated operator.  The Lagrangian for a concrete
example is given in Eq.~\eqref{eq:example-associated-op}. Here,
although the original effective operator BL~\#19 gives neutrino masses
only at the 3-loop level, the decomposition T-I-2-iii-b of BL ~\#19
also produces BL ~\#11, and it generates the 2-loop neutrino mass
diagram with the help of the SM Yukawa interactions.  Therefore, we
list the decomposition T-I-2-iii-b of BL ~\#19 as a 2-loop neutrino
mass model in Tab.~\ref{Tab:2lp}. More examples are given in the 
tables.

\begin{table}[h]
\begin{center}
\begin{tabular}{ccccccccccl}
\hline \hline 
T-I  \#  & Op.   & BL \# \ & $S$ \  \ & $\psi$ \ \   & $S^\prime$  & Diagram \  & Add.  Int.
\\
\hline 
1-i &
 $(\bar{u} d) (\bar{e}) (\bar{e}) (\bar{u} d)$ 
&
11, 12, 14 & $ (1,2)_{+1/2} $ & $(1,1)_{0}$ & $(1,2)_{-1/2}$ & type I
 &    $\bar{L} \psi_{110} H^\dag  $  \\
 \hline
1-i &
 $(\bar{u} d) (\bar{e}) (\bar{e}) (\bar{u} d)$ 
&
11, 12, 14 & $ (1,2)_{+1/2} $ & $(1,3)_{0}$ & $(1,2)_{-1/2}$ & type III
 &    $\bar{L} \psi_{130} H^\dag  $  \\
 \hline 
1-ii-a 
&
$(\bar{u} d) (\bar{u}) (d) (\bar{e} \bar{e})$
&
11, 14

& 
$ (1,2)_{+1/2} $ & $(3,3)_{+2/3}$ & $(1,3)_{+1}$ & 
type  II  
&   $S_{131} H^\dag H^\dag  $  \\
\hline 
1-ii-a 
&
$(\bar{u} d) (\bar{u}) (d) (\bar{e} \bar{e})$
&
11, 14

& 
$ (8,2)_{+1/2} $ & $(3,3)_{+2/3}$ & $(1,3)_{+1}$ & 
type  II  
&   $S_{131} H^\dag H^\dag  $  \\
\hline 
1-ii-a 
&
$(\bar{u} d) (\bar{u}) (d) (\bar{e} \bar{e})$
&
 12, 14

& 
$ (1,2)_{+1/2} $ & $(3,2)_{+7/6}$ & $(1,3)_{+1}$ & 
type  II  
&   $S_{131} H^\dag H^\dag  $  \\
\hline

1-ii-a 
&
$(\bar{u} d) (\bar{u}) (d) (\bar{e} \bar{e})$
&
12, 14

& 
$ (8,2)_{+1/2} $ & $(3,2)_{+7/6}$ & $(1,3)_{+1}$ & 
type  II  
&   $S_{131} H^\dag H^\dag  $  \\
\hline 
1-ii-b
&
$(\bar{u} d) (d) (\bar{u}) (\bar{e} \bar{e})$
 
&
12, 14
 
& 
$ (1,2)_{+1/2} $ & $(\bar{3},3)_{+1/3}$ & $(1,3)_{+1}$ & 
type  II  
& $S_{131} H^\dag H^\dag  $    \\ \hline
1-ii-b
&
$(\bar{u} d) (d) (\bar{u}) (\bar{e} \bar{e})$
 
&
12, 14
 
& 
$ (8,2)_{+1/2} $ & $(\bar{3},3)_{+1/3}$ & $(1,3)_{+1}$ & 
type  II  
& $S_{131} H^\dag H^\dag  $    \\ 
\hline
1-ii-b
&
$(\bar{u} d) (d) (\bar{u}) (\bar{e} \bar{e})$
 
&
 11, 14
 
& 
$ (1,2)_{+1/2} $ & $(\bar{3},2)_{+5/3}$ & $(1,3)_{+1}$ & 
type  II  
& $S_{131} H^\dag H^\dag  $    \\ \hline
1-ii-b
&
$(\bar{u} d) (d) (\bar{u}) (\bar{e} \bar{e})$
 
&
 11, 14
 
& 
$ (8,2)_{+1/2} $ & $(\bar{3},2)_{+5/3}$ & $(1,3)_{+1}$ & 
type  II  
& $S_{131} H^\dag H^\dag  $    \\
 \hline
  2-i-b &
 $(\bar{u} d) (\bar{e}) (d) (\bar{u} \bar{e})$  &
11, 19, 14, 20 & 
$ (1,2)_{+1/2} $ & $(1,1)_{0}$ & $(\bar{3},1)_{+1/3}$ &
 type I  & 
$\bar{L} \psi_{110} H^\dag  $        \\
 \hline
2-i-b &
 $(\bar{u} d) (\bar{e}) (d) (\bar{u} \bar{e})$  &
11, 14 & 
$ (1,2)_{+1/2} $ & $(1,3)_{0}$ & $(\bar{3},3)_{+1/3}$ &
 type III  & 
$\bar{L} \psi_{130} H^\dag  $        \\
 \hline
 2-ii-b  
&
 $(\bar{u} d) (\bar{e}) (\bar{u}) (d \bar{e})$ 
&
11, 14  
& 
$ (1,2)_{+1/2} $ & $(1,1)_{0}$ & $(3,2)_{+1/6}$ &
 type I
 &    $\bar{L} \psi_{110} H^\dag  $  
\\
\hline 
 2-ii-b  
&
 $(\bar{u} d) (\bar{e}) (\bar{u}) (d \bar{e})$ 
&
11, 14  
& 
$ (1,2)_{+1/2} $ & $(1,3)_{0}$ & $(3,2)_{+1/6}$ &
type  III
 &    $\bar{L} \psi_{130} H^\dag  $  
\\
\hline 
2-iii-a 
&
 $(d \bar{e}) (\bar{u}) (d) (\bar{u} \bar{e})$  
&
11, 19 
 & 
$ (\bar{3},2)_{-1/6} $ & $(1,1)_{0}$ & $(\bar{3},1)_{+1/3}$ &
type  I  
& $\bar{L} \psi_{110} H^\dag  $     \\
\hline
2-iii-a 
&
 $(d \bar{e}) (\bar{u}) (d) (\bar{u} \bar{e})$  
&
11 
 & 
$ (\bar{3},2)_{-1/6} $ & $(1,3)_{0}$ & $(\bar{3},3)_{+1/3}$ &
type  III  
& $\bar{L} \psi_{130} H^\dag  $     \\
 \hline
3-ii 
&
$(\bar{u} \bar{u}) (d) (d) (\bar{e} \bar{e})$
  
&
 11 
& 
$ (6,3)_{+1/3} $ & $(3,3)_{+2/3}$ & $(1,3)_{+1}$ & 
type  II  
& $S_{131} H^\dag H^\dag  $    \\ \hline
3-ii 
&
$(\bar{u} \bar{u}) (d) (d) (\bar{e} \bar{e})$
  
&
  12
 
& 
$ (6,1)_{+4/3} $ & $(3,2)_{+7/6}$ & $(1,3)_{+1}$ & 
type  II  
& $S_{131} H^\dag H^\dag  $    \\ 

\hline
  3-iii 
&
 $(dd) (\bar{u}) (\bar{u}) (\bar{e} \bar{e})$

&
12
 
& 
$ (\bar{6},3)_{-1/3} $ & $(\bar{3},3)_{+1/3}$ & $(1,3)_{+1}$ & 
type  II  
& $S_{131} H^\dag H^\dag  $    \\ 

\hline

  3-iii 
&
 $(dd) (\bar{u}) (\bar{u}) (\bar{e} \bar{e})$

&
 11
 
& 
$ (\bar{6},1)_{+2/3} $ & $(\bar{3},2)_{+5/6}$ & $(1,3)_{+1}$ & 
type  II  
& $S_{131} H^\dag H^\dag  $    \\ 
\hline
 4-i
&
$(d \bar{e}) (\bar{u}) (\bar{u}) (d \bar{e})$ 
&
 11 
 & 
$ (\bar{3},2)_{-1/6} $ & $(1,1)_{0}$ & $(3,2)_{+1/6}$ &
type  I  
& $\bar{L} \psi_{110} H^\dag  $     \\
\hline
 4-i
&
$(d \bar{e}) (\bar{u}) (\bar{u}) (d \bar{e})$ 
&
 11 
 & 
$ (\bar{3},2)_{-1/6} $ & $(1,3)_{0}$ & $(3,2)_{+1/6}$ &
type  III  
& $\bar{L} \psi_{130} H^\dag  $     \\
\hline
  5-i 
&
 $(\bar{u} \bar{e}) (d) (d) (\bar{u} \bar{e})$
 
&
11, 19, - 
& 
$ (3,1)_{-1/3} $ & $(1,1)_{0}$ & $(\bar{3},1)_{+1/3}$ &
type I   
& $\bar{L} \psi_{110} H^\dag  $     \\ \hline
  5-i 
&
 $(\bar{u} \bar{e}) (d) (d) (\bar{u} \bar{e})$
 
&
11
& 
$ (3,3)_{-1/3} $ & $(1,3)_{0}$ & $(\bar{3},3)_{+1/3}$ &
type III   
& $\bar{L} \psi_{130} H^\dag  $     \\
%
%
\hline 
\end{tabular}
\vspace{0.1cm}
\\
\begin{tabular}{cccccccccl}
%
\hline \hline 

T-II  \#  & Op.   & BL \# \ & $S$ \  \ & $S^\prime$  \ \   & $S^{\prime \prime}$    & Diagram \  & Add.  Int.
\\
\hline 
  1 
&
  $(\bar{u} d) (\bar{u}d) (\bar{e} \bar{e})$ \
&
 11, 12, 14
 
& 
$ (1,2)_{+1/2} $ & $(1,2)_{+1/2}$ & $(1,3)_{-1}$ & 
type  II  
& $S_{13-1} H H  $    \\ 

\hline 
  1 
&
  $(\bar{u} d) (\bar{u}d) (\bar{e} \bar{e})$ \
&
 11, 12, 14
 
& 
$ (8,2)_{+1/2} $ & $(8,2)_{+1/2}$ & $(1,3)_{-1}$ & 
type  II  
& $S_{13-1} H H  $    \\ 
\hline
  3
&
  $(\bar{u} 	\bar{u}) (d d) (\bar{e} \bar{e})$ \
&
 11
 
& 
$ (6,3)_{+1/3} $ & $(\bar{6},1)_{+2/3}$ & $(1,3)_{-1}$ & 
type  II  
& $S_{13-1} H H $    \\ \hline
  3
&
  $(\bar{u} 	\bar{u}) (d d) (\bar{e} \bar{e})$ \
&
 12
 
& 
$ (6,1)_{+4/3} $ & $(\bar{6},3)_{-1/3}$ & $(1,3)_{-1}$ & 
type  II  
& $S_{13-1} H H $    \\ 
\hline
\hline 
\end{tabular}
\end{center}
\vskip - .1 cm
\caption{\it 
\label{Tab:0lp}  
List of the decompositions that generate neutrino masses at tree
level.  The ID-numbers with ``T'' are assigned as in
Ref.~\cite{Bonnet:2012kh}, and the decomposition is specified in the
``Op.'' column.  We also give the ID-numbers of
lepton-number-violating effective operators, which are classified as
in Babu and Leung~\cite{Babu:2001ex}, in ``BL\#''.  The SM charges of
fields appearing in the decomposition are also given.  ``Diagram''
indicates the type of resulting tree-level neutrino mass diagrams:
``type I'' for type I seesaw mechanism, and so on.  In the column
``Add. Int.'', we give the additional interaction that is missing in
the decomposition but is necessary to generate the neutrino mass
diagram.  
For the decompositions in this table, unless some severe 
fine-tuning of parameters is done, the mass mechanism of double 
beta decay will dominate over the short-range contributions.
}
\end{table}
\begin{table}[h]
\begin{center}
\begin{tabular}{ccccccccccl}
\hline  
T-I  \#  & Op.   & BL \# \ & $S$ \  \ & $\psi$ \ \   & $S^\prime$    & Diagram \  & Add.  Int.
\\
\hline 

 1-i 
&
  $(\bar{u} d) (\bar{e}) (\bar{e}) (\bar{u} d)$
  
&
  11, 12, 14 
    &  $ (8,2)_{+1/2} $ & $(8,1)_{0}$ & $(8,2)_{-1/2}$ & 
T$\nu$-3
   
& $S_{82\frac{1}{2}} S_{82\frac{1}{2}} H^\dag H^\dag \ $\\
\hline
  1-i 
&
  $(\bar{u} d) (\bar{e}) (\bar{e}) (\bar{u} d)$
  
&
  11, 12, 14 
    &  $ (8,2)_{+1/2} $ & $(8,3)_{0}$ & $(8,2)_{-1/2}$ & 
T$\nu$-3
   
& $S_{82\frac{1}{2}} S_{82\frac{1}{2}} H^\dag H^\dag \ $\\
\hline

  2-i-a 
&
 $(\bar{u} d) (d) (\bar{e}) (\bar{u} \bar{e})$
  
&
  11, 14 
&  $ (1,2)_{+1/2} $ & $(\bar{3},2)_{+5/6}$ & $(\bar{3},1)_{+1/3}$ & 
T$\nu$-1-iii

& ${\bar{d}_R}^c \psi_{\bar{3} 2 \frac{5}{6}} H^\dag \  $\\
\hline
 2-i-a 
&
 $(\bar{u} d) (d) (\bar{e}) (\bar{u} \bar{e})$
  
&
  11, 14 
&  $ (1,2)_{+1/2} $ & $(\bar{3},2)_{+5/6}$ & $(\bar{3},3)_{+1/3}$ & 
T$\nu$-1-iii

& ${\bar{d}_R}^c \psi_{\bar{3} 2 \frac{5}{6}} H^\dag \  $\\
\hline
 2-i-a 
&
 $(\bar{u} d) (d) (\bar{e}) (\bar{u} \bar{e})$
  
&
  11, 14 
&  $ (8,2)_{+1/2} $ & $(\bar{3},2)_{+5/6}$ & $(\bar{3},1)_{+1/3}$ & 
T$\nu$-1-iii

& ${\bar{d}_R}^c \psi_{\bar{3} 2 \frac{5}{6}} H^\dag \  $\\
\hline

 2-i-a 
&
 $(\bar{u} d) (d) (\bar{e}) (\bar{u} \bar{e})$
  
&
  11, 14 
&  $ (8,2)_{+1/2} $ & $(\bar{3},2)_{+5/6}$ & $(\bar{3},3)_{+1/3}$ & 
T$\nu$-1-iii

& ${\bar{d}_R}^c \psi_{\bar{3} 2 \frac{5}{6}} H^\dag \  $\\
\hline

 
 
  2-i-b 
&
 $(\bar{u} d) (\bar{e}) (d) (\bar{u} \bar{e})$
  
&
  11, 14, 19, 20 
      &  $ (8,2)_{+1/2} $ & $(8,1)_{0}$ & $(\bar{3},1)_{+1/3}$ & 
T$\nu$-3
& $S_{82\frac{1}{2}} S_{82\frac{1}{2}} H^\dag H^\dag \ $\\
\hline

  2-i-b 
&
 $(\bar{u} d) (\bar{e}) (d) (\bar{u} \bar{e})$
  
&
 
  11, 14 
      &  $ (8,2)_{+1/2} $ & $(8,3)_{0}$ & $(\bar{3},3)_{+1/3}$ & 
T$\nu$-3
& $S_{82\frac{1}{2}} S_{82\frac{1}{2}} H^\dag H^\dag \ $\\
\hline
2-ii-a  
&
  $(\bar{u} d) (\bar{u}) (\bar{e}) (d \bar{e})$
 
&
  11, 14 
  &  $ (1,2)_{+1/2} $ & $(3,3)_{+2/3}$ & $(3,2)_{+1/6}$ & 
T$\nu$-1-iii
& $\bar{Q} \psi_{\bar{3} 3 \frac{2}{3}} H^\dag \  $\\
\hline

 2-ii-a  
&
  $(\bar{u} d) (\bar{u}) (\bar{e}) (d \bar{e})$
 
&
  11, 14 
  &  $ (8,2)_{+1/2} $ & $(3,3)_{+2/3}$ & $(3,2)_{+1/6}$ & 
T$\nu$-1-iii
& $\bar{Q} \psi_{\bar{3} 3 \frac{2}{3}} H^\dag \  $\\
\hline

   2-ii-b 
&
  $(\bar{u} d) (\bar{e}) (\bar{u}) (d \bar{e})$
  
&
  11, 14 
      &  $ (8,2)_{+1/2} $ & $(8,1)_{0}$ & $(3,2)_{+1/6}$ & 
T$\nu$-3
& $S_{82\frac{1}{2}} S_{82\frac{1}{2}} H^\dag H^\dag \ $\\
\hline

   2-ii-b 
&
  $(\bar{u} d) (\bar{e}) (\bar{u}) (d \bar{e})$
  
&
  11, 14 
      &  $ (8,2)_{+1/2} $ & $(8,3)_{0}$ & $(3,2)_{+1/6}$ & 
T$\nu$-3
& $S_{82\frac{1}{2}} S_{82\frac{1}{2}} H^\dag H^\dag \ $\\
\hline
 
  2-iii-a 
&
 $(d \bar{e}) (\bar{u}) (d) (\bar{u} \bar{e})$
  
&
  11
      &  $ (\bar{3},2)_{-1/6} $ & $(8,1)_{0}$ & $(\bar{3},1)_{+1/3}$ & 
T$\nu$-1-ii
& $S^\dag_{\bar{3}2-\frac{1}{6}} S_{\bar{3}1\frac{1}{3}} H^\dag$& \\
\hline
 2-iii-a 
&
 $(d \bar{e}) (\bar{u}) (d) (\bar{u} \bar{e})$
  
&
  11
      &  $ (\bar{3},2)_{-1/6} $ & $(8,3)_{0}$ & $(\bar{3},3)_{+1/3}$ & 
T$\nu$-1-ii
& $S^\dag_{\bar{3}2-\frac{1}{6}} S_{\bar{3}3\frac{1}{3}} H^\dag$& \\
\hline
  2-iii-a 
&
 $(d \bar{e}) (\bar{u}) (d) (\bar{u} \bar{e})$
  
&
  14 
      &  $ (\bar{3},2)_{-1/6} $ & $(1,2)_{+1/2}$ & $(\bar{3},1)_{+1/3}$ & 
T$\nu$-1-ii
& $S^\dag_{\bar{3}2-\frac{1}{6}} S_{\bar{3}1\frac{1}{3}} H^\dag$& \\
\hline
  2-iii-a 
&
 $(d \bar{e}) (\bar{u}) (d) (\bar{u} \bar{e})$
  
&
   14 
      &  $ (\bar{3},2)_{-1/6} $ & $(8,2)_{+1/2}$ & $(\bar{3},1)_{+1/3}$ & 
T$\nu$-1-ii
& $S^\dag_{\bar{3}2-\frac{1}{6}} S_{\bar{3}1\frac{1}{3}} H^\dag$& \\
\hline
  2-iii-a 
&
 $(d \bar{e}) (\bar{u}) (d) (\bar{u} \bar{e})$
  
&
  14 
      &  $ (\bar{3},2)_{-1/6} $ & $(1,2)_{+1/2}$ & $(\bar{3},3)_{+1/3}$ & 
T$\nu$-1-ii
& $S^\dag_{\bar{3}2-\frac{1}{6}} S_{\bar{3}3\frac{1}{3}} H^\dag$& \\
\hline
  2-iii-a 
&
 $(d \bar{e}) (\bar{u}) (d) (\bar{u} \bar{e})$
  
&
  14 
      &  $ (\bar{3},2)_{-1/6} $ & $(8,2)_{+1/2}$ & $(\bar{3},3)_{+1/3}$ & 
T$\nu$-1-ii
& $S^\dag_{\bar{3}2-\frac{1}{6}} S_{\bar{3}3\frac{1}{3}} H^\dag$& \\

\hline
  2-iii-b 
&
 $(d \bar{e}) (d) (\bar{u}) (\bar{u} \bar{e})$
  
&
  11
      &  $ (\bar{3},2)_{-1/6} $ & $(3,2)_{+1/6}$ & $(\bar{3},1)_{+1/3}$ & 
T$\nu$-1-ii
& $S^\dag_{\bar{3}2-\frac{1}{6}} S_{\bar{3}1\frac{1}{3}} H^\dag$& \\
\hline
  2-iii-b 
&
 $(d \bar{e}) (d) (\bar{u}) (\bar{u} \bar{e})$
  
&
  11 
      &  $ (\bar{3},2)_{-1/6} $ & $(\bar{6},2)_{+1/6}$ & $(\bar{3},1)_{+1/3}$ & 
T$\nu$-1-ii
& $S^\dag_{\bar{3}2-\frac{1}{6}} S_{\bar{3}1\frac{1}{3}} H^\dag$& \\
\hline
  2-iii-b 
&
 $(d \bar{e}) (d) (\bar{u}) (\bar{u} \bar{e})$
  
&
  11
      &  $ (\bar{3},2)_{-1/6} $ & $(3,2)_{+1/6}$ & $(\bar{3},3)_{+1/3}$ & 
T$\nu$-1-ii
& $S^\dag_{\bar{3}2-\frac{1}{6}} S_{\bar{3}3\frac{1}{3}} H^\dag$& \\
\hline
  2-iii-b 
&
 $(d \bar{e}) (d) (\bar{u}) (\bar{u} \bar{e})$
  
&
  11
      &  $ (\bar{3},2)_{-1/6} $ & $(\bar{6},2)_{+1/6}$ & $(\bar{3},3)_{+1/3}$ & 
T$\nu$-1-ii
& $S^\dag_{\bar{3}2-\frac{1}{6}} S_{\bar{3}3\frac{1}{3}} H^\dag$& \\
\hline
  2-iii-b 
&
 $(d \bar{e}) (d) (\bar{u}) (\bar{u} \bar{e})$
  
&
   14 
      &  $ (\bar{3},2)_{-1/6} $ & $(3,1)_{-1/3}$ & $(\bar{3},1)_{+1/3}$ & 
T$\nu$-1-ii
& $S^\dag_{\bar{3}2-\frac{1}{6}} S_{\bar{3}1\frac{1}{3}} H^\dag$& \\
\hline
  2-iii-b 
&
 $(d \bar{e}) (d) (\bar{u}) (\bar{u} \bar{e})$
  
&
  14 
      &  $ (\bar{3},2)_{-1/6} $ & $(\bar{6},1)_{-1/3}$ & $(\bar{3},1)_{+1/3}$ & 
T$\nu$-1-ii
& $S^\dag_{\bar{3}2-\frac{1}{6}} S_{\bar{3}1\frac{1}{3}} H^\dag$& 
\\
\hline
  2-iii-b 
&
 $(d \bar{e}) (d) (\bar{u}) (\bar{u} \bar{e})$
  
&
  14 
      &  $ (\bar{3},2)_{-1/6} $ & $(3,3)_{-1/3}$ & $(\bar{3},3)_{+1/3}$ & 
T$\nu$-1-ii
& $S^\dag_{\bar{3}2-\frac{1}{6}} S_{\bar{3}3\frac{1}{3}} H^\dag$& 
\\ 
\hline
 2-iii-b 
&
 $(d \bar{e}) (d) (\bar{u}) (\bar{u} \bar{e})$
  
&
  14 
      &  $ (\bar{3},2)_{-1/6} $ & $(\bar{6},3)_{-1/3}$ & $(\bar{3},3)_{+1/3}$ & 
T$\nu$-1-ii
& $S^\dag_{\bar{3}2-\frac{1}{6}} S_{\bar{3}3\frac{1}{3}} H^\dag$& 
\\
\hline
 3-i 
&
  $(\bar{u} \bar{u}) (\bar{e})(\bar{e}) (dd)$
  
&
  11
      &  $ (6,3)_{+1/3} $ & $(6,2)_{-1/6}$ & $(6,1)_{-2/3}$ & 
T$\nu$-3
& $S_{63\frac{1}{3}} S_{61-\frac{2}{3}}^\dag H^\dag H^\dag \ $\\
\hline 
 3-i 
&
  $(\bar{u} \bar{u}) (\bar{e})(\bar{e}) (dd)$
  
&
   12 
       &  $ (6,1)_{+4/3} $ & $(6,2)_{+5/6}$ & $(6,3)_{+1/3}$ & 
T$\nu$-3
& $S_{61\frac{4}{3}} S_{63\frac{1}{3}}^\dag H^\dag H^\dag \ $\\ \hline
  4-ii-a 
&
  $(\bar{u} \bar{u}) (d) (\bar{e}) (d \bar{e})$
 
&
  11 
  &  $ (6,3)_{+1/3} $ & $(3,3)_{+2/3}$ & $(3,2)_{+1/6}$ & 
T$\nu$-1-iii
& $\bar{Q} \psi_{3 3 \frac{2}{3}} H^\dag \  $\\
\hline
  5-ii-b 
&
 $(\bar{u} \bar{e}) (\bar{e}) (\bar{u}) (dd)$
   
&
  11 
  &  $ (3,1)_{-1/3} $ & $(3,2)_{-5/6}$ & $(6,1)_{-2/3}$ & 
T$\nu$-1-iii
& ${\bar{d}_R}^c \psi_{3 2 -\frac{5}{6}}^c H^\dag \  $\\ 
\hline
  5-ii-b 
&
 $(\bar{u} \bar{e}) (\bar{e}) (\bar{u}) (dd)$
   
&
  11 
  &  $ (3,3)_{-1/3} $ & $(3,2)_{-5/6}$ & $(6,1)_{-2/3}$ & 
T$\nu$-1-iii
& ${\bar{d}_R}^c \psi_{3 2 -\frac{5}{6}}^c H^\dag \  $\\ 
\hline

\end{tabular}

\end{center}
\vskip -.75 cm
\caption{\it 
\label{Tab:1lp} Decompositions that generate neutrino masses at 1-loop.
The naming convention of 1-loop neutrino mass diagram, which is used
in ``Diagram'' column, follows Ref.~\cite{Bonnet:2012kz} and is also
shown in fig. (\ref{fig:1lp}).   
For the decompositions in this table, unless some severe 
fine-tuning of parameters is done, the mass mechanism of double 
beta decay will dominate over the short-range contributions. }
\end{table}

\begin{table}[h]
\begin{center}
\begin{tabular}{cccccccccl}
\hline \hline 
T-II  \#  & Op.   & BL \# \ & $S$ \  \ & $S^\prime$  \ \   & $S^{\prime \prime}$    & Diagram \  & Add.  Int.
\\
\hline 

%
  2 \
&
 $(\bar{u} d) (\bar{u} \bar{e} )  (d \bar{e} )$
  
&
  11, 14  
   &  $ (1,2)_{+1/2} $ & $(3,1)_{-1/3}$ & $(\bar{3},2)_{-1/6}$ & 
T$\nu$-1-ii
& $S^\dag_{\bar{3}2-\frac{1}{6}} S_{3 1-\frac{1}{3}}^\dag H^\dag$&
\\
\hline 

%
  2 \
&
 $(\bar{u} d) (\bar{u} \bar{e} )  (d \bar{e} )$
  
&
  11, 14   
   &  $ (1,2)_{+1/2} $ & $(3,3)_{-1/3}$ & $(\bar{3},2)_{-1/6}$ & 
T$\nu$-1-ii
& $S^\dag_{\bar{3}2-\frac{1}{6}} S_{3 3-\frac{1}{3}}^\dag H^\dag$&
 \\
\hline 

%
  2 \
&
 $(\bar{u} d) (\bar{u} \bar{e} )  (d \bar{e} )$
  
&
  11, 14  
   &  $ (8,2)_{+1/2} $ & $(3,1)_{-1/3}$ & $(\bar{3},2)_{-1/6}$ & 
T$\nu$-1-ii
& $S^\dag_{\bar{3}2-\frac{1}{6}} S_{3 1-\frac{1}{3}}^\dag H^\dag$&
 \\
\hline 

%
  2 \
&
 $(\bar{u} d) (\bar{u} \bar{e} )  (d \bar{e} )$
  
&
  11, 14   
   &  $ (8,2)_{+1/2} $ & $(3,3)_{-1/3}$ & $(\bar{3},2)_{-1/6}$ & 
T$\nu$-1-ii
& $S^\dag_{\bar{3}2-\frac{1}{6}} S_{3 3-\frac{1}{3}}^\dag H^\dag$&
 \\
\hline\hline
\end{tabular}
\end{center}
\caption{\it 
\label{Tab:1lp-2} 
Decompositions (T-II) that generate neutrino mass at 1-loop,
which are continued from Tab.~\ref{Tab:1lp}.     
For the decompositions in this table, unless some severe 
fine-tuning of parameters is done, the mass mechanism of double 
beta decay will dominate over the short-range contributions.}
\end{table}

\begin{table}[h]
\begin{center}
\begin{tabular}{cccccccccccl}
\hline \hline 
T-I  \#  & Op.   & BL \# \ & $S$ \  \ & $\psi$ \ \   & $S^\prime$    & Diagram \  \\
\hline 

2-iii-a
&
 $(d \bar{e}) (\bar{u}) (d) (\bar{u} \bar{e})$ 
&
 19  
   &  $ (\bar{3},2)_{-1/6} $ & $(8,1)_{0}$ & $(\bar{3},1)_{+1/3}$ & 
PTBM-1
\\


\hline

  4-i  
&
 $(d \bar{e}) (\bar{u}) (\bar{u}) (d \bar{e})$
   
&
 11 
   &  $ (\bar{3},2)_{-1/6} $ & $(8,1)_{0}$ & $(3,2)_{+1/6}$ & 
PTBM-1

\\
\hline

  4-i  
&
 $(d \bar{e}) (\bar{u}) (\bar{u}) (d \bar{e})$

&
 11 
   &  $ (\bar{3},2)_{-1/6} $ & $(8,3)_{0}$ & $(3,2)_{+1/6}$ & 
   PTBM-1

\\
\hline

  4-ii-b  
&
 $(\bar{u} \bar{u}) (\bar{e}) (d) (d \bar{e})$
  
&
 11  
   &  $ (6,3)_{+1/3} $ & $(6,2)_{-1/6}$ & $(3,2)_{+1/6}$ & 
   PTBM-4

\\\hline
  5-i  
&
 $(\bar{u} \bar{e}) (d) (d) (\bar{u} \bar{e})$
  
&
 11, 19  
   &  $ (3,1)_{-1/3} $ & $(8,1)_{0}$ & $(\bar{3},1)_{+1/3}$ & 
PTBM-1

\\ \hline
 5-i  
&
 $(\bar{u} \bar{e}) (d) (d) (\bar{u} \bar{e})$

&
 11
   &  $ (3,3)_{-1/3} $ & $(8,3)_{0}$ & $(\bar{3},3)_{+1/3}$ & 
  PTBM-1

\\ \hline
 5-ii-a  
&
 $(\bar{u} \bar{e}) (\bar{u}) (\bar{e}) (dd)$
  
&
 11  
   &  $ (3,1)_{-1/3} $ & $(6,2)_{-1/6}$ & $(6,1)_{-2/3}$ & 
   PTBM-4

\\ 
\hline
 5-ii-a  
&
 $(\bar{u} \bar{e}) (\bar{u}) (\bar{e}) (dd)$
  
&
 11  
   &  $ (3,3)_{-1/3} $ & $(6,2)_{-1/6}$ & $(6,1)_{-2/3}$ & 
   PTBM-4

\\ 
\hline
\hline     
\end{tabular}
\vspace{0.5cm}
\\

\begin{tabular}{cccccccccl}

\hline \hline 
T-II  \#  & Op.   & BL \# \ & $S$ \  \ &$S^\prime$ \ \   & $S^{\prime \prime}$    & Diagram \\
\hline 


 4  
&
 $(\bar{u} \bar{u}) (d \bar{e}) (d \bar{e})$
  
&
 11  
 &  $ (6,3)_{+1/3} $ & $(\bar{3},2)_{-1/6}$ & $(\bar{3},2)_{-1/6}$ & 
CLBZ-1
\\
\hline
 5 
&
 $(\bar{u} \bar{e}) (\bar{u} \bar{e}) (d d)$
  
&
 11  
 &  $ (3,1)_{-1/3} $ & $(3,1)_{-1/3}$ & $(\bar{6},1)_{+2/3}$ & 
CLBZ-1

\\
\hline
 5 
&
 $(\bar{u} \bar{e}) (\bar{u} \bar{e}) (d d)$
  
&
 11  
 &  $ (3,3)_{-1/3} $ & $(3,3)_{-1/3}$ & $(\bar{6},1)_{+2/3}$ & 
CLBZ-1

\\

\hline
\hline     
\end{tabular}
\end{center}
\caption{\it 
\label{Tab:2lp} 
Decompositions that generate the $d=5$ neutrino mass operator $LLHH$
at 2-Loop.  We follow the naming convention used in
\cite{Sierra:2014rxa}.  Although the effective operator of BL~\#19 can
generate neutrino mass only at the 3-loop
level~\cite{Babu:2001ex,deGouvea:2007xp}, the decompositions of
BL~\#19 listed in this table generate not only the BL~\#19 but also
the ``associated'' BL~\#11 operator and thus are classified as 2-loop 
neutrino mass models. 
For the decompositions in this table, the mass mechanism of double 
beta decay and  the short-range contributions can be comparable.}
\end{table}

\begin{table}[h]
\begin{center}
\begin{tabular}{cccccccccccl}
\hline \hline 
T-I  \#  & Op.   & BL \# \ 
& $S$ \  \ 
& $\psi$ \ \   
& $S^\prime$    \  
& Diagram 
& Add. Int.
\\
\hline 

  2-i-a 
&
  $(\bar{u} d) (d) (\bar{e}) (\bar{u} \bar{e})$
  
&
 19, 20  
&  $ (1,2)_{+1/2} $ 
& $(\bar{3},2)_{+5/6}$ 
& $(\bar{3},1)_{+1/3}$ 
& (b)
& ${\bar{d}_R}^c \psi_{\bar{3} 2 \frac{5}{6}} H^\dag  $\\
\hline
2-i-a 
&
  $(\bar{u} d) (d) (\bar{e}) (\bar{u} \bar{e})$
  
&
 19, 20  
  &  $ (8,2)_{+1/2} $ 
& $(\bar{3},2)_{+5/6}$ 
& $(\bar{3},1)_{+1/3}$ 
& (b)
& ${\bar{d}_R}^c \psi_{\bar{3} 2 \frac{5}{6}} H^\dag  $\\
\hline

  2-ii-a 
&
  $(\bar{u} d) (\bar{u}) (\bar{e}) (d \bar{e})$
&
 19, 20  
&  $ (1,2)_{+1/2} $ 
& $(3,2)_{+7/6}$ 
& $(3,2)_{+1/6}$ 
& (a)
&$\bar{u}_R \psi_{32\frac{7}{6} }H^\dag$ & \\
\hline
  2-ii-a 
&
  $(\bar{u} d) (\bar{u}) (\bar{e}) (d \bar{e})$
&
 19, 20  
&  $ (8,2)_{+1/2} $ 
& $(3,2)_{+7/6}$ 
& $(3,2)_{+1/6}$ 
& (a)
&$\bar{u}_R \psi_{32\frac{7}{6} }H^\dag$ & \\
\hline

  2-ii-b  
&
  $(\bar{u} d) (\bar{e}) (\bar{u}) (d \bar{e})$
&
 19, 20  
   &  $ (1,2)_{+1/2} $ 
& $(1,2)_{-1/2}$ 
& $(3,2)_{+1/6}$ 
& (c)
& $\bar{e}_R \psi_{12-\frac{1}{2}}H^\dag$& \\
\hline
  2-iii-a
&
  $(d \bar{e}) (\bar{u}) (d) (\bar{u} \bar{e})$
 
&
 19 
   &  $ (\bar{3},2)_{-1/6} $ 
& $(8,1)_{0}$ 
& $(\bar{3},1)_{+1/3}$ 
& (d)
& $S^\dag_{\bar{3}2-\frac{1}{6}} S_{\bar{3}1\frac{1}{3}} H^\dag$& \\
\hline

  2-iii-a
&
  $(d \bar{e}) (\bar{u}) (d) (\bar{u} \bar{e})$ 
&
 20  
   &  $ (\bar{3},2)_{-1/6} $ 
& $(1,2)_{+1/2}$ 
& $(\bar{3},1)_{+1/3}$ 
& (d)
& $S^\dag_{\bar{3}2-\frac{1}{6}} S_{\bar{3}1\frac{1}{3}} H^\dag$& \\
\hline
 
  2-iii-a
&
  $(d \bar{e}) (\bar{u}) (d) (\bar{u} \bar{e})$ 
&
 20  
   &  $ (\bar{3},2)_{-1/6} $ 
& $(1,2)_{+1/2}$ 
& $(\bar{3},1)_{+1/3}$ 
& (c)
& $\bar{e}_R \psi_{12\frac{1}{2}}^cH^\dag$& \\
\hline
 2-iii-a
&
  $(d \bar{e}) (\bar{u}) (d) (\bar{u} \bar{e})$ 
&
 20  
   &  $ (\bar{3},2)_{-1/6} $ 
& $(8,2)_{+1/2}$ 
& $(\bar{3},1)_{+1/3}$ 
& (d)
& $S^\dag_{\bar{3}2-\frac{1}{6}} S_{\bar{3}1\frac{1}{3}} H^\dag$& \\
\hline
  2-iii-b 
&
  $(d \bar{e}) (d) (\bar{u}) (\bar{u} \bar{e})$
 
&
 19
   &  $ (\bar{3},2)_{-1/6} $ 
& $(3,2)_{+1/6}$ 
& $(\bar{3},1)_{+1/3}$ 
& (d)
& $S^\dag_{\bar{3}2-\frac{1}{6}} S_{\bar{3}1\frac{1}{3}} H^\dag$& \\
\hline
  2-iii-b 
&
  $(d \bar{e}) (d) (\bar{u}) (\bar{u} \bar{e})$
 
&
 19  
   &  $ (\bar{3},2)_{-1/6} $ 
& $(\bar{6},2)_{+1/6}$ 
& $(\bar{3},1)_{+1/3}$ 
& (d)
& $S^\dag_{\bar{3}2-\frac{1}{6}} S_{\bar{3}1\frac{1}{3}} H^\dag$& \\
\hline
 
  2-iii-b 
&
  $(d \bar{e}) (d) (\bar{u}) (\bar{u} \bar{e})$
 
&
 20  
   &  $ (\bar{3},2)_{-1/6} $ 
& $(3,1)_{-1/3}$ 
& $(\bar{3},1)_{+1/3}$ 
& (d)
& $S^\dag_{\bar{3}2-\frac{1}{6}} S_{\bar{3}1\frac{1}{3}} H^\dag$& \\
\hline
  2-iii-b 
&
  $(d \bar{e}) (d) (\bar{u}) (\bar{u} \bar{e})$
 
&
  20  
   &  $ (\bar{3},2)_{-1/6} $ 
& $(\bar{6},1)_{-1/3}$ 
& $(\bar{3},1)_{+1/3}$ 
& (d)
& $S^\dag_{\bar{3}2-\frac{1}{6}} S_{\bar{3}1\frac{1}{3}} H^\dag$& \\
\hline
  4-i 
&
  $(d \bar{e}) (\bar{u})  (\bar{u}) (d \bar{e})$
 
&
  20  
    &  $ (\bar{3},2)_{-7/6} $ 
& $(1,2)_{-1/2}$ 
& $(3,2)_{+1/6}$ 
& (c)
& $\bar{e}_R \psi_{12-\frac{1}{2}}H^\dag$& \\
\hline
  4-ii-a 
&
  $(\bar{u} \bar{u}) (d) (\bar{e}) (d \bar{e})$
 
&
  20  
    &  $ (6,1)_{+4/3} $ 
& $(3,2)_{+7/6}$ 
& $(3,2)_{+1/6}$ 
& (a)
&$\bar{u}_R \psi_{32\frac{7}{6} }H^\dag$ & \\
\hline
   5-ii-b
&
 $(\bar{u} \bar{e}) (\bar{e}) (\bar{u}) (dd)$
  
&
 19   
   &  $ (3,1)_{-1/3} $ 
& $(3,2)_{-5/6}$ 
& $(6,1)_{-2/3}$ 
& (b)
& ${\bar{d}_R}^c \psi_{3 2 -\frac{5}{6}}^c H^\dag $\\
\hline
\hline     
\end{tabular}
\vspace{0.5cm}
\\

\begin{tabular}{cccccccccl}
\hline \hline 
T-II  \#  & Op.   & BL \# \ 
& $S$ \  \ 
&$S^\prime$ \ \   
& $S^{\prime \prime}$     \  
& Diagram 
& Add. Int.
\\
\hline 
   2
&
 $(\bar{u} d) (\bar{u} \bar{e}) (d \bar{e})$
  
&
 19, 20   
   &  $ (1,2)_{+1/2} $ 
& $(3,1)_{-1/3}$ 
& $(\bar{3},2)_{-1/6}$ 
& (d)
& $S^\dag_{\bar{3}2-\frac{1}{6}} S_{31-\frac{1}{3}}^\dag H^\dag$& \\
\hline
  2
&
 $(\bar{u} d) (\bar{u} \bar{e}) (d \bar{e})$
  
&
 19, 20   
   &  $ (8,2)_{+1/2} $ 
& $(3,1)_{-1/3}$ 
& $(\bar{3},2)_{-1/6}$ 
& (d)
& $S^\dag_{\bar{3}2-\frac{1}{6}} S_{31-\frac{1}{3}}^\dag H^\dag$& \\
\hline

\hline     
\end{tabular}
\end{center}
\caption{\it 
\label{Tab:2lp-7} 
Decompositions that generate $d=7$ neutrino mass operator
$LLHHHH^{\dagger}$ at the 2 loop level.  The topologies of the
neutrino mass diagrams in the column of ``Diagram'' are shown in
Fig.~\ref{fig:2loop-dim7}.    
For the decompositions in this table, unless some severe 
fine-tuning of parameters is done, the short-range contributions will dominate over the mass mechanism of double 
beta decay.}
\end{table}

\begin{table}[h]
\begin{center}
\begin{tabular}{cccccccccccl}
\hline \hline 
T-I  \#  & Op.   & BL \# \ & $S$ \  \ & $\psi$ \ \   & $S^\prime$    \\
\hline 
%
 
 2-ii-b   
&
  $(\bar{u} d) (\bar{e}) (\bar{u}) (d \bar{e})$
 
&
  19, 20 
    &  $ (8,2)_{+1/2} $ & $(8,2)_{-1/2}$ & $(3,2)_{+1/6}$ 

\\
\hline
 4-i   
&
  $(d \bar{e}) (\bar{u}) (\bar{u}) (d \bar{e})$
 
&
  20 
    &  $ (\bar{3},2)_{-7/6} $ & $(8,2)_{-1/2}$ & $(3,2)_{+1/6}$ 

\\
\hline
 4-ii-a   
&
  $(\bar{u} \bar{u}) (d) (\bar{e}) (d \bar{e})$
 
&
  20 
    &  $ (6,1)_{+4/3} $ & $(3,1)_{+5/3}$ & $(3,2)_{+7/6}$

\\
\hline
  4-ii-b 
&
 $(\bar{u} \bar{u}) (\bar{e}) (d) (d \bar{e})$
  
&
  20 
    &  $ (6,1)_{+4/3} $ & $(6,1)_{+1/3}$ & $(3,2)_{+1/6}$ 

\\
\hline
 4-ii-b 
&
 $(\bar{u} \bar{u}) (\bar{e}) (d) (d \bar{e})$
  
&
  20 
    &  $ (6,1)_{+4/3} $ & $(6,2)_{+5/6}$ & $(3,2)_{+7/6}$ 

\\
\hline
  5-ii-a 
&
  $(\bar{u} \bar{e}) (\bar{u}) (\bar{e}) (dd)$
  
&
 19  
   &  $ (3,1)_{-1/3} $ & $(6,1)_{+1/3}$ & $(6,1)_{-2/3}$ 

\\
\hline
 5-ii-a 
&
  $(\bar{u} \bar{e}) (\bar{u}) (\bar{e}) (dd)$
  
&
 19  
   &  $ (3,1)_{-1/3} $ & $(6,2)_{-1/6}$ & $(6,1)_{-2/3}$ 

\\
\hline

  5-ii-b 
&
  $(\bar{u} \bar{e}) (\bar{e}) (\bar{u}) (dd)$
  
&
 19  
   &  $ (3,1)_{-1/3} $ & $(3,1)_{-4/3}$ & $(6,1)_{-2/3}$ 

\\
\hline\hline  
\end{tabular}
\vspace{0.5cm}
\\

\begin{tabular}{cccccccccl}
\hline \hline 
T-II  \#  & Op.   & BL \# \ & $S$ \  \ & $S^\prime$\ \   & $S^{\prime \prime}$   
\\
\hline

%
  4 
&
  $(\bar{u} \bar{u}) (d \bar{e}) (d \bar{e})$
  
&
 20  
   &  $ (6,1)_{+4/3} $ & $(\bar{3},2)_{-7/6}$ & $(\bar{3},2)_{-1/6}$

\\
\hline
  5 
&
  $(\bar{u} \bar{e}) (\bar{u} \bar{e})(dd)$
  
&
 19  
   &  $ (3,1)_{-1/3} $ & $(3,1)_{-1/3}$ & $(\bar{6},1)_{+2/3}$ 

\\
\hline\hline  
\end{tabular}
\end{center}
\caption{\it 
\label{Tab:3lp} 
Decompositions that generate neutrino masses at 3-loop. Some example 
diagrams are given in the main text.  
For the decompositions in this table, unless some severe 
fine-tuning of parameters is done, the short-range contributions will dominate over the mass mechanism of double 
beta decay.}
\end{table}

\begin{table}[h]
\begin{center}
\begin{tabular}{cccccccccccl}
\hline \hline 
T-I  \#  & Op.   & BL \# \ & $S$ \  \ & $\psi$ \ \   & $S^\prime$    
\\
\hline 
 3-i  
&
 $(\bar{u} \bar{u}) (\bar{e})(\bar{e}) (dd)$
   
&
  -  
    &  $ (6,1)_{+4/3} $ & $(6,1)_{+1/3}$ & $(6,1)_{-2/3}$ 

\\
\hline
 3-ii  
&
  $(\bar{u} \bar{u}) (d) (d) (\bar{e} \bar{e})$
  
&
  -  
    &  $ (6,1)_{+4/3} $ & $(3,1)_{+5/3}$ & $(1,1)_{+2}$ 

\\
\hline
  3-iii 
&
  $(dd) (\bar{u}) (\bar{u}) (\bar{e} \bar{e})$
  
&
  -  
    &  $ (\bar{6},1)_{+2/3} $ & $(\bar{3},1)_{+4/3}$ & $(1,1)_{+2}$ 

\\
\hline
 5-i  
&
 $(\bar{u} \bar{e}) (d) (d) (\bar{u} \bar{e})$
  
&
  -  
    &  $ (3,1)_{-1/3} $ & $(8,1)_{0}$ & $(\bar{3},1)_{+1/3}$ 

\\
\hline
  5-ii-a 
&
  $(\bar{u} \bar{e}) (\bar{u}) (\bar{e}) (dd)$
  
&
  -  
    &  $ (3,1)_{-1/3} $ & $(6,1)_{+1/3}$ & $(6,1)_{-2/3}$ 

\\
\hline
  5-ii-b  
&
  $(\bar{u} \bar{e}) (\bar{e}) (\bar{u}) (dd)$
  
&
  -  
    &  $ (3,1)_{-1/3} $ & $(3,1)_{-4/3}$ & $(6,1)_{-2/3}$ 

\\
\hline\hline  
\end{tabular}
\vspace{0.5cm}
\\

\begin{tabular}{cccccccccl}

\hline \hline 
T-II  \#  & Op.   & BL \# \ & $S$ \  \ & $S^\prime$ \ \   & $S^{\prime\prime}$    
\\
\hline 


  3  
&
  $(\bar{u} \bar{u})  (dd) (\bar{e}\bar{e})$
  
&
  -  
    &  $ (6,1)_{+4/3} $ & $(\bar{6},1)_{+2/3}$ & $(1,1)_{-2}$ 

\\
\hline
  5  
&
  $(\bar{u} \bar{e}) (\bar{u}\bar{e}) (dd)$
  
&
  -  
    &  $ (3,1)_{-1/3} $ & $(3,1)_{-1/3}$ & $(\bar{6},1)_{+2/3}$ 

\\
\hline\hline  
\end{tabular}
\end{center}
\caption{\it 
\label{Tab:4lp} Decompositions that generate neutrino masses at 4-loop.  
For the decompositions in this table,  the short-range contributions will dominate over the mass mechanism of double 
beta decay.}
\end{table}

\begin{figure}[tbh]
\centering
\includegraphics[width=1.1\linewidth]{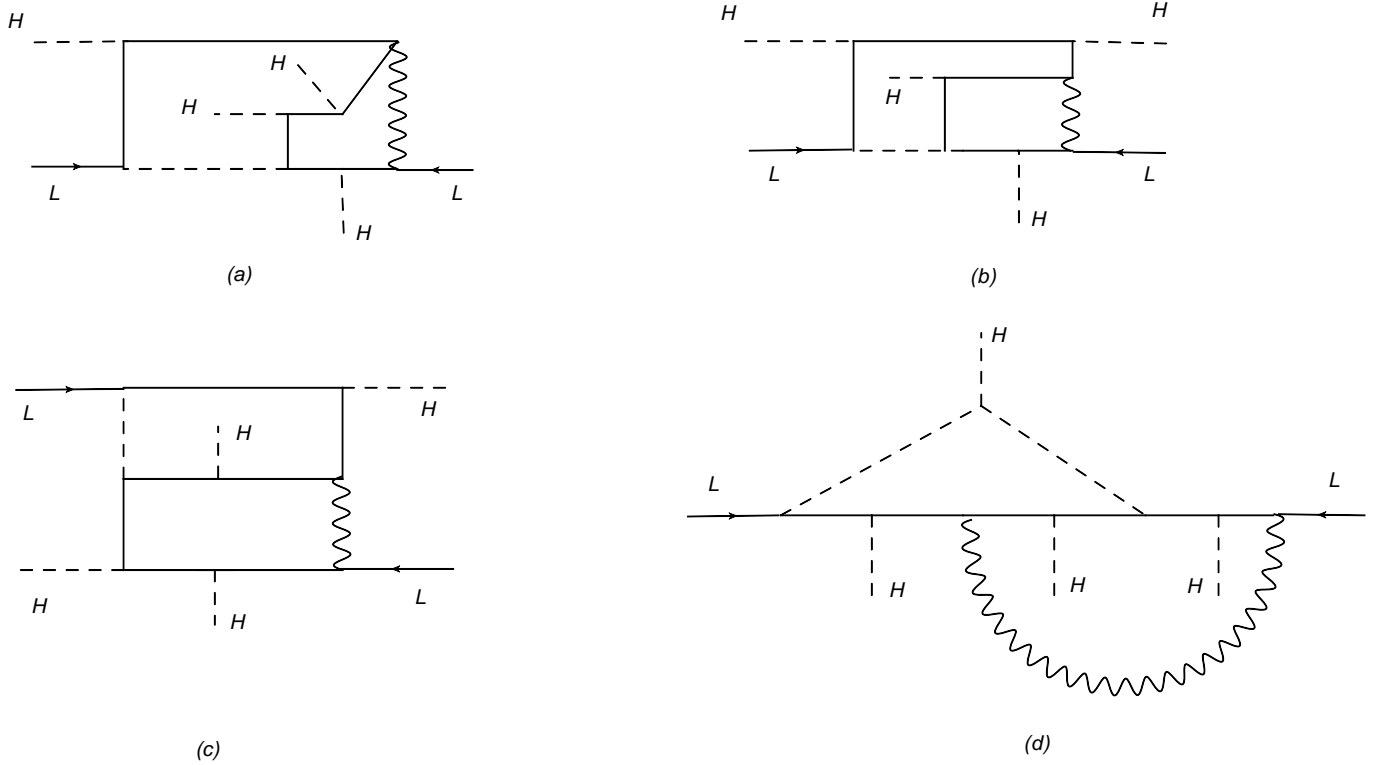}
\caption{ Dimension 7 ($d=7$) neutrino mass diagrams generated by the
  decompositions listed in Table \ref{Tab:2lp-7}. }
\label{fig:2loop-dim7}
\end{figure}

\bibliography{references_0nubb}
\bibliographystyle{h-physrev5}

\end{document}